%% file: main.tex
\PassOptionsToPackage{dvipsnames}{xcolor}
\documentclass[conference]{IEEEtran}
\IEEEoverridecommandlockouts

\usepackage{amsmath,amsfonts}
\usepackage{graphicx}
\usepackage{textcomp}
\usepackage{xcolor}

\usepackage{epsfig,endnotes}
\usepackage{def}
\usepackage{enumitem}
\usepackage{algpseudocode}
\usepackage{multirow}
\usepackage{soul}
\usepackage{tikz}
\usepackage{bm}
\usepackage{comment}
\usepackage{multicol}

\usepackage{algorithmicx}
\usepackage[ruled,linesnumbered]{algorithm2e} % For algorithms
\usepackage{booktabs}
\usepackage{array}
\usepackage{nicefrac}
\usepackage{tablefootnote}
\usepackage{pifont}% http://ctan.org/pkg/{graphicx,pifont}
\let\oldding\ding% Store old \ding in \oldding
\renewcommand{\ding}[2][1]{\scalebox{#1.1}{\oldding{#2}}}% Scale \oldding via optional argument

\usepackage{setspace}

\usepackage{fancyvrb}

\usepackage[hidelinks]{hyperref}
\usepackage{cleveref}

\begin{document}

%%%%%%%%%%%---SETME-----%%%%%%%%%%%%%
\title{Exploring the Efficiency of 3D-Stacked AI Chip Architecture for LLM Inference with \pname{}}
%%%%%%%%%%%%%%%%%%%%%%%%%%%%%%%%%%%%

\author{
\IEEEauthorblockN{Yiqi Liu\IEEEauthorrefmark{1}, Noelle Crawford\IEEEauthorrefmark{1},
Michael Wang\IEEEauthorrefmark{1}, Jilong Xue, Jian Huang\IEEEauthorrefmark{1}}
\IEEEauthorblockA{\IEEEauthorrefmark{1}\textit{University of Illinois Urbana-Champaign}\\
\{yiqiliu2, noellec3, mjwang6, jianh\}@illinois.edu}
\IEEEauthorblockA{xuejilong@gmail.com}
}

%% IEEE copyright notice: fill in after completing the IEEE copyright
%% process on the CPS submission site (the ISBN/exact notice is issued
%% there). Typical form:
%% \IEEEpubid{\makebox[\columnwidth]{978-1-XXXX-XXXX-X/26/\$31.00~\copyright~2026 IEEE \hfill} \hspace{\columnsep}\makebox[\columnwidth]{ }}

\maketitle

\input{abstract}

% \begin{IEEEkeywords}
% \end{IEEEkeywords}

%%%%%%%%%%%%%%%%%%%%%%%%%%%%%%%%%%%%%%%%
%%%%%%%% -- PAPER CONTENT STARTS -- %%%%%%%%%

\input{intro}
\input{background}
\input{motivation}

\input{design}

\input{explore}
\input{related}
\input{conclusion}

\input{appendix}

%%%%%%% -- PAPER CONTENT ENDS -- %%%%%%%%

%%%%%%%%% -- BIB STYLE AND FILE -- %%%%%%%%
\bibliographystyle{IEEEtranS}
\bibliography{refs}
%%%%%%%%%%%%%%%%%%%%%%%%%%%%%%%%%%%%

\end{document}

%% file: abstract.tex
\begin{abstract}
% \vspace{-0.3em}
To overcome the well-known memory bottleneck of AI chips, 3D-stacked architectures that employ advanced packaging technology with high-density through-silicon vias (TSVs) pins have proven to be a promising solution. The 3D-stacked AI chip enables ultra-high memory bandwidth between compute and memory by stacking numerous DRAM banks atop many AI cores in a distributed manner. However, it is not easy to explore the efficiency of the 3D-stacked AI chip, due to its unique distributed nature. And we need to carefully consider multiple intertwined factors that range from upper-level computing paradigm to machine learning (ML) compiler optimizations, and to the underlying hardware architecture.  

In this paper, we develop \pname{}, a fast and compiler-aware end-to-end simulation framework to facilitate exploring the efficiency of 3D-stacked AI chips for large language model (LLM) inference. \pname{} enables the software/hardware co-exploration by employing a programming interface that allows ML compilers to customize the model execution plans. After validating the results of \pname{} with an emulator on real silicon, we thoroughly examine the impact and correlation of different aspects of 3D-stacked AI chips, 
%Our exploration involves multiple dimensions that 
including state-of-the-art compute paradigms, tile-to-core mapping, {tensor-to-bank} mapping, NoC topologies and link bandwidth, DRAM bank bandwidth, per-core SRAM capacity, and energy/thermal constraints. Our findings disclose that the end-to-end efficiency of a 3D stacked AI chip not only is determined by the cooperative function of these factors, but also significantly depends on the mappings from tiles to AI core and DRAM banks. We report our findings in the paper, expecting that they will shed light on the development of 3D-stacked AI chip ecosystem. We open source \pname{}\footnote{\url{https://github.com/yiqiliu2/voxelsim}} for public research.

\end{abstract}


%% file: intro.tex
\section{Introduction}
\label{sec:intro}

The recent development of large language models (LLMs)~\cite{megatron, vllm} exacerbates the memory bottleneck of AI chips~\cite{h100, tpuv4-optical, deepseek-isca, gholami2024aimemorywall}, as LLM inference increasingly demands fast access to an unprecedented volume of data.
% Ingi: these drift a bit, I attempt some revision
To overcome the memory bottleneck, recent initiatives demonstrate that employing 3D packaging technology to stack DRAM banks on top of the logic die is a promising approach~\cite{nvidia3dpatent, amd3dpatent}.
3D architectures use high-density through-silicon vias (TSVs) to achieve scalable bandwidth between compute and memory.
Since pin count in a 3D package scales with chip area rather than perimeter, memory bandwidth scales directly with compute capability, enabling scalable designs.

\begin{figure}[t]
    \centering
    \vspace{-1.5ex}
    \includegraphics[width=.99\linewidth]{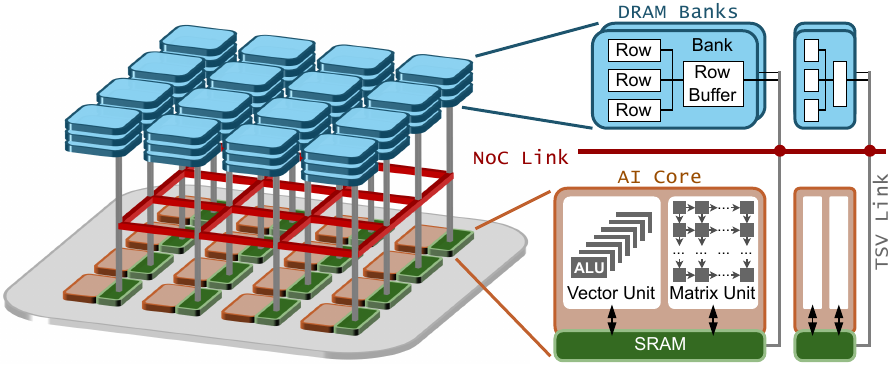}
    \vspace{-3.9ex}
    \caption{A typical architecture of 3D-stacked AI chips.}
    \label{fig:3d-arch}
    \vspace{-1ex}
\end{figure}

A typical 3D-stacked AI chip consists of a layer of AI cores interconnected via a network-on-chip (NoC) layer, topped with a grid of stacked DRAM banks, as shown in Figure~\ref{fig:3d-arch}. 
Each AI core is directly connected to other AI cores via NoC links. 
%Each core can access data from DRAM module(s) directly via TSV links. 
%In this paper, our study focuses on such an architecture and refers to this type of chips as 3D-stacked AI chips or 3D AI chip. 
Unlike AI chips that use 2.5D packaging technology (e.g., H100~\cite{h100} and TPU~\cite{tpu_v4i}), the 3D AI chip has numerous dedicated TSV buses between the compute units and DRAM banks distributed across the chip area. %With this distributed architecture, 

\looseness=-1
Although the 3D AI chip enables ultra-high memory bandwidth between AI cores and DRAM banks, \textit{it requires careful exploration of both software and hardware to maximize the chip efficiency, due to its unique distributed nature} (see $\S$\ref{sec:background}). To achieve the peak memory bandwidth of a 3D AI chip, we need to fully utilize its dedicated DRAM buses and keep all DRAM banks busy.
This requires careful planning of tensor placement and accesses. To fully utilize AI cores while minimizing NoC contention, we need to trade off design factors such as the compute paradigm (e.g., single-program-multiple-data (SPMD)~\cite{alpa,xla}, dataflow~\cite{samba-whitepaper,inter-layer}, compute-shift~\cite{t10,waferllm:osdi2025}), DRAM bandwidth, {core size v.s core count,} NoC topology, link bandwidth, and per-core SRAM capacity (see $\S$\ref{sec:explore}). 

Exploring the expansive optimization space of 3D AI chips requires simulators that consider the complex trade-offs across both the software and hardware stack. 
Unfortunately, no existing open-source simulation tools meet our requirements (see Table~\ref{tab:compare_sim}). Most simulation frameworks for 3D stacked computing target conventional CPU architectures~\cite{kurshan20253d, siddhu2022comet, ProcessorDesignin3D, DesignandManagementof3DMultiprocessors} -- they lack support for AI cores and the essential compute paradigms for parallel tensor executions. State-of-the-art AI chip simulators mostly focus on 2.5D designs and assume a monolithic, uniform memory architecture~\cite{parashar2019timeloop, accelsim, onnxim, npusim, llmcompass}, 
without support for distributed memory. 
%To drive the design space exploration of 3D AI chips, it is essential to develop an end-to-end simulation framework that can accurately reflect the 3D architecture, while enabling the exploration of both hardware and software optimization techniques. 
%However, existing simulators designed for traditional monolithic memory systems, like those in GPUs and TPUs, are ill-equipped to model the complex interplay between hundreds of distributed cores and DRAM modules across a spatial network topology.
% The design space for 3D AI chips is large, and can be represented by many inter-related parameters architects must trade-off between, with the ideal values being dependent on chip budget and the optimization targets. 
% The performance tradeoffs associated with these parameters can be quite complicated to reason about due to the challenges discussed above.
% Existing simulation tools, designed for traditional monolithic memory systems like those in GPUs and TPUs, are ill-equipped to model the complex interplay between hundreds of distributed cores and DRAM modules across a spatial network topology. % from gemini but I like it
% \hl{This results in a gap between...}
%Thus, chip architects lack the tools to effectively explore and evaluate the hardware and software tradeoffs of 3D AI chips.

In this paper, we present \pname{}, a compiler-aware end-to-end simulation framework that can accurately reflect the 3D architecture while enabling co-exploration of hardware and software.
% optimization techniques. 
With a simple yet expressive API,  \pname{} allows ML compilers to specify different compute paradigms and execution plans. Thus, \pname{} includes the performance impact of software optimizations in its results. From the execution plans, \pname{} constructs execution graphs to track the end-to-end progress through an event-driven simulation of all hardware components (e.g., AI cores, DRAM banks, NoC topology/link, and per-core SRAM), while enforcing thermal constraints. To accelerate the simulation of LLMs with thousands of operator tiles, \pname{} caches repetitive memory access patterns (when cores execute repeated operators), and reuses the previously simulated results across structurally equivalent sequences. Since there are no commodity 3D AI chips available, we validate \pname{} with an emulator built on a real AI chip (Graphcore IPU \cite{ipu2}) that has thousands of connected cores and distributed on-chip memory. 
%~\footnote{To the best of our knowledge, there are no commodity 3D AI chips on the market. Therefore, we use an emulator built on a real AI chip to validate the simulation results of \pname{}.}. 
The results of \pname{} are within 6.8\% of the emulation performance for various LLM models (see Figure~\ref{fig:verify_ipu}). We will open source \pname{} to enable further 3D AI chip research in our community.

With \pname{}, we explore the design space of 3D AI chips, and thoroughly analyze the key factors limiting their performance and scalability for LLM workloads. We examine the performance impact of compute paradigms, tile-to-core and tensor-to-bank mapping policies, as well as various design tradeoffs in their core hardware components. We summarize our key findings below, with the expectation that they help the development of the 3D AI chip ecosystem.
% shed light on the development of 3D AI chip ecosystem. 

\begin{itemize}[leftmargin=*]
% \item Compute paradigms are critical to the performance of 3D AI chips, our experiments show that the performance difference between different compute paradigms can reach to 2.12$\times$. An efficient compute paradigm should maximize the overlaps among the tile computation, NoC communication, and DRAM bank accesses. Among the existing compute paradigms, compute-shift performs the best.   
\item Compute paradigms are critical to 3D AI chip performance. We find that the performance difference between different compute paradigms can reach 1.84$\times$. An efficient compute paradigm should maximally overlap tile computation, NoC communication, and DRAM accesses. Among existing compute paradigms, compute-shift performs best (\S\ref{sec:explore:compute-paradigm}).   
%Current popular compute paradigms such as SPMD and dataflow may not work well for 3D AI chips.

% \item Efficient tile-to-core mapping reduces NoC hops and overhead across topologies; combined with a mesh NoC, it achieves near-optimal 3D AI chip performance (\S\ref{sec:explore:comm}).

\item Efficient tile-to-core mapping significantly reduces the number of NoC hops per data transfer. We find that dimension-ordered mapping can minimize the NoC overhead for different NoC topologies. Among popular NoC topologies (mesh, torus, all-to-all), a mesh with dimension-ordered mapping delivers near-optimal performance for 3D AI chips (\S\ref{sec:explore:comm}).

\item Scaling the DRAM bandwidth of a 3D AI chip cannot always benefit end-to-end performance, as it will be more difficult to hide row-buffer conflicts. Tensor-to-bank placement is critical to reduce row-buffer conflicts. Our experiments show that a software-aware placement strategy can reduce performance overhead by up to 80.7\%, compared to a uniform placement strategy (\S\ref{sec:explore:dram-bw}).

\item Simply adding more cores on a 3D AI chip may cause more frequent row-buffer conflicts and exacerbate DRAM bandwidth underutilization. 
% By grouping physically adjacent cores (core groups) and synchronize memory accesses between cores from the same group, we can improve the utilization of both AI cores and DRAM banks.
{To improve utilization of both AI cores and DRAM banks, we group physically adjacent cores into \textit{core groups} and synchronize their DRAM accesses within each group via a hardware tracker.}
The helps scale 3D AI chip designs (\S\ref{sec:explore:flops}).

% \hl{Increasing the per-core SRAM capacity} may not always bring performance benefits to 3D AI chips.
\item For memory-bound workloads like LLM decode, larger SRAM size helps improve overall performance by enabling a larger window of data prefetching. However, the benefit diminishes when DRAM bandwidth is saturated. For compute/NoC-bound workloads like LLM prefill, larger SRAM brings limited performance benefit, as AI cores have reached high FLOPS utilization (\S\ref{sec:explore:sram}).

\item As we scale DRAM bandwidth for memory-bound workloads (e.g., LLM decoding), the energy efficiency of 3D AI chips improves due to significantly reduced workload execution time. However, energy efficiency improvements are limited when scaling AI core count for compute-bound workloads (e.g., LLM prefill), since the increased chip power outweighs the performance improvement (\S\ref{sec:explore:energy}).
%As we scale the DRAM bandwidth in a 3D AI chip, its energy efficiency will be improved when running memory-bound workloads (e.g., LLM decoding), as the workload execution time is significantly reduced. 

\end{itemize}

%For hardware, we study various scalability challenges faced by 3D AI chips; for example, the performance tradeoffs between increasing core size and core count.
%For hardware, we study various scalability challenges faced by 3D AI chips, such as the performance tradeoffs between increasing core size and core count. 
%\hl{thi is the core contributions of this paper, we need to elaborate them well here.}

%Based on our findings, we propose new compiler and hardware optimizations for 3D AI chips. \hl{we should claim compiler/harware co-design.}
%First, we enable \textit{program-aware tensor-to-DRAM-bank mapping} to allocate tensor parts likely to be accessed concurrently to different DRAM banks, reducing the DRAM row conflict overhead.
%Second, we propose the \textit{core group} optimization to enable nearby cores to coalesce their memory requests.
%This coarsens the DRAM access granularity as core count scales, improving the DRAM performance of large 3D AI chips. \hl{for both optimizations, we need to highlight their novelty.}
% We envision \pname{} to be a valuable tool to guide the exploration of the promising new architecture of 3D AI chips.
%We envision \pname{} as a valuable tool to help architects unleash the potential of 3D AI chips. 
% \hl{Need to elaborate this paragraph to highlight our core contributions and insights.}

%\begin{comment}
In addition to our findings, we summarize the technical contributions of this paper as follows: 

\begin{itemize}[leftmargin=*]

\item We identify the efficiency challenges of the 3D AI chip caused by the distributed nature of the architecture (\S\ref{sec:background}).
%\vspace{1ex}

\item We develop \pname{}, a compiler-aware simulation framework for 3D AI chips, which enables hardware/software co-exploration of different design tradeoffs and choices (\S\ref{sec:design}).
%\vspace{1pt}
%\vspace{1ex}

\item We {explore the design} of 3D AI chips in multiple dimensions ranging from compute paradigms to hardware design, and report our findings throughout the paper (\S\ref{sec:explore}).
%\vspace{1ex}
% \vspace{.1pt}

% \item We conduct the design space exploration of 3D AI chips in multiple dimensions that range from upper-level compute paradigms to the lower-level hardware design, and report our findings throughout the paper (\S\ref{sec:explore}).
% \vspace{1pt}

\item We propose a few optimizations to address efficiency challenges of 3D AI chip architecture, including "core group" and software-aware {tensor-to-bank} mapping (\S\ref{sec:explore}).  
%\vspace{1ex}
% \item We propose a few optimization techniques to address the efficiency challenge of the 3D AI chip architecture, including the "core group" technique and the software-aware tile-to-bank mapping scheme (\S\ref{sec:explore}).  
%\item We propose the "core group" idea to scale the 3D AI chip by grouping physically co-located cores via simple crossbar switches and enabling memory coalescing. We demonstrate its scalability benefit in our design exploration (\S\ref{sec:explore:flops}).  
%\vspace{1pt}

\end{itemize}

%% file: background.tex
\section{Background and Motivation}
\label{sec:background}
%We first present the essential technical background of 3D-stacked AI chips and their typical hardware architecture. After that, we investigate their efficiency challenges.  
% As  today's Large Language Models (LLMs) continue to scale in their model sizes \cite{}, they demand larger memory capacity and much higher memory bandwidth in order to finish generation in time.

% For instance, to generate each token, which is one word in a sentence, the accelerator has to read the entire model for multiple times. Thus, for large model inferences, the available memory bandwidth on accelerator usually bottlenecks the minimum possible per-token latency.

% \hl{compare hbm and pim}

\subsection{Development of 3D-stacked Technology}
% \noindent\textbf{LLM inference demands high memory bandwidth.}
% As today's large language models (LLMs) continue to scale in model size, their inference demands higher memory bandwidth than ever.
% To generate each new token, the AI chip(s) need to read through all model parameters of the LLM and all KV cache of the previously generated tokens \cite{}.
% Thus, generating long text with thousands of tokens requires reading all LLM parameters and KV cache from memory thousands of times, which demands very high DRAM bandwidth (e.g., tens of TB/s) for the generation to finish in time.
% However, it is challenging for AI hardware to satisfy the high DRAM bandwidth demand of LLM decoding.

% TODO need numbers -> how big is a tsv / what can density be? what is density of perimiter pins
% What's a normal chip size?

%Over the past years, we have seen the decline of Moore's law as transistor sizes become more difficult to shrink, leading to slower scaling of performance per area. 
%However, emerging workloads demand increasingly capable chips, necessitating new approaches to improve performance. \hl{this sentense is not needed.}
%One approach is the exploration of non-standard packaging techniques such as \textit{3D-stacked integration using through-silicon vias (TSVs)}.

% add small title!!!!!!!!!

% \noindent\textbf{3D-stacked Integration using TSVs.}
3D-stacked integration is an advanced packaging technology that stacks multiple silicon dies vertically to achieve ultra-high bandwidth connectivity~\cite{die-stacking:micro2006,3dstack:isca2008} with Through-Silicon Vias (TSVs). 
TSVs are vertical electrical interconnects fabricated as copper-filled channels encased in insulation layers, with bumps exposed at the die surfaces to form connections.
These bumps occupy minimal area and can be densely packed for fast, high-bandwidth communication.
TSVs' dense packaging provides high connectivity in a compact area, enabling ultra-high bandwidth between 3D-integrated dies.
% Second, unlike traditional 2.5D designs where dies are placed side-by-side and the bandwidth is limited by chip perimeter, 3D integration stacks dies vertically, allowing the number of connections to scale with chip area rather than edge length. 
% They also enables superior bandwidth scalability.
Importantly, the scalability of this bandwidth is also superior.
Traditional 2.5D designs place dies side by side \cite{2.5d}, limiting the connection bandwidth between dies by the die perimeter.
In 3D integration, TSV connections can be provisioned across the die surface, allowing the number of connections to instead scale with die area.
Using current fabrication technology, 3D integration can achieve a bandwidth density of 400 GB/s per 0.02 mm$^2$~\cite{nvidia3dpatent}.

\subsection{3D AI Chip with Stacked DRAM}
\label{sec:background:3d-arch}

% \hl{yiqi: revise by 8/12 ddl 1}

% no need to explain mem wall again!!!!!!!!!!!!!!

%Today's large-scale AI workloads have become increasingly memory-intensive.
%Every time a large language model (LLM) generates a new token, the AI chip must read the model weights and the KV-cache from memory \cite{vllm}, so the performance is highly bound by the memory bandwidth.
To overcome the well-known memory bottleneck of AI chips, recent studies have suggested using 3D technology to scale the memory bandwidth~\cite{deepseek-isca,3d-moe}.
The industry has been exploring new AI chip designs that vertically stack DRAM modules on top of computation dies \cite{nvidia3dpatent, amd3dpatent}. We define this architecture as 3D AI chip in this paper.

\noindent\textbf{Architecture of 3D AI chip.}
We show the architecture of a 3D AI chip with stacked DRAM in \Cref{fig:3d-arch}.
The chip contains a layer of AI cores interconnected by a network on-chip (NoC) layer.
A grid of DRAM banks is stacked on top of the cores and NoC, with multiple layers of DRAM banks to scale capacity.
Each DRAM bank contains numerous rows, whose contents are accessed via the row buffer.
To provide high aggregated memory bandwidth, vertical buses made of densely packed TSV links allow each AI core to access DRAM banks stacked directly on top of it.
Inside each AI core, a fast local SRAM buffers the data from DRAM, a vector unit performs generic vector computations using the data in SRAM,
% \hl{Jian: why it is SIMD? other tensor cores do not work?}
and a matrix unit (e.g., systolic array) handles large matrix multiplication (MatMul) operations at high throughput.
The horizontal NoC links allow an AI core to access data from the local SRAM of other cores.
% However, to optimize the execution performance of this new architecture, we need to consider its unique challenges that are not encountered by conventional AI chips.

% \hl{Change ALL mentions of dedicated DRAM bus to new term.}
\noindent\textbf{Unique features of 3D AI chips.}
%The 3D AI chip enables unprecedented memory bandwidth, but it does so by introducing distinguishing features that differ significantly from conventional 2.5D AI chips.
% These features create new tradeoffs and require careful co-design of hardware and software to maximize performance.
%The two most notable features are the dedicated buses for DRAM banks and the distributed DRAM banks.
Unlike conventional AI chips with 2.5D architecture, the 3D AI chip has two notable features: (1) dedicated TSV buses (i.e., channels) for DRAM banks and (2) the distributed characteristic of DRAM banks.

% One unique architectural feature of the 3D AI chip is its dedicated DRAM buses.
% As illustrated in \Cref{fig:dram-diff} (a), conventional 2D chips have limited pin connectivity, forcing many DRAM banks to share a single data bus. While this shared bus has limited total bandwidth, it is easily kept busy as long as any one of its connected banks is ready to transmit data. In contrast, 3D integration provides massive connectivity, allowing for many more data buses—potentially even a dedicated bus for each DRAM bank. This design offers extremely high peak bandwidth, but achieving it in practice is difficult. Any time a bank is not ready to transmit (e.g., it is activating a new row), its dedicated bus stalls, leading to underutilization.

\begin{figure}[t]
    \centering
    \includegraphics[width=.99\linewidth]{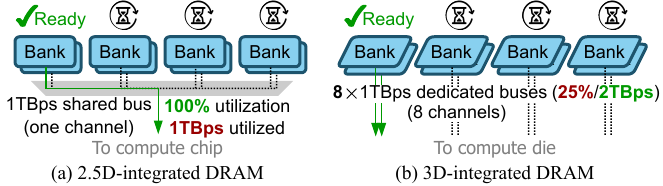}
    \vspace{-3ex}
    \caption{2.5D-integrated DRAM has limited bandwidth but high utilization. 3D-integrated DRAM offers high bandwidth, but may suffer low utilization.}
    \label{fig:dram-diff}
    \vspace{-1ex}
\end{figure}

\begin{figure}[t]
    \centering
    \includegraphics[width=0.93\linewidth]{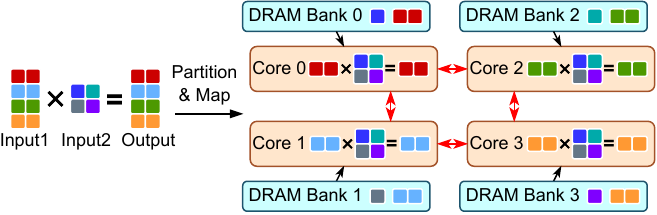}
    \vspace{-1ex}
    \caption{Partition a MatMul operator into 4 tiles and map them to 4 AI cores.}
    % \hl{revise this figure to show how the tiles are distributed across AI cores and DRAM banks, make it specific for 3D AI chip. ADD banks}
    \label{fig:tile}
    \vspace{-1ex}
\end{figure}

% \textit{First, the high connectivity provided by 3D integration forms numerous dedicated data buses between DRAM and compute units.} 
% Unlike 2.5D-integrated design where limited connectivity forces many banks to share a few data buses and therefore provide lower aggregate DRAM bandwidth (\Cref{fig:dram-diff} (a)), the 3D-integrated design provides numerous data buses and each bus is dedicated to one bank, as shown in \Cref{fig:dram-diff} (b).
% It provides high aggregated bandwidth.
% However, to realize this bandwidth, we must keep all buses fully utilized. 
% This is challenging as a dedicated bus will stall whenever its memory bank is not ready or a row-buffer conflict happens. 

\textit{First, the high connectivity provided by 3D integration forms numerous dedicated data buses between DRAM and compute units, which makes bandwidth utilization a new challenge.}
In 2.5D-integrated designs (\mbox{\Cref{fig:dram-diff}} (a)), limited connectivity forces multiple DRAM banks to share a few data buses.
To improve utilization, existing studies exploit inter-bank bus sharing and optimize memory scheduling to maximize bank interleaving \mbox{\cite{threadclustermemoryscheduling,Parallelism-AwareBatchScheduling}}, keeping each bus busy so long as one of its banks is ready to transmit data, even if others banks are not.
% This keeps the bus utilized even when other banks are not ready.
In contrast, 3D-integrated design provides numerous data buses and each bus is dedicated to one bank (\mbox{\Cref{fig:dram-diff}} (b)). 
Although this provides high aggregate bandwidth, it is challenging to fully utilize the bandwidth by keeping every bus busy.
This is because a dedicated bus will stall whenever its memory bank is not ready (e.g., due to an ongoing row-buffer conflict).
And due to insufficient bus sharing, conventional inter-bank bus sharing and scheduling techniques fall short in addressing the bandwidth utilization challenge with 3D-integrated DRAM in AI chips.

\textit{Second, a 3D AI chip distributes its numerous DRAM modules and buses across its area to enable scalable memory bandwidth.}
With this distributed memory architecture, each AI core is connected to the DRAM {banks} directly on top of it via TSVs.
However, during computation, each core may need to access data from a distant DRAM module, depending on the applied computing parallelisms for the deployed AI workloads {(e.g., with tensor parallelism \mbox{\cite{efficient_scale_trans_infer}}, activation inputs and outputs are shared between cores)}. 
As these remote accesses are serviced via the NoC, their latencies vary widely based on the factors like NoC contention, and physical proximity of the core and DRAM module. 
% Without carefully considering these factors, a 3D AI chip may suffer from long data access latencies.
{All these factors significantly impact the data access latencies experienced by 3D AI chips.}
This differs from a 2.5D AI chip \cite{h100,tpu_v4i} whose DRAM modules form a global memory space with near-uniform access latency.
\subsection{Efficiency Challenges with 3D AI Chips}
\label{sec:challenges}

\looseness=-1
\noindent\textbf{{Underutilization of dedicated DRAM buses.}} 
As discussed in $\S$\ref{sec:background:3d-arch}, achieving peak memory bandwidth of a 3D AI chip requires keeping all banks busy to saturate its data buses (Figure~\ref{fig:dram-diff})
and minimizing row-buffer conflicts, which is not easy.

First, a single core's execution will cause frequent row-buffer conflicts depending on (1) tensor placement and (2) tensor access pattern induced by computation.
% a compiler partitions each tensor evenly across all DRAM banks on the chip 
% To place tensors in a way that maximizes the aggregate access throughput of any single tensor, 
To saturate DRAM bandwidth, each tensor should be evenly partitioned and placed across DRAM banks.
% This placement strategy can saturate all DRAM buses when any single tensor is accessed, but it also collocates tiles of all tensors in each bank.
However, this collocates tiles of all tensors in each bank, incurring frequent row-buffer conflicts when multiple tensors are concurrently accessed. %as a bank needs to constantly switch between accessing different tensors.
 % By default, a compiler partitions each tensor evenly across all DRAM banks on the chip to maximize the aggregated access throughput of any single tensor; tiles of different tensors will be mapped to the same bank.
% However, this mapping strategy will map tiles from different tensors to the same bank.
% Accessing different tensors concurrently is very frequent during tensor computation.
For instance, when executing a tensor operator like \textstt{out = matmul(input1, input2)}, a core will repeatedly read input tiles 
% from input tensors and 
and write output tiles.
{While writing}, a core may concurrently prefetch new input tiles, causing row-buffer conflicts.
% \hl{This conflict will worsen if an operator needs more input tensors (e.g., a fused operator).}
% \hl{This effect will worsen as the number of input tensors an operator consumes increases (e.g., after fusion).}
{This worsens as the number of concurrently accessed tensors grows (e.g., a fused operator accessing 3+ inputs).}

Second, inter-core data sharing causes frequent row-buffer conflicts. 
%due to varying communication latencies between the cores and the DRAM bank. %containing the shared data.
In \Cref{fig:tile}, when an AI model is executed, its tensor operators are partitioned into tiles that run on different cores, and tensor data is sharded across DRAM banks.
To compute a tile, each core accesses required tensor parts from corresponding DRAM banks, and multiple cores may share the same data (e.g., ``Input 2'' tensor in \Cref{fig:tile}).
As the DRAM banks storing the shared data are inevitably closer to some cores than others, cores experience different access latencies. This desynchronization can cause interleaved accesses to different rows in a bank, and thus unnecessary row-buffer conflicts.
%The non-uniform accesses to the shared data exacerbate row conflicts 
%\hl{(see our detailed study in \mbox{$\S$\ref{sec:explore:dram-bw}})}.

\begin{table*}[t]
\caption{List of state-of-the-art simulators and their comparison with \pname{}.}
\vspace{-1ex}
\footnotesize
\centering
\setlength{\tabcolsep}{1.35pt}
\begin{tabular}{@{}ccr|cccc|c|ccccc|c|@{}}
\cline{4-14}
\multicolumn{1}{l}{} & \multicolumn{1}{l}{} & \textbf{} & \multicolumn{4}{c|}{\textbf{Generic AI Chip}} & \textbf{GPU} & \multicolumn{5}{c|}{\textbf{Processing-In/Near-Memory Arch.}} & \textbf{3D AI Chip} \\ \cline{4-14} 
\multicolumn{2}{c}{\textbf{}} & \textbf{} & \multicolumn{1}{c|}{\multirow{2}{*}{\begin{tabular}[c]{@{}c@{}}TimeLoop\\ \cite{parashar2019timeloop}\end{tabular}}} & \multicolumn{1}{c|}{\multirow{2}{*}{\begin{tabular}[c]{@{}c@{}}ScaleSim\\ \cite{scalesim3}\end{tabular}}} & \multicolumn{1}{c|}{\multirow{2}{*}{\begin{tabular}[c]{@{}c@{}}ONNXim\\ \cite{onnxim}\end{tabular}}} & \multirow{2}{*}{\begin{tabular}[c]{@{}c@{}}LLMCompass\\ \cite{llmcompass}\end{tabular}} & \multirow{2}{*}{\begin{tabular}[c]{@{}c@{}}AccelSim\\ \cite{accelsim}\end{tabular}} & \multicolumn{1}{c|}{\multirow{2}{*}{\begin{tabular}[c]{@{}c@{}}Neurocube\\ \cite{kim2016neurocube}\end{tabular}}} & \multicolumn{1}{c|}{\multirow{2}{*}{\begin{tabular}[c]{@{}c@{}}NicePIM\\ \cite{wang2023nicepim}\end{tabular}}} & \multicolumn{1}{c|}{\multirow{2}{*}{\begin{tabular}[c]{@{}c@{}}NeuroSim\\ \cite{read2025neurosimv15}\end{tabular}}} & \multicolumn{1}{c|}{\multirow{2}{*}{\begin{tabular}[c]{@{}c@{}}H2-LLM\\ \cite{h2llm}\end{tabular}}} & \multirow{2}{*}{\begin{tabular}[c]{@{}c@{}}\hlcommon{MPU-Sim}\\\hlcommon{\mbox{\cite{xie2023mpu,xie2022mpusim}}}\end{tabular}} & \multirow{2}{*}{\textbf{\pname{}}} \\
\textbf{} & \textbf{} & \multicolumn{1}{l|}{} & \multicolumn{1}{c|}{} & \multicolumn{1}{c|}{} & \multicolumn{1}{c|}{} &  &  & \multicolumn{1}{c|}{} & \multicolumn{1}{c|}{} & \multicolumn{1}{c|}{} &  &  &  \\ \hline
\multicolumn{1}{|c|}{\multirow{3}{*}{\rotatebox{90}{\textbf{Hardware}}}} & \multicolumn{2}{c|}{\begin{tabular}[c]{@{}c@{}}Parameterizable\\Cores/DRAM\end{tabular}} & \multicolumn{1}{c|}{\textcolor{teal}{\ding{52}}} & \multicolumn{1}{c|}{\textcolor{teal}{\ding{52}}} & \multicolumn{1}{c|}{\textcolor{teal}{\ding{52}}} & \textcolor{teal}{\ding{52}} & \textcolor{teal}{\ding{52}} & \multicolumn{1}{c|}{\textcolor{teal}{\ding{52}}} & \multicolumn{1}{c|}{\textcolor{teal}{\ding{52}}} & \multicolumn{1}{c|}{\textcolor{teal}{\ding{52}}} & \multicolumn{1}{c|}{\textcolor{teal}{\ding{52}}} & \textcolor{teal}{\ding{52}} & \textcolor{teal}{\ding{52}} \\ \cline{2-14} 
\multicolumn{1}{|c|}{} & \multicolumn{2}{c|}{Diverse NoC Topologies} & \multicolumn{1}{c|}{\textcolor{teal}{\ding{52}}} & \multicolumn{1}{c|}{\textcolor{teal}{\ding{52}}} & \multicolumn{1}{c|}{\textcolor{teal}{\ding{52}}} & \textcolor{teal}{\ding{52}} & \textcolor{teal}{\ding{52}} & \multicolumn{1}{c|}{\textcolor{purple}{\ding{55}}} & \multicolumn{1}{c|}{\textcolor{purple}{\ding{55}}} & \multicolumn{1}{c|}{\textcolor{purple}{\ding{55}}} & \multicolumn{1}{c|}{\textcolor{purple}{\ding{55}}} & \textcolor{purple}{\ding{55}} & \textcolor{teal}{\ding{52}} \\ \cline{2-14} 
\multicolumn{1}{|c|}{} & \multicolumn{2}{c|}{\begin{tabular}[c]{@{}c@{}}Non-Uniform Access to\\ Distributed DRAM Banks\end{tabular}} & \multicolumn{1}{c|}{\textcolor{purple}{\ding{55}}} & \multicolumn{1}{c|}{\textcolor{purple}{\ding{55}}} & \multicolumn{1}{c|}{\textcolor{purple}{\ding{55}}} & \textcolor{purple}{\ding{55}} & \textcolor{purple}{\ding{55}} & \multicolumn{1}{c|}{\textcolor{purple}{\ding{55}}} & \multicolumn{1}{c|}{\textcolor{purple}{\ding{55}}} & \multicolumn{1}{c|}{\textcolor{purple}{\ding{55}}} & \multicolumn{1}{c|}{\textcolor{purple}{\ding{55}}} & \textcolor{teal}{\ding{52}} & \textcolor{teal}{\ding{52}} \\ \hline
\multicolumn{1}{|c|}{\multirow{8}{*}{\rotatebox{90}{\textbf{Software}}}} & \multicolumn{2}{c|}{Custom Tensor Tiling} & \multicolumn{1}{c|}{\textcolor{teal}{\ding{52}}} & \multicolumn{1}{c|}{\textcolor{teal}{\ding{52}}} & \multicolumn{1}{c|}{\textcolor{teal}{\ding{52}}} & \textcolor{teal}{\ding{52}} & \textcolor{teal}{\ding{52}} & \multicolumn{1}{c|}{\textcolor{teal}{\ding{52}}} & \multicolumn{1}{c|}{\textcolor{teal}{\ding{52}}} & \multicolumn{1}{c|}{\textcolor{teal}{\ding{52}}} & \multicolumn{1}{c|}{\textcolor{teal}{\ding{52}}} & \textcolor{teal}{\ding{52}} & \textcolor{teal}{\ding{52}} \\ \cline{2-14} 
\multicolumn{1}{|c|}{} & \multicolumn{2}{c|}{Custom Compute Paradigms} & \multicolumn{1}{c|}{\textcolor{purple}{\ding{55}}} & \multicolumn{1}{c|}{\textcolor{purple}{\ding{55}}} & \multicolumn{1}{c|}{\textcolor{purple}{\ding{55}}} & \textcolor{purple}{\ding{55}} & \textcolor{teal}{\ding{52}} & \multicolumn{1}{c|}{\textcolor{purple}{\ding{55}}} & \multicolumn{1}{c|}{\textcolor{purple}{\ding{55}}} & \multicolumn{1}{c|}{\textcolor{purple}{\ding{55}}} & \multicolumn{1}{c|}{\textcolor{teal}{\ding{52}}} & \multicolumn{1}{c|}{\textcolor{purple}{\ding{55}}} & \textcolor{teal}{\ding{52}} \\ \cline{2-14} 
\multicolumn{1}{|c|}{} & \multicolumn{2}{c|}{\begin{tabular}[c]{@{}c@{}}Custom Tile-to-Core\\Mapping\end{tabular}} & \multicolumn{1}{c|}{\textcolor{teal}{\ding{52}}} & \multicolumn{1}{c|}{\textcolor{teal}{\ding{52}}} & \multicolumn{1}{c|}{\textcolor{purple}{\ding{55}}} & \textcolor{teal}{\ding{52}} & \textcolor{purple}{\ding{55}} & \multicolumn{1}{c|}{\textcolor{teal}{\ding{52}}} & \multicolumn{1}{c|}{\textcolor{teal}{\ding{52}}} & \multicolumn{1}{c|}{\textcolor{teal}{\ding{52}}} & \multicolumn{1}{c|}{\textcolor{teal}{\ding{52}}} & \textcolor{teal}{\ding{52}} & \textcolor{teal}{\ding{52}} \\ \cline{2-14} 
\multicolumn{1}{|c|}{} & \multicolumn{2}{c|}{\begin{tabular}[c]{@{}c@{}}Custom Tensor-to-Bank\\Mapping\end{tabular}} & \multicolumn{1}{c|}{\textcolor{purple}{\ding{55}}} & \multicolumn{1}{c|}{\textcolor{purple}{\ding{55}}} & \multicolumn{1}{c|}{\textcolor{purple}{\ding{55}}} & \textcolor{purple}{\ding{55}} & \textcolor{purple}{\ding{55}} & \multicolumn{1}{c|}{\textcolor{purple}{\ding{55}}} & \multicolumn{1}{c|}{\textcolor{purple}{\ding{55}}} & \multicolumn{1}{c|}{\textcolor{purple}{\ding{55}}} & \multicolumn{1}{c|}{\textcolor{purple}{\ding{55}}} & \textcolor{teal}{\ding{52}} & \textcolor{teal}{\ding{52}} \\ \cline{2-14} 
\multicolumn{1}{|c|}{} & \multicolumn{2}{c|}{\begin{tabular}[c]{@{}c@{}}\hlcommon{Simulation Speed is Practical}\\\hlcommon{for Full-LLM Simulation}\end{tabular}} & \multicolumn{1}{c|}{\textcolor{teal}{\ding{52}}} & \multicolumn{1}{c|}{\textcolor{teal}{\ding{52}}} & \multicolumn{1}{c|}{\textcolor{teal}{\ding{52}}} & \textcolor{teal}{\ding{52}} & \textcolor{purple}{\ding{55}} & \multicolumn{1}{c|}{\textcolor{purple}{\ding{55}}} & \multicolumn{1}{c|}{\textcolor{teal}{\ding{52}}} & \multicolumn{1}{c|}{\textcolor{teal}{\ding{52}}} & \multicolumn{1}{c|}{\textcolor{teal}{\ding{52}}} & \textcolor{purple}{\ding{55}} & \textcolor{teal}{\ding{52}} \\ \hline
\end{tabular}
\label{tab:compare_sim}
\vspace{-2ex}
\end{table*}

\noindent\textbf{{NoC contention and data transfer overhead.}}
Due to the distributed nature of 3D AI chips, AI cores may need to access shared data from remote DRAM banks or from other cores via the NoC.
The latency of these data transfers depends on both the distance between source and destination, and the NoC congestion. 
When multiple distant NoC transfers occur concurrently, NoC contention will worsen, increasing the data transfer overhead. Our experiments with various LLM workloads show that, without careful planning for data placement and sharing across DRAM banks, NoC congestion will cause up to a 1.3$\times$ performance slowdown (see $\S$\ref{sec:explore:comm}). 
\noindent\textbf{{Underutilization of AI Cores.}}
The compute of 3D AI chips cannot be well utilized without working coordinately with other hardware components, such as DRAM banks, NoC, and per-core SRAM.
When DRAM bandwidth and/or NoC are inefficiently utilized, slow data access and
data transfer may block AI core execution. Similarly, the
compute of AI cores must be carefully configured
to avoid bottlenecking other hardware components. 
%AI cores
%should work coordinately with other hardware components (i.e., DRAM banks, NoC, and per-core SRAM) to maximize the whole system efficiency. 
This further motivates our design
space exploration. 
%\hl{High compute utilization on 3D AI chips depends on the balanced provisioning of many on-chip resources.
%, including DRAM bandwidth, NoC bandwidth, and per-core SRAM capacity. 
%If the DRAM bandwidth or NoC are underprovisioned, they will bottleneck data movement and limit AI core performance.
%However, the core's computational resource must also be appropriately sized -- oversized systolic arrays waste FLOPS, while undersized SRAM increases DRAM overhead. 
%Coordinating the selection of these components is therefore essential to maximize system efficiency.}
We present a quantitative study of these design trade-offs and discuss their implications in \S\ref{sec:explore}.
\vspace{-1ex}

\subsection{Challenges to Maximize the Efficiency}
\label{subsec:challenge}
Exploring the efficiency of 3D AI chips requires a concerted effort between hardware and software. 
%Without considering appropriate software optimizations, the actual capabilities of the hardware cannot be unleashed, leading to misleading conclusions about a design's efficiency.
%Designing chips which avoid the performance pitfalls described above is challenging, as doing so requires a concerted effort between both hardware and software.
%
First, to alleviate row-buffer conflicts ($\S$\ref{sec:challenges}), we should allow software to intelligently map tensors to DRAM banks based on expected access patterns. To maximize the utilization of data buses, ML compilers and 3D AI chip hardware should work coordinately to generate predictable memory access patterns and minimize unnecessary row-buffer conflicts.  
%
%As simultaneous accesses to shared tensors by multiple cores may become staggered due to the non-uniform access latencies on a 3D AI chip, the desynchronization will cause unnecessary bank conflicts,  
%
%when each core concurrently accesses multiple tensors, 
%we want to allow compilers to prevent this access pattern by intelligently mapping tensors to DRAM banks based on the expected tensor access order.
%% to improve the DRAM bandwidth utilization by ensuring minimal row-changes,
%% At the same time, due to the non-uniform communication cost on a 3D AI chip, multiple cores may access a shared tensor from one DRAM bank at different paces, causing an interleaved row access pattern.
%At the same time, simultaneous accesses to a shared tensor by multiple cores may become staggered due to the non-uniform communication cost on a 3D AI chip. 
%This desynchronization will cause interleaved accesses to different rows in the bank, causing unnecessary row-change overhead.
%Thus, we want to carefully examine this inefficient access pattern and explore how to efficiently synchronize accesses to shared data with hardware techniques.
% , to prevent interleaved accesses to multiple rows from thrashing the row buffer.
%
%to minimize the de-synchronization overhead. 
\hlE{Second, to mitigate the NoC overhead, 
%we want to explore various NoC topologies, 
we should allow software to map tiles to cores in a way that minimizes communication distances and traffic for different NoC topologies, and identify suitable compute paradigms (e.g., SPMD \mbox{\cite{gspmd}}, dataflow \mbox{\cite{samba-whitepaper}}, and compute-shift \mbox{\cite{waferllm:osdi2025}}) that can best overlap communication with computation.}
%For instance, the single-program-multiple-data (SPMD) computing paradigam works well for AI chips that organize the memory in a global shared space~\cite{roller,alpa,jax2018github,gspmd}, while the dataflow-based computing paradigm performs well for AI chips that enable direct core-to-core communication via NoC~\cite{t10, samba-whitepaper, inter-layer}.  
%
%For example, many compilers follow the single-program-multiple-data (SPMD) compute paradigm, which shards each operator in the model into multiple parallel sub-tasks and appends a separate communication stage (e.g, AllReduce) to merge the sharded results when necessary \cite{roller,alpa,jax2018github,gspmd}.
%While SPMD works well on an AI chip with a global memory space, some compilers for AI chips with distributed interconnected cores observe that dataflow-style compute paradigms can significantly outperform SPMD on these AI chips \cite{t10, samba-whitepaper, inter-layer}, since these paradigms can better overlap communication with computation.
%Thus, improving the communication performance on 3D AI chips requires a co-exploration between hardware and software factors.
% \hl{maybe instead of citing other chips cite our evaluation}
% \hl{These paradigms allow cores to access tensor data from their peers via the fast NoC instead of accessing it from the slower DRAM, so tensor data will flow across cores. TODO: Need to motivate via NoC contention instead of DRAM being slow}
% above has 3, 4 ^^
%
Third, even with efficient DRAM bank and NoC utilization, achieving high compute FLOPS of AI cores requires coordinated hardware design choices.
Thus, a thorough search of the large hardware design space is essential to maximize the 3D-stacked AI chip efficiency. 
%with minimum hardware resource wastage.
% To reap the memory bandwidth benefits of 3D AI chips, we must ensure long sequential accesses, which requires software to intelligently determine the tensor placement based on the hardware characteristics and the tensor access patterns.
% (1) Software need to smartly map tensor in DRAM banks
% (2) Hardware topology needs to minimize the non-uniform for some core group
% (3) compiler tile-to-core mapping
% (4) compute paradigm
% (5) hardware configuration space exploration
%
% Therefore, maximizing the efficiency of 3D AI chips requires navigating a complex, interdependent design space where hardware and software decisions are deeply intertwined. Consequently, a co-simulation framework that enables rapid exploration of both hardware architectures and appropriate software optimizations is essential for identifying efficient 3D AI chip designs that achieve high performance in realistic settings.
%Therefore, maximizing the efficiency of 3D AI chips requires navigating a complex, interdependent design space where hardware and software decisions are deeply intertwined. 
In summary, 
\textit{without considering appropriate software or compiler optimizations, %performance bottlenecks from generic compilation strategies will mask 
the capabilities of 3D AI chips cannot be fully realized.} 

%% file: motivation.tex
\subsection{Limitations of State-of-the-Art Tools}
%\subsection{Exploring the Design Space of 3D AI Chips}
\label{sec:motivation}

% \hl{ADDRESSED -- it is unclear what is the definition of this section. Are you talking about the related work? or are you talking about the design goals of TSim? It is unclear how existing studies cannot address the challenges you mentioned in Section 2.2}

% \hl{Jian: this section should focus on discussing the limitations of existing approaches.}

% Considering the unique bandwidth benefits of 3D stacked DRAM, AI chip vendors have proposed various 3D AI chip designs \cite{nvidia3dpatent,amd3dpatent}.
% why existing designs cannot achieve good performance
% need a way to fix this -> simulator
% why existing simulators cannot work

% \hl{CHANGE, I moved to previous section: To explore the performance implications of various 3D AI chip designs, architects need efficient and accurate simulation tools.}
% We want to do design space exploration without too much extra work ? can we use existing simulators? no!
% To design efficient, performant 3D AI chips, we need to explore a vast design space of intertwined hardware and software decisions.
While the need for hardware-software co-design of 3D AI chips is clear, existing simulators are insufficient due to incompatible architecture and program assumptions (Table\mbox{~\ref{tab:compare_sim}}). 
%which obscure true chip potential.
% Current simulators face fundamental tradeoffs between...
% Navigating the design space of 3D AI chips is difficult due to the need to consider a vast array of unique hardware and software factors and their interactions.
% Re-purposing existing simulators to explore the design space cannot provide sufficient insights to help us address the challenges described in \S\ref{sec:challenges}.

Most simulation frameworks for 3D integrated circuits target CPU architectures with stacked cores and caches~\cite{kurshan20253d, siddhu2022comet, ProcessorDesignin3D, DesignandManagementof3DMultiprocessors}. 
They focus on general-purpose workloads, and cannot execute large AI workloads. % with massive tensor computations. 
%and are too slow to feasibly simulate the end-to-end performance of large AI workloads with massive tensor computations.
Some PIM \hlcommon{and NMP} implementations share architectural features with 3D AI chips, especially regarding the distributed memory architecture. However, these designs cannot faithfully model the 3D AI chip architecture~\cite{wang2023nicepim,kim2016neurocube,xie2023mpu,xie2022mpusim}. They usually assume a fixed SPMD paradigm and cannot reveal the performance implications of various computing paradigms and other ML compiler optimizations. 
\looseness=-1
\hlcommon{Finally, many simulation frameworks for AI chips assume a monolithic, uniform memory architecture where all cores access a global memory space with near-uniform latency and bandwidth~\mbox{\cite{parashar2019timeloop, accelsim, onnxim, npusim, llmcompass}}}. They cannot track interactions between distributed cores and DRAM banks, and cannot capture performance of 3D AI chips.
% Moreover, many simulation frameworks for AI chips assume a monolithic, uniform memory architecture where all cores access a global memory space with near-uniform latency and bandwidth~\cite{parashar2019timeloop, accelsim, onnxim, npusim, llmcompass}. When applied to 3D AI chips, they cannot efficiently track the interaction between numerous distributed AI cores and DRAM banks, and cannot capture the performance behaviors of 3D AI chips.
% Simulators for these designs \cite{wang2023nicepim,kim2016neurocube} can consider a distributed memory space, but do not support fine-grained software decisions, as they assume a fixed execution model.

% Existing 3D CPU simulators do not support 3D AI chips with stacked DRAM, and are often too slow to feasibly simulate the massive tensor computations found in large AI models.
% While simulators for 3D integrated circuits support fine-grained, flexible simulation, they cannot efficiently simulate large AI workloads.
% They are built on the assumption of a monolithic, uniform memory system where all cores access a global memory with uniform latency and bandwidth. 

% , as they do not model the full system performance of 3D AI chips.

% The limitations of existing simulators necessitate a new simulator designed to accurately and efficiently model the full system performance of 3D AI chips, by taking both their unique hardware characteristics and the effect of software decisions into account.

\looseness=-1
These limitations prevent existing approaches from accurately and efficiently modeling the full system performance of 3D AI chips. Thus, a new simulation framework is needed to capture the unique hardware characteristics of 3D AI chips and the effect of software/compiler decisions on end-to-end performance.

%% file: design.tex
\section{Design and Implementation of \pname{}}
\label{sec:design}

% \subsection{Design Goals of \pname{}}

To systematically explore the design space of 3D AI chips, we propose a new simulator called \pname{}.
Based on the discussion in \S\ref{sec:background}, we set four primary goals for \pname{}:
\begin{itemize}[leftmargin=*]
    \item \textbf{Software awareness:} 
    \pname{} should account for the significant performance impact of software factors such as tiling, mapping, and compute paradigms, such that it can reveal the hardware potential of a 3D AI chip. 
% \pname{} should consider the performance benefit brought by various compiler optimizations.
% On a 3D AI chip with distributed compute, communication, and DRAM access patterns, advanced compiler support and novel compute paradigms can yield significant performance benefits, as noted by existing works \cite{t10,samba-sn40}.
% Thus, to reveal the true potential of a 3D AI chip design, \pname{} needs to consider the impacts of compiler decisions.

    \item \textbf{Accurate behaviors:}
    To accurately model the performance of distributed DRAM and AI cores, \pname{} should capture the performance behaviors of each DRAM bank and AI core as well as NoC architecture at fine granularity. 

    \item \textbf{Fast simulation:}
    % However, the \pname{} should maintain low simulation overhead.
    \pname{} should maintain low simulation overhead to facilitate the rapid evaluation of large chip designs with hundreds of AI cores and DRAM banks.
% Otherwise, we either obtain inaccurate performance results or need multiple weeks to complete a simulation.

    \item \textbf{Reliable performance statistics:} \pname{} should provide reliable results verifiable by experiments with real AI chips.
    % However, since there is no existing 3D AI chip on the market, we need to develop an alternative approach using other AI chips to validate \pname{}.
\end{itemize}

\subsection{\pname{} Overview}

% In \Cref{fig:overview}, we overview the design of \pname{}, whose workflow is listed below:
% \begin{enumerate}[leftmargin=*]
%     \item Given a DL model and an AI chip, we allow a DL compiler to decide how the DL model is partitioned and executed on the chip. 
% Thus, \pname{} can consider the performance benefits brought by compiler optimizations, by taking the compiler-optimized execution plan as input, via \pname{}'s frontend compiler interface (\S\ref{sec:design:interface}).

%     \item Based on the execution plan, \pname{} constructs an execution graph to track the execution progress. Each graph node represents an execution event (e.g., tile compute, data access) of a chip component (e.g., core, DRAM, NoC link), and each edge represents the data dependency between two execution events (\S\ref{sec:design:interface}).

%     \item To simulate the execution, \pname{} \ding{172} obtains each event's latency by querying sub-level simulators that can efficiently capture each chip component's behavior, and \ding{173} inserts new, runtime-generated events into the execution graph (\S\ref{sec:design:components}).

%     \item \pname{} analyzes the execution graph to obtain the simulation results, which include total execution time, hardware utilization, and other performance information (\S\ref{sec:design:together}).

%     \item Finally, to validate \pname{}'s results, we built a hardware emulator using existing AI chips (\S\ref{sec:design:validation}).
% \end{enumerate}

\begin{figure}[t]
    \centering
    \includegraphics[width=.99\linewidth]{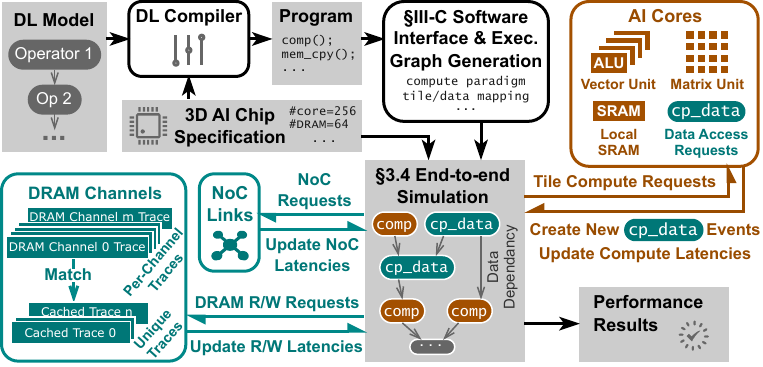}
    \vspace{-3.9ex}
    \caption{System overview of \pname{}.}
    \label{fig:overview}
    \vspace{-1ex}
\end{figure}

We present the design overview of \pname{} in \Cref{fig:overview}.
\textbf{First}, given an AI model, \pname{}'s software interface allows an ML compiler to decide how the model is partitioned, optimized, and executed on the chip. These decisions are used to construct a graph of execution events (\S\ref{sec:design:interface}).
% \textbf{Second}, based on the compiler decisions, \pname{} constructs a graph of execution events to track the execution, enabling \pname{} to consider the effect of diverse software optimizations (\S\ref{sec:design:compiler-optimization}).
% Thus, the construction of \pname{}'s execution graph can be informed by various software optimizations.
% by taking the compiler-optimized execution plan as input via the frontend compiler interface .
% \textbf{Second}, based on the execution plan, \pname{} constructs an execution graph to track the end-to-end simulation of numerous execution events (e.g., tile compute, data access) on different chip components (e.g., core, DRAM, NoC link) (\S\ref{sec:design:graph_gen}).
\textbf{Second}, to enable fast yet accurate simulation of numerous execution events on distributed hardware components, \pname{} creates and executes the events across corresponding hardware components (\S\ref{sec:design:e2e_simulation}).
% \textbf{Third}, to simulate the execution, \pname{} \ding{172} obtains each event's latency by querying sub-level simulators that can efficiently capture each chip component's behavior, and \ding{173} inserts new, runtime-generated events into the execution graph. \hl{yiqi: fast simulation} (\S\ref{sec:design:components}).
% \textbf{Fourth}, \pname{} applies thermal threshold  (\S\ref{sec:design:thermal}).
\textbf{Third}, to ensure that \pname{} produces reliable simulation results, we validate \pname{} with a hardware emulator using state-of-the-art AI chips (\S\ref{sec:design:validation}).
To our knowledge, \pname{} is the first simulator enabling fast and accurate simulation of 3D AI chips with distributed stacked DRAM. It extensively supports software-hardware co-exploration.
% as compared in \mbox{\Cref{tab:compare_sim}}.}
%We show the limitations of prior work in \mbox{\Cref{tab:compare_sim}}.

%\hlJ{Add one paragraph here to summarize the new contributions of your Tsim (address Reviewer A's last question).}

\begin{figure*}
\centering
% \vspace{-0.5ex}
\includegraphics[width=0.9\linewidth]{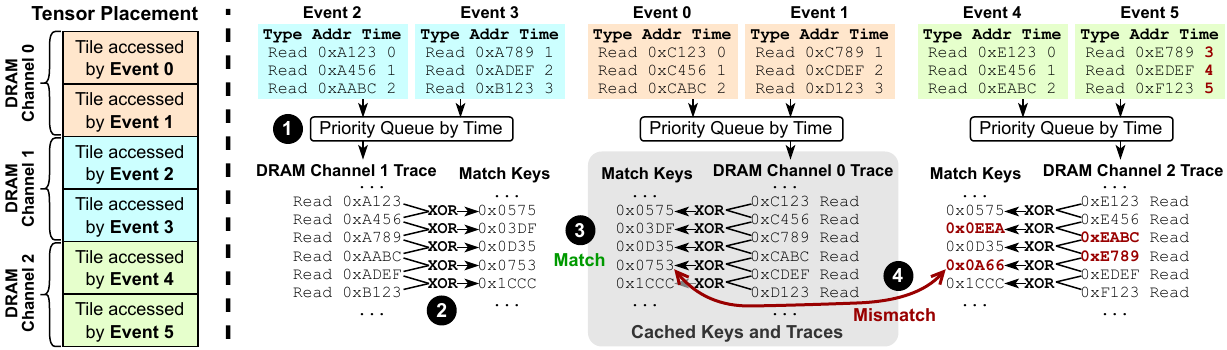}
    \vspace{-1ex}
    \caption{Coalesce identical DRAM access traces across DRAM channels to accelerate the simulation of 3D AI chips.}
    \label{fig:dram_coalesce}
    \vspace{-1ex}
\end{figure*}

\subsection{From Software Interface to Execution Graph}
\label{sec:design:interface}

% 3D AI chips inherently feature an interconnected, distributed architecture and compute pattern, unlocking new opportunities for DL compilers to optimize how tensor data is placed, computed, and moved across the chip (see \S\ref{sec:challenges}).
% Existing compiler works~\cite{t10,roller,inter-layer} show that novel compiler optimizations can bring significant performance benefits for existing AI chips.
%To estimate the true potential of a 3D AI chip design, 
\looseness=-1
For software optimizations, \pname{}'s programming interface allows ML compilers to customize the model partition and execution plan, from which \pname{} constructs an execution-event graph to track execution progress on the simulated 3D AI chip.

\noindent\textbf{Software interface.}
% \label{sec:design:interface-func}
To let ML compilers specify execution plans, we design a flexible interface with three basic functions:

\begin{itemize}[leftmargin=*]
\item \looseness=-1 \textbf{\comp{op\_tile,core\_id}} lets ML compilers specify which operator tile runs on which core.
\textstt{op\_tile} is a partitioned tile of a tensor operator, such as MatMul, elementwise, or a fused operator. 
It also specifies its input/output tensor parts.
Inputs not present in the core's SRAM are accessed on demand from DRAM banks.
\textstt{core\_id} is optional and specifies which core executes \textstt{op\_tile}.

\item \looseness=-1 \textbf{\mvdata{src\_tensor,dest\_tensor}} defines the initial placement or a runtime copy (prefetch) of a tensor part.
% If \textstt{copy\_data} is not used to fetch data, data is to be accessed on demand.
\textstt{dest\_tensor} specifies the data type (e.g., BF16), shape, and location (e.g., a DRAM bank or a core's SRAM) of the tensor part to be placed or copied to.
\textstt{src\_tensor} is either a same-shape tensor part (if runtime copy) or \textstt{NULL} (if initial placement).
% This function places \textbf{\textstt{tensor}} to the \textbf{\textstt{destination}} memory space, where \textbf{\textstt{destination}} can be any specific DRAM module or any core's local SRAM. 
% \textstt{op\_id} is an optional parameter that specifies the first operator that will access \textstt{dest\_tensor}. 
%it is used to avoid mapping concurrently accessed tensors to the same bank of a DRAM bank when applicable.
% \textbf{\textstt{op\_id}} is an optional parameter that specifies the first operator that will access \textbf{\textstt{tensor}}, and it is used to avoid mapping concurrently accessed tensors to the same bank in the \textbf{\textstt{destination}} DRAM module.
% \hl{We will elaborate more on this parameter in the ``\textbf{Tensor-to-DRAM bank mapping}'' paragraph}.

\item \textbf{\texttt{sync()}} synchronizes functions via a barrier.
% All functions called after a \textbf{\textstt{sync()}} must wait until the completion of all functions called before the \textbf{\texttt{sync()}}. 
\end{itemize}

{\pname{} also provides compound functions\footnote{\textftt{allReduce}, \textftt{allGather}, \textftt{broadcast}, \textftt{reduceScatter}, and extensible to more \cite{nvidia_nccl}.} for inter-core collective communication.
\hlE{For example, \mbox{\textstt{allReduce()}} comprises multiple \mbox{\mvdata{}} functions for moving partial results among cores in a ring-based pattern \mbox{\cite{nvidia_nccl}}, and multiple \mbox{\comp{}} functions for reducing partial results locally on each core;}
\mbox{\textstt{reduceScatter()}} is similar to \mbox{\textstt{allReduce()}}, but it copies partial results to a designated core instead of all cores; \mbox{\textstt{allGather()}} copies each input tensor via \mvdata{} to consecutive tiles of the concatenated output.}
% \hlcommon{For collectives we model the ring-based algorithms used by production libraries \mbox{\cite{nvidia_nccl}}, (e.g., ring \mbox{allReduce} and \mbox{reduceScatter}), so their NoC traffic and its overlap with computation are captured at request granularity.}
% \mbox{\textcolor{red}{XXX.}}
% \hl{this interface is core independent, and is compatible with various compilers, we bulid example compute paradigm follow T10 and roller}
% Using the interface functions, the compiler can deploy various optimizations to improve the execution performance, including tile and tensor mapping optimization and different compute paradigms.

The \textbf{interface} gives ML compilers granular control over how a program is mapped to and executed on the simulated device, enabling exploration of the software optimization space, including tile and tensor mapping strategies as well as different compute paradigms and
% \hlB{Because a tile may itself be a fused operator, the same interface also supports kernel-level studies, e.g., comparing a fused kernel against an unfused baseline.}
% In \pname{}, operators may be fused by combining their inputs and computational operations. 
% I, tiles may be fused by combining their inputs and computational operations.
% \hlB{It also enables exploration of kernel fusion optimizations. 
% In \pname{} a tile may itself be a fused operator; compilers can easily express kernel fusions by combining the inputs and computational operations of two tiles. \pname{} can then evaluate the benefits of these optimizations as it provides both end-to-end LLM inference performance, and provides chip-component-level performance breakdowns for each operator.
% } 
\hlB{
kernel optimizations such as kernel fusion.
In \pname{}, a tile can be a fused operator. Compilers can express kernel fusions simply by combining the inputs and computations of two operators.
\pname{} can evaluate the benefits of these optimizations, by providing both end-to-end performance and operator-level breakdowns (see the FlashAttention \mbox{\cite{flashatt}} covered in \mbox{\Cref{fig:op_break}}).
}

\noindent\looseness=-1 \textbf{Generation of execution graphs.}
% \label{sec:design:compiler-optimization}
Based on compiler decisions, \pname{} constructs an execution graph to track end-to-end execution progress.
Each graph node represents an execution event (\comp{}, \mvdata{}, \textstt{sync()}) on a core, DRAM bank, or NoC link. Each edge represents a dependency between events.
%To accurately capture the execution patterns of numerous chip components, \pname{} will traverse the graph to drive the simulation of each chip component and obtain the latency of each execution event.
% To maximize usability, \pname{} can produce a valid execution graph as long as the compiler defines the tile partitioning for all operators through the \textstt{op\_tile} parameter in the \comp{} function.
%
\pname{} produces a valid execution graph so long as \comp{} is invoked for each operator tile,
enabling easy simulation of compiler-produced programs.
At minimum, \mbox{\pname{}} needs (1) a list of operators (parsed from tensor expression \cite{tensor_comprehensions,t10} or StableHLO \mbox{\cite{hlo}}), and per operator, (2) its partitioned tile shape and (3) input/output tensor identifiers.
Thus, \pname{} only needs simple modifications to work with ML compilers like XLA \cite{roller,xla,ansor,t10,tvm,inter-layer} that can specify this information.

\subsection{End-to-end Simulation with Exec. Graph}
\label{sec:design:e2e_simulation}

% maybe try to make it clear why this is a hard thing to do -- what is being addressed
A 3D AI chip has hundreds to thousands of components (e.g., cores and DRAM banks), and an LLM program involves hundreds of operators, and millions of execution events. 
% A 3D AI chip has hundreds to thousands of hardware components (e.g., AI cores and DRAM banks), and an LLM program involves hundreds of operators, yielding millions of total execution events. 
 % it impractical to explicitly simulate every event on every component. 
Explicitly simulating every event is impractical. To reduce simulation time without sacrificing accuracy, \pname{} coalesces identical events and components.
% It is impractical to explicitly simulate every event. To reduce simulation time without sacrificing accuracy, \pname{} coalesces identical events and hardware components.

% To complete the simulation of all events on all chip components within a reasonable time, \pname{} coalesces identical execution events or chip components with identical execution patterns when possible. 

\noindent\textbf{Execution graph traversal.}
% \label{sec:design:graph_gen}
% Based on the execution plan specified by compiler and autocompleted by \pname{}, \pname{} constructs an execution graph to track the end-to-end execution progress.
% In this graph, nodes represent execution events (\comp{}, \mvdata{}, and \texttt{sync()}), and directed edges represent data dependencies between events.
% \pname{} traverses this graph to drive the simulation of each chip component and obtain the latency of each execution event.
% To simulate execution events on their corresponding chip components while respecting data dependencies,
\looseness=-1
\pname{} traverses the event graph chronologically, issuing each event to its designated component %as soon at the earliest time
as soon as its data dependencies have resolved.
% simulates the end-to-end execution of an AI model by traversing the graph of execution events.
% Each event is issued to its component at the earliest time when its data dependencies have resolved.
% to capture the execution \hl{behavior/flow} of each chip component. 
% It 
% During use the dependency edge connections to enforce inter-event data dependencies.
% To track the execution of numerous events on the chip components and consider the inter-event data dependencies, 
% To track the data dependencies of numerous events and the event executions on numerous chip components.
% To track the execution of events on numerous chip components while respecting the data dependencies between events...
% \pname{} issues execution events from the graph to the corresponding chip components chronologically, based on the earliest time at which all previous events it depends on are completed.
% \pname{} issues execution events from the graph to the corresponding chip components ordered by each event's start time, where the start time of an event is the time point when all its previous events with data dependencies have completed. 
A \comp{} is issued to an AI core, while a \mvdata{} is issued to the NoC, and if applicable, the corresponding DRAM channel.
After processing all events, \pname{} reports end-to-end performance statistics. % of the AI workload execution.

\noindent
\textbf{AI core simulation.}
% All \comp{} events will be routed to AI cores, which will estimate the computation costs of the tiles.
% An AI core will check if the tile's required tensors are ready in the core's local memory.
% If the compiler did not schedule a prefetch in \S\ref{sec:design:interface}, \pname{} creates new \mvdata{} events to load data from DRAM.
%Inside an AI core, 
\pname{} adapts the implementations of an existing NPU simulator~\cite{scalesim3} to simulate an AI core's vector unit, matrix unit, and SRAM.
% When a core executes a tile \comp{} whose inputs are not present in its local SRAM, \pname{} immediately sends a sequence of data access requests for the complete input tile, ordered by their consumption time in the computation, and routes the requests to the appropriate components, at the rate of one request per cycle.
When a core executes a \comp{} without inputs present in local SRAM, \pname{} creates concurrent \mvdata{} events that access the input tiles,
% following the data consumption order
and routes the events to appropriate hardware components.
\hlE{Each AI core executes one \mbox{\comp{}} at a time}.
% following the data consumption order during computation 
%ordered by their consumption time in the computation, ordered by their usage in the computation/operator.
\pname{} reuses computation costs of identical tiles.
% with identical shapes, given that an ML model often has many identical operators partitioned into identical tiles. 
% \hl{core to core communication -- TODO}

\noindent
\textbf{NoC simulation.}
% A 3D AI chip leverages its fast NoC links to transmit the data access requests and move tensor data across the chip.
The NoC handles core-to-core and core-to-DRAM communication.
For each data transfer, \pname{} determines the involved links;
% based on the user-specified routing policy (e.g., negative/positive first, XY, and odd-even policy \cite{xy-oddeven}).
\hlE{if at any time multiple transfers take the same link, each will experience extended latency as the link bandwidth is shared.}
% Then, following the roofline model \cite{roofline}, \pname{} computes the transfer's required number of cycles based on the transfer volume, available bandwidth, and number of NoC hops.
\pname{} computes the transfer overhead using the volume, available bandwidth, and number of NoC hops.
For core-to-core communication, we assume the NoC bandwidth is lower than SRAM read bandwidth. 
\hlC{For TSVs, \pname{} uses an existing 3D IO timing model \mbox{\cite{cacti3dd}} to estimate TSV path delay based on the wire characteristics.}
%therefore, the transfer latency depends solely on the NoC.  
For core-to-DRAM communication, \pname{} reports overall performance using both NoC and DRAM simulations.
% \hlcommon{Because concurrent transfers that share a link split its bandwidth, and concurrent requests at a channel are merged into a single per-channel queue \mbox{(\Cref{fig:dram_coalesce}\ding{182})}, \mbox{\pname{}} captures the fine-grained contention between concurrent execution-graph events rather than simulating each event in isolation.}
% For core-to-DRAM communication, the NoC simulation determines the overall latency in conjunction with the DRAM simulation.

\noindent
\looseness=-1\textbf{Distributed DRAM simulation.}
% enqueue based on order of serving? -> 
% can either be issued via graph
% or can be transmitted over noc (added to queue at end of noc) 
On a 3D AI chip, a DRAM channel contains one or more banks that share a TSV bus.
When a \mvdata{} event arrives at a DRAM channel, it is decomposed into a series of memory requests, each with an address, operation type (read or write), and arrival timestamp.
Each request accesses one DRAM burst. % (128 bytes in our evaluation).
% Arrival time is defined as the sum of the time at which the request was sent and the NoC latency it experienced.
% As each DRAM bank will serve access requests from multiple cores and multiple \mvdata{} events, \pname{} needs to determine the order of DRAM access requests before sending them to a DRAM simulator \cite{ramulator2} instance. 
\hlE{Since a channel may serve overlapping requests concurrently, \mbox{\pname{}} maintains a per-channel priority queue of outstanding requests ordered by arrival time, as shown in \mbox{\Cref{fig:dram_coalesce} \ding{182}}.
If multiple requests have the same arrival time, they are ordered by event index.}
% To determine the order of requests from multiple overlapping events, each DRAM bank in \pname{} maintains a priority queue of outstanding DRAM requests ordered by their arrival times.
\hlC{\pname{} uses the DRAM simulator Ramulator2 \mbox{\cite{ramulator2}} to simulate arrived requests, configuring it to provision more DRAM channels with fewer banks per channel, accounting for more TSVs (\mbox{\Cref{fig:dram-diff}}).} % in the queue. % and simulate DRAM activities.
% \hlcommon{By default, we use FR-FCFS policy and an all-bank, fixed-interval refresh. While all above options are user-configurable, there is one soft exception in refresh policy: an adaptive refresh policy may lead to extended simulation time, as it requires \pname{} to partially disable its caching approach as introduced below.}
% Arrival time is the sum of the AI core's request issuance timestamp and the NoC latency experienced to reach the module.
% To do so, \pname{} maintains a priority queue of outstanding DRAM requests ordered by arrival time at the DRAM bank.
% Arrival time is defined as the sum of the time when an AI core sent the request and the NoC latency experienced by the request.
% Then, the DRAM simulator will return the latency of each access in the queue.

% By default, to serve all requests on $m$ DRAM banks, \pname{} needs to maintain $m$ DRAM simulator instances, one for each bank.
% This is because each DRAM bank on a 3D AI chip has a dedicated data bus, and supporting the hardware-isolated buses requires one simulator instance per bus, using existing DRAM simulators \cite{ramulator2}. 
% However, if \pname{} sends all DRAM access requests for all DRAM banks to the DRAM simulator, we need to maintain $m$ DRAM simulator instances for a 3D AI chip with $m$ DRAM banks (by default, $m$=16 in our evaluation).
% To best capture the performances of all DRAM banks 
% To accurately capture the non-uniform behaviors of distributed DRAMs, \pname{} should simulate all requests generated during the execution of an AI workload.
Simulating at request granularity allows \pname{} to accurately capture the non-uniform behaviors of distributed DRAM.
However, simulating every request explicitly is prohibitively expensive:
a 30 billion parameter BF16 precision LLM produces $\approx$500 million DRAM reads assuming a 128-byte DRAM burst width. %Individually 
% -- reading the weights of a medium-sized LLM with 30 billion parameters at BF16 precision produces around 500 million DRAM reads, assuming a 128-byte DRAM burst width. %Individually 
% Simulating all the requests would take weeks on a modern 64-thread server.
This would take weeks on a modern 64-thread server.
% Thus, \pname{} reduces DRAM simulation overhead by (1) coalescing DRAM banks with similar input traces, and (2) reusing .
% To accurately reflect the DRAM overhead of executing the memory-bandwidth-bound decode stage of an LLM, we must  
% For a medium-sized LLM with around 30 billion parameters (30B), the DRAM simulator needs to serve around 70 million distinct DRAM reads in total, assuming the interface width of DRAM is 128 bytes.
% In this case, the end-to-end simulation may take weeks to finish.,
Thus, \pname{} accelerates DRAM simulation by exploiting repetitive access patterns (\Cref{fig:dram_coalesce}).
%access large tensor chunks or
% In \Cref{fig:dram_coalesce}, we show how \pname{} identifies repetitive access patterns.
% Two sequences of DRAM requests are considered to have identical access patterns if the corresponding requests in each sequence have matching operation types and address bits (excluding the DRAM bank and module selection bits).

From a channel's trace, \pname{} produces a list of \textit{match keys} by computing the bit-wise XOR of each request's address with its predecessor's, as shown in \Cref{fig:dram_coalesce} \ding{183}.
Each match key encodes row and column transitions by tracking which bits change between requests. %, including their magnitude.
This is sufficient information to simulate DRAM behavior, 
since intra-channel DRAM behavior depends solely on the pattern of bank, row, and column changes, not on the specific rows or columns accessed.
Thus, two traces with identical match key lists, such as in \Cref{fig:dram_coalesce} \ding{184}, exhibit identical timing, enabling \pname{} to reuse cached results across structurally equivalent access patterns.

When the current and reference traces mismatch (\Cref{fig:dram_coalesce} \ding{185}), \pname{} tags the divergent requests and a window of $N$ surrounding requests, where $N$ is the DRAM internal queue size ($N=32$ by default in \pname{}).
This covers the performance impact of divergent requests on their neighbors.
Once all requests are examined, \pname{} simulates all tagged requests in the DRAM simulator
% For each contiguous tagged block, the first $N$ requests warm up the DRAM state.
and uses the results to update the latencies of the remaining requests in the block. 
\pname{} reuses cached latency results for non-tagged requests.
% \hl{When simulating a 256-core chip that runs models that each has 40-80 repeated layers, only 0.09\% of requests cannot reuse cached results.}
%
The cache hit rate is usually high. For example, 
when running an LLM with 40 repetitive layers on a 3D AI chip with 64 DRAM channels, each access pattern repeats 40$\times$64=2,560 times.
Within the access of each tile on each channel, the pattern will repeat as the tile is scanned through.
%Multiplying them together, the maximum hit rate can far exceed $10^4$.
Our experiments with LLM workload traces show a hit rate of 99.91\%.
% (i.e., each result is reused $\approx$1100 times).
% \hlD{Because the cache reuses results across structurally identical layers while still re-simulating divergent requests, simulating all layers is nearly as cheap as simulating a single representative layer, yet avoids the inaccuracy of blindly extrapolating from one layer; the hit rate stays comparably high within each layer.}

\hlD{We further use the hit rate as an indicator of how many identical layers we should simulate in a model.
Specifically, as identical layers usually experience similar execution cost, we do not have to repeatedly calculate all their costs to get an accurate full-model estimation.
However, it is also insufficient to derive only the cost of one layer (e.g., the first one) and multiply by number of layers, since the performance of a subsequent layer depends on its predecessor (e.g., the predecessor may allow or not allow an operator fusion or DRAM prefetch opportunity that is cross layer boundary).
Thus, to safely simulate fewer layers, we want \pname{} to automatically detect if the per-layer execution pattern reaches a steady state, and we observe that the hit rate is a good indicator.
The hit rate is comparatively low for the first layer, but extremely close to 100\% at the last few layers.
As high hit rate directly indicates repeated data access and execution pattern, we allow \pname{} to skip remaining identical layers after the hit rate reaches 99.9\%.
Thus, \pname{} only needs to simulate 3-4 identical layers in most cases, almost halving the total simulation time compared to simulating the full model.
% However, the reduction in total simulation overhead is much less significant than the reduction in number of layer simulated.
% This is because each skipped layer would only introduce a small simulation overhead even if not skipped, since its simulation expects to have extremely high hit rate.
Note that this optimization is not compatible with those emerging models \mbox{\cite{deepseek-isca,agarwal2025gpt}} that employ heterogeneous layers.}

% \hlD{Because the cache reuses results across layers while still re-simulating divergent requests, simulating all layers does not incur significant overhead compared to simulating a single layer. 
% While all layers are structurally identical in our experiments, we find it necessary to simulate them explicitly, since each layer depends on the timings of its immediate predecessor, which can introduce small differences. In particular, the first layer has no predecessor, so it has no opportunity to prefetch data, unlike later layers. The effects of this startup propagate through the next few layers before reaching a steady state.
% Additionally, many emerging models \mbox{\cite{deepseek-isca,agarwal2025gpt}} employ heterogeneous layers with different attention mechanisms and FFN types (i.e., dense vs. MoE). Simulating all layers means \pname{} can support these architectures natively.}

%In our experiments, only 0.009\% of requests experience cache misses and thus need to be fully simulated.

% For the simple example in \Cref{fig:dram_coalesce}, when DRAM banks 0 and 1 have received identical request sequences, \pname{} computes the corresponding match keys. If the keys are identical, \pname{} reuses the latency outputs.

While cached results can capture the performance of most DRAM operations, they cannot capture the impact of DRAM refreshes. Thus, \pname{} tracks the memory address range that is currently undergoing a refresh \hlE{based on a static refresh policy by default}. If an incoming request targets an address that matches the active refresh set, \pname{} {delays the request to the end time of the ongoing refresh.}
% \textcolor{red}{how penalty applied?}
\hlE{Note that, for adaptive refresh policies, they may incur altering refresh patterns and disrupt trace similarity. Thus, they may decrease the efficiency of our caching mechanism.}

  \begin{table}[t]
  \footnotesize
  \centering
  \caption{\hlC{Key parameters used for our thermal simulation.}}
  \label{tab:thermal-params}
  \vspace{-1ex}
  \begin{tabular}{@{}>{\raggedright\arraybackslash}p{0.4\linewidth}p{0.5\linewidth}@{}}
  \toprule
  Parameter & Value \\
  \midrule

  Cooling $R_{\mathrm{convec}}$
    & 0.12 K/W~\cite{ats_cooling_high_power} \\

  Ambient temperature
    & 25$^\circ$C~\cite{ashrae2021thermal_guidelines} \\

  Stack
    & Logic + 8 HBM dies \\

  Die thickness, thermal conductivity $k_d$
    & 50 $\mu$m logic~\cite{lau2010critical_tsv, chien2012thermal},
      30 $\mu$m DRAM~\cite{skhynix2021hbm3};
      100 W/mK~\cite{meng2011thermal3d,chalise2023dissertation} \\

  Underfill thickness, thermal conductivity $k_u^\dagger$
    & 10 $\mu$m~\cite{miwa2020ncf7um,imaps2016ncf,colgan2012microbumps};
      2 W/mK~\cite{chalise2023dissertation} \\

  \bottomrule
  \end{tabular}

  \vspace{0.35ex}
  \begin{minipage}{0.95\linewidth}
  \footnotesize
  $^\dagger$ 2 W/mK models the effective through-plane conductivity of the
  combined TSV/underfill layer, not the intrinsic conductivity of pure polymer
  underfill. We perform thermal modeling with underfill between each layer.
  \end{minipage}

  \vspace{-1ex}
  \end{table}

\noindent\textbf{Power and thermal modeling.}
\label{sec:design:thermal}
Stacking DRAM above compute provides substantial bandwidth benefits, but can exacerbate thermal issues.
% by DRAM, SRAM, NoC, and compute power within a single package footprint.
\hlcommon{To enable realistic simulation, \mbox{\pname{}} can track thermal effects at two fidelity levels.
The first is fine-grained spatial-temporal thermal simulation for detailed thermal studies.
The second is a lightweight power-density threshold throttling execution when stacked power density exceeds a user-set limit, enabling rapid design-space exploration.}
% users can calibrate this limit against the fine-grained simulation.

Both modes use the power and area data generated by \pname{}.
For power traces, \pname{} records each chip component's activity start time, duration, and energy for every execution event during simulation.
% component-level
Per-component energy comes from models for the AI core~\cite{scalesim3}, DRAM~\cite{regate}, SRAM~\cite{regate}, and NoC~\cite{kahng2009orion}.
For area, \pname{} estimates each component's footprint using established models for AI cores~\cite{llmcompass}, SRAM~\cite{openram}, NoC~\cite{pat-noxim}, and TSVs~\cite{jiao2024tsvdensity}. These cost models are exposed via a modular interface, so users can easily use other models. \hlE{For instance, a more recent NoC power model 
% such as \mbox{DSENT}
\mbox{\cite{sun2012dsent}} could be substituted. Since NoC power is only a small fraction of the total energy consumption \mbox{(\Cref{fig:energy-breakdown})}, this choice does not affect the study results.}
\looseness=-1\hlcommon{In terms of the component areas and the 3D chip specification, \mbox{\pname{}} constructs a per-die floorplan and layer configuration.
% , such as the one in \mbox{\Cref{fig:peak_temps}}.
Power traces, floorplan, and layer configuration are exported as standard \texttt{.ptrace}, \texttt{.flp}, and \texttt{.lcf} files that are readily consumable by most external thermal simulators.}

\hlcommon{For fine-grained thermal studies, \mbox{\pname{}} integrates dedicated thermal simulation backends.
By default, it supports \mbox{HotSpot~\cite{hotspot}} and \mbox{3D-ICE~\cite{3d-ice}}, and exposes an interface for integrating other thermal solvers.
These detailed solvers can model effects such as lateral heat spreading, vertical thermal coupling across dies, and package/cooling boundary conditions.
We use this detailed simulation to validate the power-density threshold used for our design-space exploration in \mbox{\S\ref{sec:explore}}. 
The high-fidelity simulation may also be useful for users specifically interested in thermal behavior of 3D AI chips. }
% Studies addressing the interactions between performance and thermal behavior (e.g., different cooling solutions, material stacks, underfill thicknesses, floorplans, or thermal-aware data placement policies) may use the HotSpot/3D-ICE backend directly.}

% \begin{table}[t]
% \footnotesize
% \centering
% \caption{\hlC{Key parameters used for our thermal simulation.}}
% \label{tab:thermal-params}
% \vspace{-4ex}
% \begin{tabular}{p{0.55\linewidth}p{0.4\linewidth}}
% \toprule
% Parameter & Value \\
% \midrule

% Cooling $R_{\mathrm{convec}}$
% & 0.12 K/W~\cite{ats_cooling_high_power} \\

% Ambient temperature
% & 25$^\circ$C~\cite{ashrae2021thermal_guidelines} \\

% Stack
% & Logic + 8 HBM dies \\

% Die thickness, thermal conductivity $k$
% & 50 $\mu$m logic~\cite{lau2010critical_tsv},
%   30 $\mu$m DRAM~\cite{skhynix2021hbm3};
%   100 W/mK~\cite{meng2011thermal3d,chalise2023dissertation} \\

% Underfill thickness, thermal conductivity $k$\tablefootnote{$2$ W/mK models the effective through-plane
% conductivity of the combined TSV/underfill layer, not the intrinsic conductivity of pure polymer underfill. We
% perform thermal modeling with underfill between each layer.}
% & 10 $\mu$m~\cite{miwa2020ncf7um,imaps2016ncf,colgan2012microbumps};
%   2 W/mK~\cite{chalise2023dissertation} \\

% \bottomrule
% \end{tabular}
% \vspace{-2.5ex}
% \end{table}

For fast, realistic performance evaluations, \pname{} allows users to specify the chip's maximum power density.    
During simulation, \pname{} tracks the aggregate power density of chip components. For events exceeding the threshold, 
% During simulation, \pname{} tracks the aggregate power density of stacked chip components, and identifies events that exceed the power density limit.
% \pname{} slows down their execution by adjusting the frequency of AI cores by the ratio of exceeded power over the maximum power, and reports the updated execution time. 
\pname{} uses frequency scaling to slow their execution 
% by reducing the AI cores' frequency 
by the ratio of exceeded power to maximum power, and report the updated time. 
Since this paper focuses on architectural and compiler optimizations, we use this approach to reduce simulation overhead. 
\hlcommon{For our evaluation, we set a conservative default power density limit of $0.7~W/mm^2$ 
to ensure the hardware operates within a safe thermal envelope below $85^\circ C$, and thereby avoid temperature-induced DRAM refresh penalties that would complicate the DRAM performance pattern \mbox{\cite{jedec2020hbm}}.} 

% For performance-oriented evaluations such as ours, which focus on architecture and compiler-level optimization opportunities, the threshold-based approach can be employed to reduce simulation overheads.
\begin{figure}
    \centering
    \vspace{-1ex}
    \includegraphics[width=0.95\linewidth]{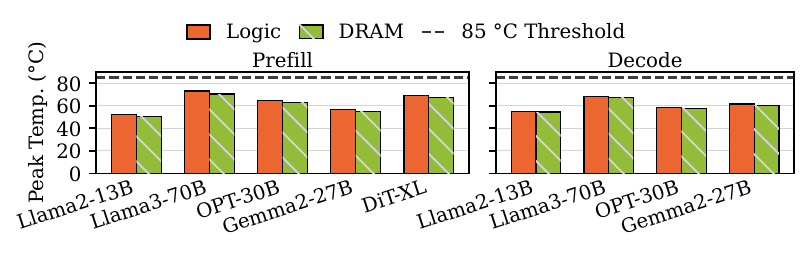}
    \vspace{-2.5ex}
    \caption{\hlcommon{Peak temperatures of default 3D AI chip configuration with \mbox{ 0.7W/$\text{mm}^2$} power density threshold.}}
    \label{fig:peak_temps}
    \vspace{-1ex}
\end{figure}

\hlcommon{For the selected chip configurations (\mbox{\Cref{tab:parameter_overview}}), we validate our power density threshold using the 3D-ICE thermal simulation backend with the parameters shown in \mbox{\Cref{tab:thermal-params}}. 
With the power density threshold in place, the peak logic and DRAM temperatures remain below 73 and 70 degrees Celsius, respectively, as shown in \mbox{\Cref{fig:peak_temps}}.}

%\hlJ{Reviewer E.2: Add one paragraph to discuss "put it all together" and go through the workflow of your simulator.}

\noindent\textbf{\hlE{Putting it all together.}} \hlE{We summarize our entire simulation workflow as follows: 
First, an ML compiler translates an AI model into a sequence of device-specific tiled operations. 
These computations are then mapped to \mbox{\pname{}} primitives (i.e., \mbox{\comp{}}, \mbox{\mvdata{}}, and \mbox{\textstt{sync()}} events), and assembled into an execution graph which represents inter-primitive dependencies.
\mbox{\pname{}} then traverses this graph. For each event, it uses the per-component execution models to obtain the corresponding event latency.
For dependent events, they either execute sequentially and can be simulated individually, or they execute in an overlapped manner and require back-to-back simulation.
For example, when a \mbox{\comp{}} accesses DRAM data via a \mbox{\mvdata{}}, \pname{} executes these per-component simulators together.
% (\mbox{\S\ref{sec:design:e2e_simulation}})
When events occur concurrently on one hardware component (e.g., two \mbox{\mvdata{}} events hit one DRAM bank), \mbox{\pname{}}'s simulation models reflect the resulting resource contention as specified in \mbox{\S\ref{sec:design:e2e_simulation}}.
Throughout execution, \mbox{\pname{}} continuously monitors the energy consumption of each hardware component, producing a power trace that can be used for thermal simulation and to enforce a power-density limit.}

\begin{figure}[t]
    \centering
    \includegraphics[width=.95\linewidth]{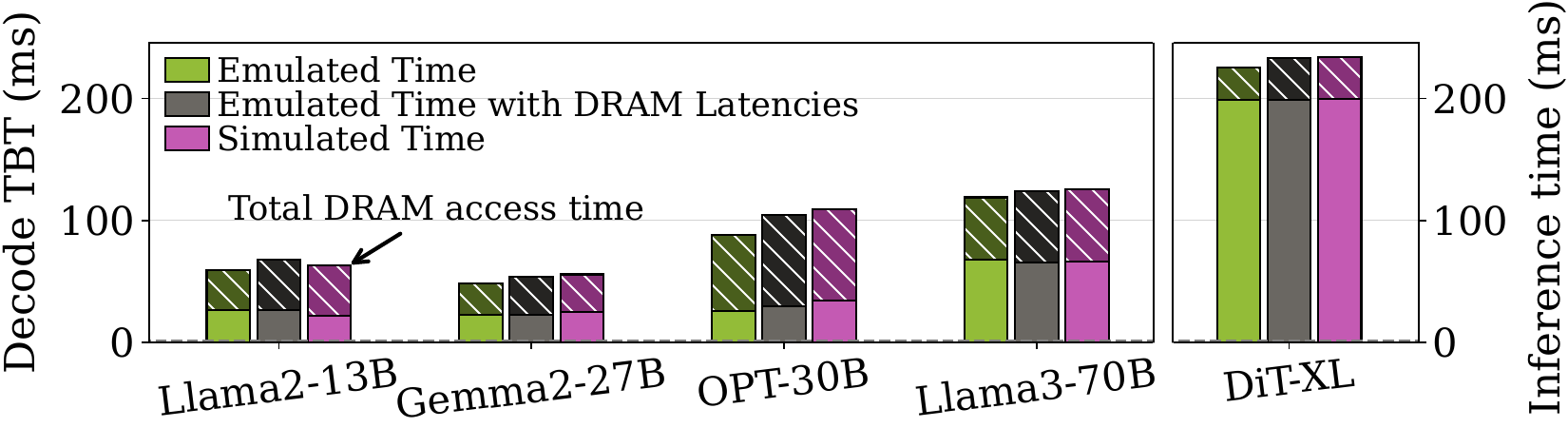}
    \vspace{-1ex}
    \caption{Validation of \pname{} on real chip. We compare \textbf{Emulated Time} -- execution time on a real IPU chip by loading ``DRAM data'' from SRAM, \textbf{Emulated Time with DRAM} -- replayed IPU execution traces with added DRAM latency, and \textbf{Simulated Time} with \pname{}.
    % \hl{\mbox{\yiqi{TODO:Breakdown}}}
    {Shaded parts of each bar shows the time when DRAM is busy accessing data.}}
    \label{fig:verify_ipu}
    \vspace{-1ex}
\end{figure}

\subsection{Simulator Validation}
\label{sec:design:validation}

% \hl{to validate we nned chip with ... based on our resaech we find IPU}

% \hl{yiqi: write by 8/14 ddl 2}
To ensure the reliability of \pname{}, we cross-validate its results with real silicon.
% However, although 3D AI chips with stacked DRAM represent a promising approach for breaking the memory bandwidth wall, and the industry has proposed various designs \cite{nvidia3dpatent,amd3dpatent}, 
There are no commercially available 3D AI chips, thus, we build an emulator from the Graphcore IPU \cite{ipu2}, an existing AI chip.
% A suitable emulator chip needs interconnected cores to allow emulation of distributed compute patterns and a distributed memory space to resemble the distributed stacked DRAM on a 3D AI chip.
% Candidate emulator chips should meet at least two criteria.
% First, its cores should be interconnected to allow emulation of the distributed compute patterns on a 3D AI chip.
% Second, the chip should have distributed memory spaces with ultra-high bandwidth, which resemble the distributed stacked DRAM on the 3D AI chip.
%Finally, we must be able to access the chip.
% We find the Graphcore IPU \cite{ipu2} to be the most suitable: 
An IPU Mk2 chip fully connects 1,472 cores with an interconnect of 7.8 TB/s total bandwidth, 
% NOTE: IPU interconnect is all-to-all. The 3D ring is applied between multi-IPU, which is not our case.
and the per-core SRAMs form a distributed memory space with 62 TB/s bandwidth and 896 MB capacity in total, which can store an entire LLM tensor operator.

% \hl{mention we compare decode because NoC latency is negligible}
To emulate the execution of a 3D AI chip, we use 960 cores as ``AI cores''. \hlcommon{Since the IPU chip does not have the distributed DRAM, we use its distributed SRAM scratchpads from the remaining 512 cores to emulate low-latency distributed ``DRAM banks''. We use the default software-aware tensor-to-DRAM bank mapping strategy (see \mbox{\Cref{tab:parameter_overview}}) to assign which addresses are served by each emulated bank.}
% (see \S\ref{sec:explore:dram-bw})
%\hlcommon{and use the remaining 512 cores to emulate low-latency distributed ``DRAM banks'' with 2.7TB/s of total bandwidth.}
% \hl{explain noc can't overlap things on IPU}
The emulated AI cores then access data over the NoC from the assigned ``DRAM banks'' at burst granularity.  
%Because the SRAM capacity is much lower than the actual DRAM capacity and cannot fit a full model, dummy data is returned instead of real model weights.
% \hl{bc entire model doesn't fit on chip, and Mkii doesn't have HBM connection}
% The 736 cores acting as DRAM can emulate a total memory bandwidth of 4 TB/s.
We configure \pname{} to match the hardware specifications of the emulator chip {(e.g., hardware parameters and overlap constraints)} and compare its reported performance (Simulated Time) with that of the emulator (Emulated Time).
In \Cref{fig:verify_ipu}, the performance results of the emulator are 12.7\% faster than those of \pname{} on average, because the SRAM emulating DRAM reaches full bandwidth utilization with any data access pattern.
% \hlcommon{explain no sram bank conflicts}
\hlcommon{
To address this, for each model, we extract the core transformer block that is repeated in the model, and run its full DRAM trace on the DRAM simulator \mbox{\cite{ramulator2}}.
We intentionally disable the coalesce technique depicted in \mbox{\Cref{fig:dram_coalesce}} to ensure the current DRAM simulation relies only on the previously validated simulation \mbox{\cite{ramulator2}}.}
% We intentionally disable the coalesce technique depicted in \mbox{\Cref{fig:dram_coalesce}} to ensure the DRAM simulation relies only on previously validated .
\hlcommon{Specifically, we first run the hardware emulator to obtain the memory access trace, which reports the traffic between ``AI cores'' and ``DRAM banks'', and then run the trace in the DRAM simulator to obtain the DRAM access latencies.}
\hlC{To account for the stacked DRAM, we use the simulator to obtain the internal access latency of the DRAM die, and add the TSV transfer latency calculated using the timing model \mbox{\cite{cacti3dd}}.}
% \hl{Thus, for each model, we run the original non-coalesced DRAM trace of one transformer block that is repeated in the model on the DRAM simulator \mbox{\cite{ramulator2}}.}
\hlcommon{And then, we rerun the emulator to apply DRAM latencies by inserting a delay before returning data for every ``AI core''-to-``DRAM Bank'' access, where the delay is the difference between the simulated DRAM time and the original SRAM access time.
We show that their performance results (Emulated Time with DRAM Latencies) match closely with \mbox{\pname{}} in both the total time and the DRAM access time breakdown}.
The error rate is 0.24\%--6.8\% for the evaluated models.

\hlcommon{To further validate 
% \mbox{\pname{}}, we carefully examine 
the new components (i.e., AI core and NoC) simulated in \mbox{\pname{}}, we validate the compute and communication time with hardware profiling traces from the emulator.  
%To validate the individual new components modeled by \pname{} (i.e., the AI core and NoC model), we further examine the communication and computation time from the emulator's hardware profiling trace. 
% We find that these latencies match well with those of the corresponding \mbox{\comp{}} and \mbox{\mvdata{}} events in \pname{}.
% This validation is designed to isolate the AI-core and NoC models from other sources of simulator error. 
To compare the AI core and NoC models against the emulator in isolation, we decompose the emulator trace into periods dominated by local computation and periods dominated by inter-core data movement. 
We then compare these measured intervals with the corresponding \mbox{\comp{}} and \mbox{\mvdata{}} events generated by \pname{} for the same execution. 
We find that the emulated and simulated execution times match well. This further validates that \mbox{\pname{}} can accurately capture the behaviors of 3D AI chip components.
% Unlike end-to-end runtime validation, this event-level comparison reduces the possibility that errors in one component are hidden by compensating errors in another. 
}

\setlength{\tabcolsep}{3.6pt}
\begin{table*}[t]
\vspace{-.5ex}
\caption{A summary of design decisions investigated in our study.}
\vspace{-1ex}
\footnotesize
\begin{tabular}{c|c|c|c|}
\cline{2-4} 
& \textbf{Name} & \textbf{Description} & \textbf{Default Value} \\ \hline

\multicolumn{1}{|c|}{\multirow{3}{*}{\textbf{\begin{tabular}[c]{@{}c@{}}Software\\Factors\end{tabular}}}} 
& Computation Paradigm & The per-core compute and inter-core communication patterns used by the compiler. & Compute-shift \\ 
\cline{2-4} \multicolumn{1}{|c|}{} 
& Tile-to-Core Mapping & How each partitioned operator tile is mapped to each specific core. & Dim.-ordered \\ 
\cline{2-4} \multicolumn{1}{|c|}{} 
& Tensor-to-DRAM Bank Mapping & How each partitioned tensor tile is mapped to each specific DRAM bank. & Software-aware \\ \hline

\multicolumn{1}{|c|}{\multirow{7}{*}{\textbf{\begin{tabular}[c]{@{}c@{}}Hardware\\Factors\end{tabular}}}} 
& NoC Topology & The topology of the on-chip inter-core interconnect. & 2D mesh \\ 
\cline{2-4} \multicolumn{1}{|c|}{} 
& DRAM Bandwidth & Total DRAM bandwidth on the 3D AI chip. & 12 TB/s \\ 
\cline{2-4} \multicolumn{1}{|c|}{} 
& Number of Cores & Total number of cores on the 3D AI chip. & 256 \\ 
\cline{2-4} \multicolumn{1}{|c|}{} 
& Systolic Array Size & The width of the systolic array on each core. & 32 \\ 
\cline{2-4} \multicolumn{1}{|c|}{} 
& Core Group Size & Core count of each group (cores in the same group can coalesce their DRAM accesses). & 8 \\ 
\cline{2-4} \multicolumn{1}{|c|}{} 
& NoC Link Bandwidth & The bandwidth of each NoC link. & 16 B/cycle \\ 
\cline{2-4} \multicolumn{1}{|c|}{} 
& Per-Core SRAM Size & The capacity of the SRAM scratchpad on each core. & 2048 KB \\ \hline
\end{tabular}
\label{tab:parameter_overview}
\vspace{-2ex}
\end{table*}

%% file: explore.tex
\begin{table}[t]
\caption{Other design parameters configurable in \pname{}.}
\vspace{-1ex}
\footnotesize
\setlength{\tabcolsep}{0.5pt}
% \begin{tabular}{|c|c|c|c|c|}
% \cline{1-2} \cline{4-5}
% \textbf{Parameter} & \textbf{Default} & \textbf{} & \textbf{Parameter} & \textbf{Default} \\ \cline{1-2} \cline{4-5} 
% DRAM layer count & 8 &  & DRAM Capacity & 192 GB \\ \cline{1-2} \cline{4-5} 
% \multirow{2}{*}{\begin{tabular}[c]{@{}c@{}}Number of DRAM\\banks per layer\end{tabular}} & \multirow{2}{*}{16} &  & \begin{tabular}[c]{@{}c@{}}Freq. (DRAM \& AI core)\end{tabular} & 1.6 GHz \\ \cline{1-2} \cline{4-5} 
% \multirow{2}{*}{\begin{tabular}[c]{@{}c@{}}DRAM timing\\ (tCL-tRCD-tRP-tRAS)\end{tabular}} & \multirow{2}{*}{14-14-14-34} &  & Power Density Limit & 0.7W/mm$^2$ \\ \cline{4-5} 
%  &  & \multicolumn{1}{l|}{} & \hlcommon{Batch size (Prefill-Decode)} & \hlcommon{1-32} \\ \cline{1-2} \cline{4-5} 
% DRAM interface size & 128 B &  & Sequence length & 2048 \\ \cline{1-2} \cline{4-5} 
% DRAM burst length & Vary with BW &  & Precision & BF16 \\ \cline{1-2} \cline{4-5} 
% \end{tabular}
\begin{tabular}{@{}|c|c|c|c|c|@{}}
\cline{1-2} \cline{4-5}
\textbf{Parameter} & \textbf{Default} & \textbf{} & \textbf{Parameter} & \textbf{Default} \\ \cline{1-2} \cline{4-5} 
DRAM layer count & 8 &  & DRAM Capacity & 192 GB \\ \cline{1-2} \cline{4-5} 
\multirow{2}{*}{\begin{tabular}[c]{@{}c@{}}Number of DRAM\\banks per layer\end{tabular}} & \multirow{2}{*}{16} &  & \begin{tabular}[c]{@{}c@{}}Freq. (DRAM \& AI core)\end{tabular} & 1.6 GHz \\ \cline{4-5} 
 &  &  & Power Density Limit & 0.7W/mm$^2$ \\ \cline{1-2} \cline{4-5} 
\multirow{2}{*}{\begin{tabular}[c]{@{}c@{}}DRAM timing\\ (tCL-tRCD-tRP-tRAS)\end{tabular}} & \multirow{2}{*}{14-14-14-34} &  & \hlA{Batch size (Prefill-Decode)} & \hlA{1-32} \\ \cline{4-5} 
 &  & \multicolumn{1}{l|}{} & Sequence length & 2048 \\ \cline{1-2} \cline{4-5} 
DRAM interface size & 128 B &  & Precision & BF16 \\ \cline{1-2} \cline{4-5} 
DRAM burst length & Vary with BW &  & \hlE{DRAM Scheduling Policy} & \hlE{FR-FCFS} \\ \cline{1-2} \cline{4-5} 
\end{tabular}
\label{tab:other_parameter}
\vspace{-2.5ex}
\end{table}

\setlength{\tabcolsep}{8pt}
\begin{table}[t]
\caption{Area breakdown of the chip's bottom die.}
\vspace{-1ex}
\footnotesize
\centering
\begin{tabular}{|c|c|c|c|c|}
\cline{1-2} \cline{4-5}
\textbf{Component} & \textbf{Total area}             &  & \textbf{Component} & \textbf{Total area}             \\ \cline{1-2} \cline{4-5} 
Systolic arrays     & 260mm$^2$  &  & TSVs                & 18.4mm$^2$ \\ \cline{1-2} \cline{4-5} 
SRAMs               & 433mm$^2$ &  & Other        & 91.2mm$^2$  \\ \cline{1-2} \cline{4-5} 
\end{tabular}
\label{tab:area}
\vspace{-1ex}
\end{table}

\newcounter{mycounter}
\newcommand\showmycounter{\stepcounter{mycounter}\themycounter}

\section{Exploring the Efficiency of 3D AI Chips}
\label{sec:explore}
We use \pname{} to explore the performance impact of various software and hardware design decisions on a 3D AI chip.
We list the design decisions we explored in \Cref{tab:parameter_overview}, 
and the default values of \pname{}'s other configurable parameters in \mbox{\Cref{tab:other_parameter}}. We break down the default chip configuration's area in \Cref{tab:area}.
% \hlE{For our experiments, we use an FR-FCFS scheduling policy and static refresh in the DRAM controller.}
We simulate the prefill and decode of 4 different LLMs (Llama2-13B \cite{llama2}, Gemma2-27B \cite{gemma}, OPT-30B \cite{opt}, and Llama3-70B \cite{grattafiori2024llama3herdmodels}), and the inference of a vision transformer (DiT-XL \cite{dit}).
% Each model can fit into the DRAM of a default 3D AI chip.
% \hlA{The prefill and decode latency represents the two most important LLM service performance metrics experienced by LLM customer: the Time to First Token (TTFT) and the Time Per Output Token (TPOT).}
In our tests, the model weights always fit in the 3D AI chip's DRAM.
% also not sure if this is going in the right direction -- maybe 
% Stratum~\mbox{\cite{3d-moe}} targets 3D-stacked MoE serving.} -- probably don't say this
% The model execution plans are optimized following the SOTA approach for each tested compute paradigm \cite{welder:osdi23,t10, inter-layer}.
For evaluated compute paradigms, model execution plans are optimized using SOTA approaches \cite{welder:osdi23, t10, inter-layer, samba-whitepaper}.

\begin{figure}[t]
    \centering
    \vspace{-0.5ex}
\includegraphics[width=0.99\linewidth]{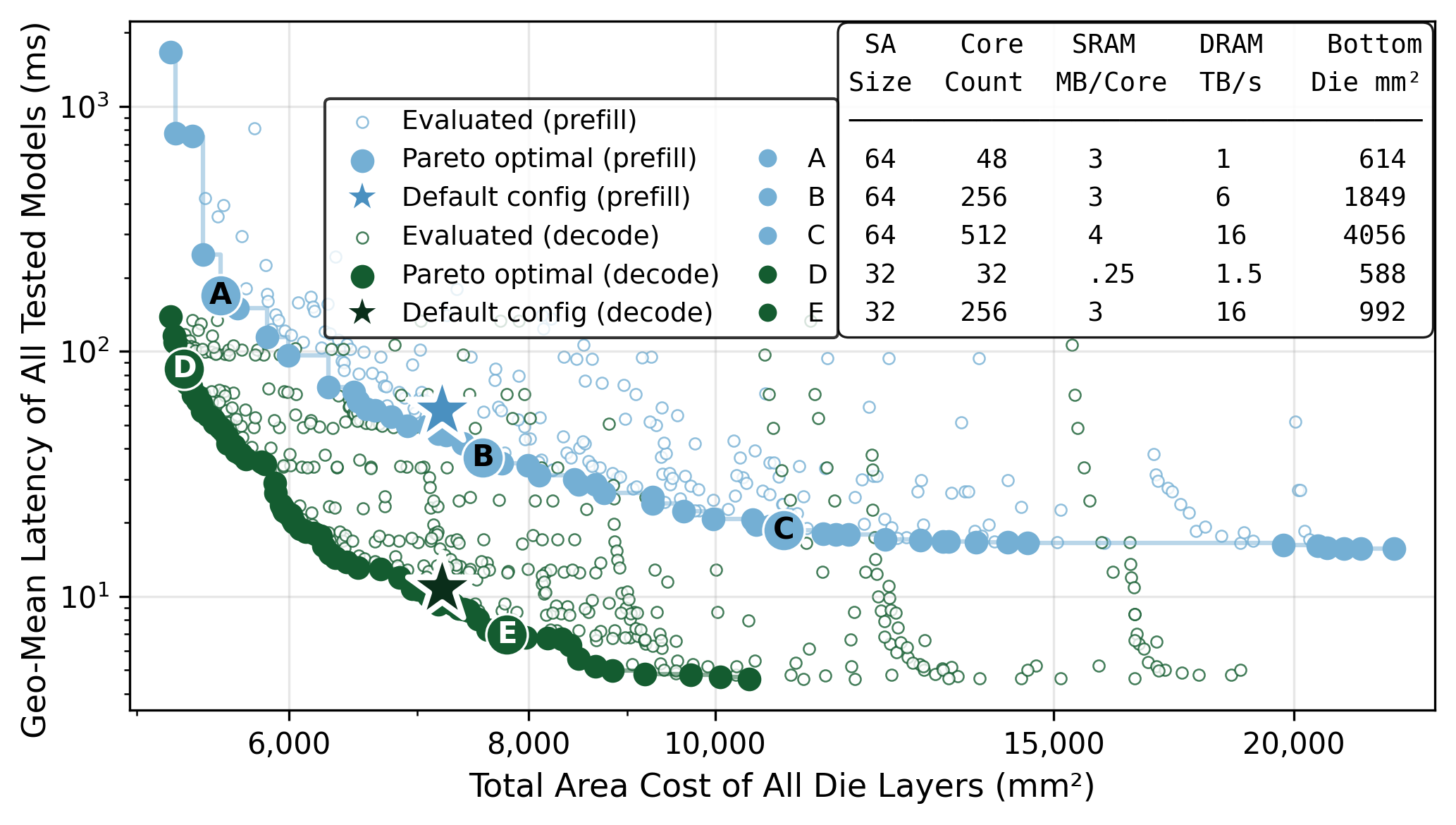}
    \vspace{-5ex}
    \caption{Pareto frontier of chip configurations with lower total area cost and higher performance (prefill and decode).}
    \label{fig:pareto}
    \vspace{-1ex}
\end{figure}

\noindent\textbf{Exploration approach.}
{The low simulation overhead of \pname{} allows us to rapidly evaluate the expansive hardware design space. \mbox{\Cref{fig:pareto}} demonstrates the ability of \pname{} to identify
% this capability by identifying
the Pareto area-performance frontier for LLM serving via a multi-level area-constrained coordinate descent search,
% To efficiently approximate the Pareto frontier for this design space, 
in which we discretize the area constraint into multiple geometric thresholds, and then use
% At each threshold, we use 
coordinate descent \mbox{\cite{wright2015coordinatedescentalgorithms}} to minimize the geometric mean of AI models' latencies.}

% The algorithm changes one architectural parameter each time (e.g., SA size or core count)
% , while warm-starting subsequent levels to accelerate convergence.}

% CHANGE(Ingi):
% To break down insights and showcase innovations uncovered from our exploration, we pick a default configuration (marked with stars) as our baseline. 
% REWORDED:
To highlight insights and discoveries from exploration, we select default prefill and decode configurations as baselines (marked with stars on \Cref{fig:pareto}).
Baseline configurations respect an 850mm$^2$ per-die area limitation\mbox{~\cite{asml_nxe3600d}}
% The configuration respects a 850mm$^2$ per-die area limitation\mbox{~\cite{asml_nxe3600d}}
% Area seems to not match with the figure?
% We deliberately prioritize decode performance, 
and prioritize decode performance, since memory-bound LLM decoding is the primary domain of 3D AI chips. {As we provision a high 12TB/s DRAM bandwidth, this configuration does not lie on the Pareto optimal curve for prefill.}

\noindent\hlA{\textbf{Exploration overview.} Because our compiler optimizations enable significant overlaps among the executions of individual hardware units in \mbox{\pname{}}, we break down 
the overhead of each hardware component throughout our exploration.
We begin with the NoC, examining its overhead under different compute paradigms (\mbox{\S\ref{sec:explore:compute-paradigm}}) and tile-to-core mapping strategies (\mbox{\S\ref{sec:explore:comm}}).
We next explore DRAM, analyzing the DRAM row-buffer conflict overhead and the overall DRAM access overhead across different tensor-to-bank placement policies (\mbox{\S\ref{sec:explore:dram-bw}}).
We then study how core count and systolic array size affect serving latency (\mbox{\S\ref{sec:explore:flops}}). We also study how SRAM size affects prefill and decode latency (\mbox{\S\ref{sec:explore:sram}}).
Next, we break down the energy consumption of each hardware unit (\mbox{\S\ref{sec:explore:energy}}).
Finally, \mbox{\S\ref{sec:explore:operator-breakdown}} reports how individual LLM operators contribute to prefill and decode latency.}

\subsection{Performance Impact of Compute Paradigms}
\label{sec:explore:compute-paradigm}

\begin{figure*}[t]
    \centering
    % \vspace{-0.5ex}
    \includegraphics[width=0.95\linewidth]{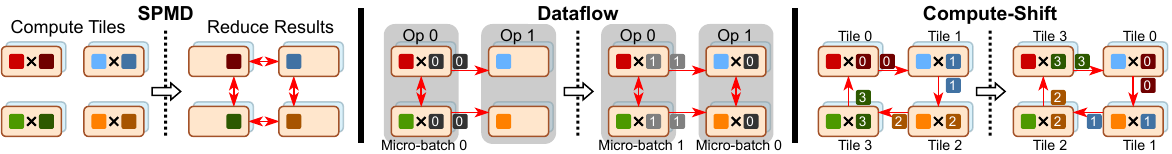}
    \vspace{-1.5ex}
    \caption{Three representative compute paradigms explored in \pname{}.}
    % \hl{map them to 3D AI chip.}}
    \label{fig:comp_para}
    \vspace{-2ex}
\end{figure*}

\begin{figure}[t]
    \centering
\includegraphics[width=.95\linewidth]{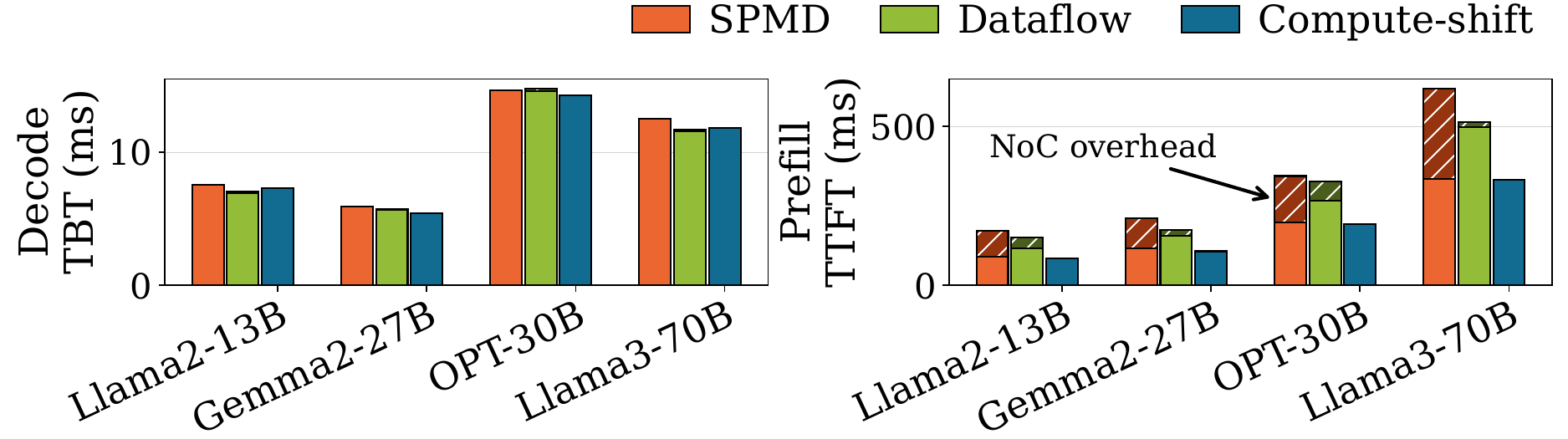}
    \vspace{-2ex}
    \caption{LLM serving latencies with different compute paradigms. Shaded parts represent communication overhead.}
    % \hl{legend is confusing, use an arrow or something instead for NoC overhead, also need to fix y-axis for decode}}
    \label{fig:sw_compiler}
    \vspace{-2ex}
\end{figure}

As discussed in \S\ref{sec:design:interface}, \pname{}'s software interface enables the deployment of custom compute paradigms, 
% Compute paradigms %(\Cref{fig:comp_para})
which impact 3D AI chip performance via different overlap opportunities.
% s they enable different overlap opportunities (e.g., compute-communication or compute-prefetch). 
%or per-core tile size.
\looseness=-1 We examine three representative compute paradigms (\Cref{fig:comp_para}).
% \Cref{fig:sw_compiler}).
% We implement LLM decode and prefill using the three representative compute paradigms (\Cref{fig:comp_para}), and explore their impact in \Cref{fig:sw_compiler}.
% Different paradigms impact performance by enabling different compute-communication overlap opportunities, per-core tile sizes, and core utilization.

First, the conventional SPMD paradigm \cite{welder:osdi23,alpa,gspmd,ansor,xla,wscllm} divides each tensor operator into many independent tasks across cores, each producing a partial result.
\hlE{
Once produced, partial results are combined in a separate reduction step.
%Once all partial results have been produced, they are combined via a separate reduction step.
We model the compute and reduction phases as non-overlapping 
%for the SPMD execution model, 
for SPMD execution, 
following the bulk-synchronous model used by many existing SPMD frameworks \mbox{\cite{nccl,gspmd,jax2018github}}.}
The reduction step incurs high NoC overhead
--- up to 47\% of the total SPMD execution time (\Cref{fig:sw_compiler}) ---
which can block %distributed execution on 3D AI chips,
execution, especially during % compute-intensive LLM % the compute-intensive LLM
prefill.
%The reduction frequently bottlenecks the distributed execution on 3D AI chips, especially during LLM prefill, which causes high, non-overlappable NoC overhead.
% \hl{Michael: changed wording} 
% First, the conventional SPMD compute paradigm \cite{welder:osdi23,alpa,jax2018github,gspmd,ansor,xla} divides tensor operators into many independent tasks, which are distributed across all cores. Cores produce partial results and combine them in a separate reduction step, which  frequently bottlenecks the distributed execution on 3D AI chips, especially during LLM prefill, which causes high, non-overlappable NoC overhead.
% In \Cref{fig:sw_compiler}, NoC overhead constitutes up to 31\% of the total SPMD execution time.
%NoC overhead can be up to 31\% of the total SPMD execution time (\Cref{fig:sw_compiler}).
% the SPMD execution is bottlenecked by the NoC during up to \hl{X}\% of the total execution time.

% \ding{173} Dataflow paradigms~\cite{samba-sn40} take an alternative approach, mapping multiple operators on-chip and allowing the results to flow between operators as defined by the computational graph as a pipeline, with each set of cores executing its assigned operator on incoming data and streaming results to downstream operators, reducing DRAM traffic. 
% Second, dataflow paradigms~\cite{samba-whitepaper,inter-layer} map each operator to only a few cores, but schedule multiple operators to reside on a chip at any given time.
Second, dataflow paradigms~\cite{samba-whitepaper,inter-layer} map each operator to a few cores, and schedule multiple operators to reside on the chip simultaneously.
The partial results flow across operators as a pipeline,
with each set of cores executing its operator on the incoming tensor and forwarding the result downstream.
% A pipeline is formed, with each core group executing on tensor inputs before streaming results to the downstream operator.
This on-chip dataflow reduces DRAM traffic and facilitates compute-communication overlap, as cores can start processing the next input tensor while transmitting %the current
results over the NoC.
% \hl{A little unclear: how do we transition new operators in?}
% In our experiment, as SRAM on a 3D AI chip cannot host an entire LLM, we implement the dataflow program to pipeline only a few operators on the chip at a time.
Because a 3D AI chip's SRAM cannot host an entire LLM, only a few operators reside at a time;
%Then, when executing a set of operators, 
while they execute,
% During execution,
we concurrently load the next operators' data from DRAM.
During execution, we divide each input batch into micro-batches and flow them across operators via \mvdata{} (\S\ref{sec:design:interface}).
In \Cref{fig:sw_compiler}, dataflow outperforms SPMD by 12.6\% on average during prefill due to lower NoC overhead. 
%by better overlapping compute and communication.

% -- we load $n$ operators that can fit into the SRAM, divide the batch into $n$ microbatches, and reduce partial results as they move between operators.
% Unfortunately, true dataflow requires all operators to fit on chip, which is impossible for very large models, where only a few operators can fit on chip at a time. Therefore, we use a dataflow \textit{inspired} paradigm for our evaluation -- 
% we load $n$ operators to the chip and divide the batch into $n$ microbatches.
% We also reduce the partial results as they move between operators.
% On \pname{}, All required inter-core data movements can use \mvdata{} to specify any inter-core data movement.
% Third, the compute-shift paradigm \cite{t10, waferllm} represents another way for adapting dataflow to large models that cannot fit entirely in SRAM.
Third, the compute-shift paradigm \cite{t10, waferllm:osdi2025} uses the full chip to execute each operator, but organizes each operator's tiled computation as a circular dataflow:
% organizes operator execution following a dataflow pattern.
tensors shared by multiple operator tiles are partitioned over a ring of cores and circularly shifted during execution.
%where each tensor shared by multiple operator tiles is partitioned across a ring of cores and the shared tensor is circularly shifted during execution.
% to \hl{iteratively read through the entire tensor}.
In \Cref{fig:sw_compiler}, compute-shift outperforms SPMD by 47.4\% on average during prefill, %as it 
almost eliminating the NoC overhead
by fully overlapping compute and shift. %by using a double buffer.
% Compute-shift also better overlaps DRAM prefetch with compute.
% Since it does not duplicate shared tensors across cores, the saved SRAM space can be used to prefetch more future data from DRAM.
Compute-shift also outperforms dataflow by an average of 39.7\% during prefill.
Since shared tensors are not duplicated across cores, more SRAM is free to host prefetched data, 
% As compute-shift does not duplicate shared tensors across cores, but instead shifts them, it saves more SRAM space on each core to host prefetched data.
which allows 
%The additional prefetch opportunities allow 
cores to better overlap DRAM accesses with computation, decreasing latency. 
% Also, compared with dataflow, compute-shift performs slightly better in most cases, since it can better overlap compute with DRAM accesses.
%However, dataflow still outperforms compute-shift in some cases (e.g., OPT-30B prefill).
%As dataflow runs multiple operators on the chip as a pipeline, each operator runs on fewer cores, enabling larger tiles and higher FLOPS on each core.

% In conclusion, while NoC is a significant performance bottleneck for SPMD execution, the dataflow paradigm mitigates the bottleneck with better compute-communication overlapping, improving the average performance by \hl{X}\%.
% Then, compute-shift further reduces the DRAM access bottleneck over dataflow by enabling better compute-DRAM access overlapping, further improving the average performance by \hl{X}\%.

\observation{A1}{
Compute paradigms are critical to performance of 3D AI chips. We reveal that the performance difference across compute paradigms can reach 1.84$\times$.
}

\observation{A2}{
An efficient compute paradigm should maximize the overlaps among tile computation, NoC communication, and DRAM accesses. Among the three representative paradigms, compute-shift performs best.
}

\observation{A3}{
Simply applying existing paradigms such as SPMD and dataflow to 3D AI chips cannot deliver optimized performance. For instance, SPMD cannot overlap its compute and reduction, resulting in high NoC overhead that contributes up to 47\% of the total execution time.
}

\subsection{Co-Optimize On-Chip Network with Software/Hardware}
\label{sec:explore:comm}
% In this subsection, we explore how software and hardware factors
% (i.e., tile-to-core mapping, NoC topology, NoC bandwidth) collectively affect on-chip communication during LLM execution.

% We study the impact of varying these parameters in the 3D AI chip.

% In the 3D AI chip, communication costs increase with NoC hop count and network contention.

\begin{figure} 
\vspace{-1ex}
    \includegraphics[width=.97\linewidth]{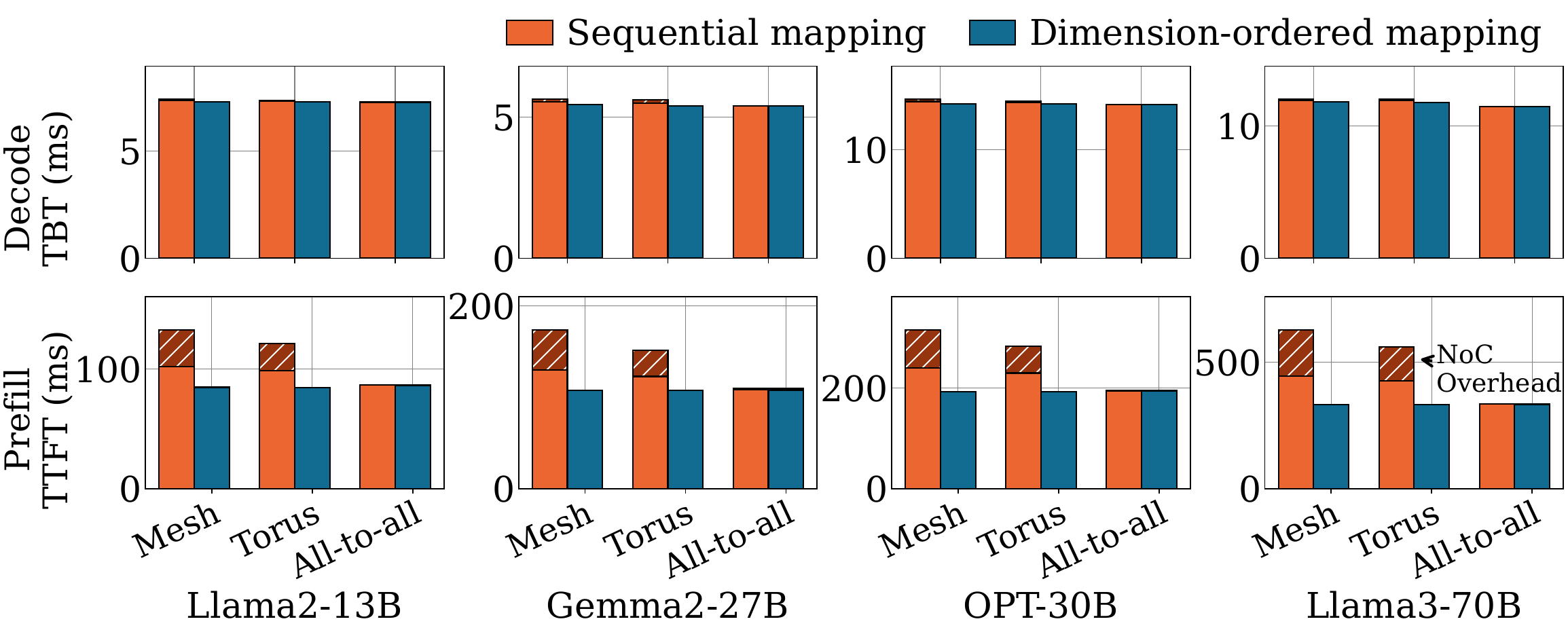}
\vspace{-2ex}
  \caption{LLM serving latencies with various tile-to-core mapping policies and NoC topologies. Shaded parts represent communication overhead.}
  \vspace{-1ex}
  \label{fig:eval_noc_topo}
\end{figure}

% A 3D AI chip experiences variable on-chip communication costs based on the number of NoC hops traversed and the NoC contention.
The 3D AI chip's on-chip communication cost varies with NoC hop count and contention.
% Efficient tile-to-core mapping can improve communication performance by reducing the hops required by each data transfer.
Efficient tile-to-core mapping reduces hops per transfer, improving communication performance.
\Cref{fig:eval_noc_topo} compares sequential and dimension-ordered \cite{waferllm:osdi2025,tpuv4-optical} mapping across three popular NoC topologies (2D mesh, 2D torus, and all-to-all) connecting all cores on the compute layer.
Sequential mapping assigns each tile the next available core.
Dimension-ordered mapping {(on 2D mesh, MeshGEMM 
%its variant on 2D-mesh NoC is called MeshGEMM 
\mbox{\cite{waferllm:osdi2025}})} assigns tiles that use the same shared data to cores sharing a row or column, {minimizing the data transfer distance during a ring reduce or circular shift.}

% 2D layer of cores on our 3D AI chip's compute layer: 2D mesh, 2D torus, and all-to-all. These represent the most common interconnect architectures for 2D core arrays.
% For each policy, we test different NoC topologies connecting the 2D grid of cores on the compute layer of our 3D AI chip. 
% connecting a rectangular 2D grid of cores.
% For the 2D mesh topology, using dimension-ordered mapping nearly eliminates the NoC overhead for both prefill and decode.
% \hl{For 2D mesh, using dimension-ordered mapping with compute-shift can hide most communication behind compute during both prefill and decode, as shown in \mbox{\Cref{fig:eval_noc_topo}}.}
{With 2D mesh, dimension-ordered mapping significantly reduces total NoC time over sequential mapping by minimizing the transfer distance.
This reduced NoC time is more easily hidden by compute or DRAM accesses,
% in an easier manner, 
leading to low NoC overhead as shown in \mbox{\Cref{fig:eval_noc_topo}}.}
% It maps tiles operating on a shared tensor to a connected line of cores in the mesh, minimizing data transfer distance during a ring reduce or circular shift.
The benefit is especially obvious for LLM prefill (40\% faster on average across models), 
% as it requires more data sharing on NoC.
% results in a higher volume of NoC traffic due to shared data.
which transfers a high volume of data over NoC.
% as it needs to transfer a large  data on the NoC.
% due to its high data transfer volume.
% as it requires a higher total data transfer volume on the NoC.
%
%The 2D torus topology 
2D torus adds wraparound connections on the grid edges, halving average inter-core distance and improving performance under the baseline mapping.
% The extra connections reduce the average distance between any two cores by half, enabling better performance even with the baseline mapping.
The dimension-ordered mapping still provides large benefits for prefill (33\% faster on average). 
For all-to-all NoCs, mapping policies have no impact, as all cores are equidistant (i.e., 1 hop).
%
% \hl{
% Among the three topologies, dimension-ordered mapping delivers similar performances.
% Using dimension-ordered mapping, all topologies deliver similar performance.
% This is because the mapping strategy is able to effectively reduce the average transfer distance to be very short (i.e., $< 2$ hops on average).
Dimension-ordered mapping greatly shortens average transfer distance ($\leq2$ hops~\cite{waferllm:osdi2025}), so transfer time varies little across topologies. 
%Therefore, we find 
A suitable mapping policy 
lets chip designers opt for a NoC topology with low wiring cost (2D mesh), without sacrificing performance.

%
% Once a topology has been selected, it is... \Cref{fig:eval_all_decode} (e) and \Cref{fig:eval_all_prefill} (e)
% \hl{something  about first mitigating effect of topology, then using a simple topology find minimum bandwidth to get a cheap but sufficient noc?}
% % \hl{focus on different prefill decode}
%
% \hl{done above, not done below

Using dimension-ordered mapping and 2D mesh, we %further
examine the performance impact of NoC bandwidth across LLM serving stages (\Cref{fig:eval_all_combined} (a)).
Decode performance is insensitive to NoC bandwidth, as it is memory-bound.
However, for prefill, decreasing 
per-link NoC bandwidth
%the bandwidth per NoC link 
below 32 bytes/cycle increases the execution time.
% However, increasing the NoC bandwidth can effectively decrease the prefill latency.
This is because prefill requires intensive inter-core data sharing, 
reusing each model weight tensor across numerous input tokens.
%as each model weight tensor is reused to process numerous tokens from the input.
If NoC bandwidth is too low, even an efficient tile-to-core mapping cannot fully hide communication overhead.
% increasing the total latency.
% Notably, this benefit will diminish when the per-core NoC bandwidth is over 32 bytes/cycle, since the execution will then be bounded by compute instead of communication.

% Different LLM inference stages exhibit different sensitivity to NoC bandwidth, as shown in \Cref{fig:eval_all_decode} (e) and \Cref{fig:eval_all_prefill} (e). While decode performance is not significantly affected by NoC bandwidth, prefill performance improves with greater bandwidth. This is due to the different communication intensities of prefill and decode. Since LLM decoding only produces at most a few tokens each step, the activation tensors of decode operators are small and can be designated as shared tensors to minimize inter-core communication volume. Thus, even with low NoC bandwidth, the communication can easily be hidden by compute operations or DRAM accesses, so increasing NoC bandwidth does not improve performance. On the other hand, activation tensors in prefill operators scale in proportion to the input sequence length, so they are typically large, causing higher communication overhead to share them across cores during prefill. 
% % Thus, low NoC bandwidth will expose inter-core communication on the critical path. 
% Therefore, increasing NoC bandwidth effectively decreases prefill latency until the execution is fully bottlenecked by compute.

% \hl{cannot only target prefill}
% \hl{be specific, name which noc topologies}
% \hl{need similar format}
% \hl{we need conclusions, not summarizing the subsection}

% When designing the NoC for a 3D AI chip, low-cost topologies can be as performant as high-cost topologies, given a suitable tile-to-core mapping strategy. In our experiments, we find that the \hl{XX}\% difference in latency between 2D-mesh and all-to-all NoCs under a standard, sequential mapping strategy can be reduced to just \hl{XX}\% using dimension-ordered mapping.
\observation{B1}{
% Efficient tile-to-core mapping can bring superior performance and bridge the potentially massive performance gap between different NoC topologies, as it minimizes the number of NoC hops traversed per data transfer. 
%To reduce the NoC overhead when using spatial NoC topologies such as mesh and torus, 
An efficient tile-to-core mapping can minimize the number of NoC hops per data transfer and reduce the NoC overhead, when using spatial NoC topologies (e.g., mesh, torus).
% significantly reducing the NoC overhead. 
% In our experiments, the dimension-ordered mapping averages between 1 to 2 hops per transfer across all tested topologies.
In our experiments, dimension-ordered mapping effectively mitigates the NoC overhead, reducing end-to-end latency by up to 39\% for NoC-bound workloads.
% Thus, the performance difference between chips using different topologies is below \hl{XX}\% in our experiment.
% , achieving up to \hl{X}$\times$ speedup in total execution time compared to a generic sequential mapping strategy.
% 1. sw bridges gap
}

\observation{B2}{
% Since an efficient tile-to-core mapping can 
% \hl{Version A}
With an efficient tile-to-core mapping, all NoC topologies (mesh, torus, and all-to-all) examined in our experiments have low NoC overhead in a 3D AI chip, further indicating that an efficient tile-to-core mapping can help tolerate the limitation of different NoC topologies. 
% Thus, the selection of NoC topology is not important for the performance of a 3D AI chip.
%Thus, the selection of NoC topology will not significantly affect the performance of a 3D AI chip.
}

% \observation{B2}{
% \hl{Version B}
% With an efficient tile-to-core mapping, the selection of NoC topology does not significantly impact the performance of a 3D AI chip. 
% Our experiments show that a dimension-ordered mapping can achieve near-ideal NoC overhead across all tested NoC topologies. 
% }

\observation{B3}{
Among these popular NoC topologies (mesh, torus, and all-to-all), our experiments suggest that mesh NoC plus dimension-ordered mapping will be an optimized setting for 3D AI chips. They deliver near-optimal performance with the lowest chip area overhead.
%recommend 3D AI chips to adopt mesh NoC and dimension-ordered mapping, since they enable near-optimal performance while requiring the lowest chip area.
}

% \begin{figure} 
% % \vspace{-1ex}
%     \centering
%   \subfloat[\footnotesize{LLM per-token decode latency.}\label{fig:eval_noc_topo_decode}]{%
%        \includegraphics[width=\linewidth]{figures/eval_noc_topo_decode.pdf}\vspace{-0.3em}}
% \\
%     \vspace{0.5ex}
%   \subfloat[\footnotesize{LLM prefill latency.}\label{fig:eval_noc_topo_prefill}]{%
%         \includegraphics[width=\linewidth]{figures/eval_noc_topo_prefill.pdf}\vspace{-0.3em}}
% \vspace{-0ex}
%   \caption{LLM serving latencies with various tile-to-core mapping policies and NoC topologies. The dark upper parts of columns represent inter-core communication overhead.}
%   \vspace{-2ex}
%   \label{fig:eval_noc_topo}
% \end{figure}

% \observation{\showmycounter}{
% % \hl{TODO} High NoC bandwidth allows faster prefill, as prefill is more communication-intensive than decode.
% % and noc topology
% what is the insight here
% }

% Execution method -- uses for noc.
% prefill results, reason
% decode results, reason

\subsection{Improving DRAM Bandwidth Utilization}
\label{sec:explore:dram-bw}

\begin{figure} 
% \vspace{-1ex}
    \centering
       \includegraphics[width=.93\linewidth]{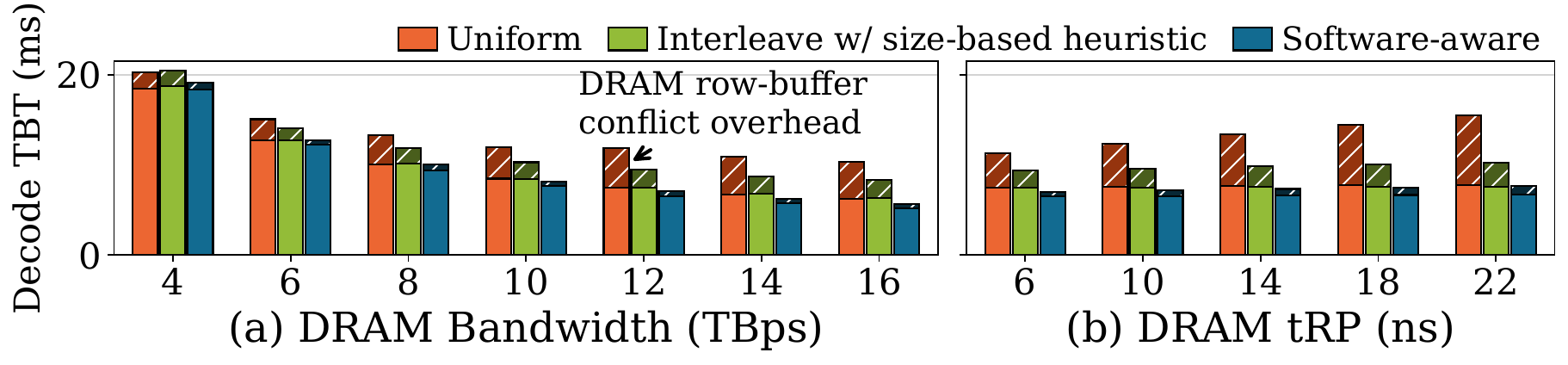}
\vspace{-2ex}
  \caption{\hlE{Llama2-13B} decode time with various tensor-to-bank placement policies. Shaded parts represent DRAM row-buffer conflict overhead.}
  % \vspace{-1ex}
  \label{fig:dram_bw}
\end{figure}

% In this section, we examine how various hardware parameters of the 3D AI chip influence its performance.
% Once we have optimized on-chip communication, we need to also make use of the provided DRAM bandwidth to maximize data movement, since large models cannot fit in on chip bandwidth.
% data movement, as they must continuously fetch data from off chip DRAM.
%In this subsection, 

% Even with optimized on-chip communication and ultra-high DRAM bandwidth, 3D AI chips may still suffer low DRAM bandwidth utilization.
% We now examine the factors that affect DRAM bandwidth utilization as we scale the bandwidth.

% 2.5D chip HBM bandwidth -> 3Tb/s H100, 4.8 TB/s H200
\looseness=-1
Even with optimized on-chip communication, scaling DRAM bandwidth in 3D AI chips without considering tensor-to-bank mappings yields diminishing returns, due to low DRAM bandwidth utilization.
We first evaluate bandwidth utilization under uniform tensor placement, splitting each tensor evenly across all DRAM banks. 
% \hlA{We sweep the DRAM bandwidth by changing the number of TSVs per bank without changing the number of banks. 
\hlA{To isolate the impact of per-bank bandwidth, we sweep total DRAM bandwidth while keeping the 
number of banks
fixed by varying TSV count.
\mbox{\pname{}} also supports raising TSV data-rates,
but this faces significant power, signal-integrity, and timing challenges.
Thus, scaling the TSV count is more practical.} %a more practical mechanism.}
% Alternatively, \mbox{\pname{}} can support increasing the bandwidth by increasing the data rate of existing TSVs; however, would introduce additional power, signal-integrity, and timing constraints. 
% Alternatively, we can increase the DRAM bandwidth by increasing the data rate of existing TSVs. However, 
% this would encounter significant power, signal-integrity and timing challenges. 
% Thus, updating the number of TSVs is a more practical mechanism.}
%to maximize per-tensor throughput.
\Cref{fig:dram_bw} (a) shows the resulting LLM decode latencies across various DRAM bandwidths, 
%with the latencies 
breaking down the DRAM row-buffer conflict overhead.

% When DRAM bandwidth matches a conventional 2.5D design
For bandwidths comparable to conventional 2.5D AI chips 
(e.g., 4 TBps~\cite{h100}), the row conflict overhead is low because few TSV data buses are provisioned:
% only a limited number of TSV data buses are provisioned.  
%but we need numerous DRAM banks to provide sufficient capacity, so each bus is shared by many DRAM banks.
% \hl{As we need a large number of DRAM banks to provide sufficient capacity, many DRAM banks need to share a single bus.}
% Thus, the TSVs act similarly to traditional 2.5D DRAM busses, where multiple banks must share a single TSV, 
as explained in \S\ref{sec:background:3d-arch}, temporal bus sharing hides individual banks' row buffer conflicts, enabling high bandwidth utilization.
% When bank conflicts occur, the temporal sharing of the bus can naturally mask the resulting idleness of the affected bank, maintaining high bandwidth utilization.
% When bank conflicts occur, the temporal sharing of the bus can naturally mask the resulting idleness of the affected banks.
% The temporal sharing of the bus naturally masks the idleness of any individual bank, ensuring high bandwidth utilization even in the presence of bank conflicts.
% This natural multiplexing makes compiler-memory optimizations have little effect, as the DRAM overhead comes mainly from the sharing of the bus, not from the sharing between rows in a bank.
% \hl{However, as DRAM bandwidth continues to scale?} 
% However, as we increase the DRAM bandwidth, we add more TSV buses without changing the bank count \hl{(i.e., keep capacity invariant)}, so each bus is shared by fewer banks.
{Adding DRAM bandwidth without scaling capacity increases the number of TSV buses, so each bus serves fewer banks.}
With limited bus sharing, row-buffer conflicts are not
% can no longer be
hidden across banks, harming per-bank performance and plateauing the benefit of increased bandwidth at $\approx$10TB/s.
%For a bandwidth of 
At 16TB/s
% When available bandwidth scales to 16TB/s,
this overhead
constitutes
% contributes up to
43.35\% of total decode latency.
% \hl{While this overhead could be lower on a DRAM with lower $\frac{tRP}{tCL}$ ratio, which makes the overhead of accessing a new row versus the current row less obvious, this kind of DRAM is rare on market.}
% Address DRAM timing -- explicitly say why we don't adjust/study timing.
% Explain the impact of timing; explain more clearly why it's not feasible
\hlA{
Notably, 
%this overhead could be mitigated by using 
DRAM with a lower $\frac{tRP+tRCD}{tCL}$ ratio}\footnote{tRP: row precharge time, tRCD: row-to-column delay, tCL: column address strobe time.}\hlA{
mitigates this overhead by reducing the
%, which reduces the 
relative latency penalty of a row miss. We show this in \mbox{\Cref{fig:dram_bw}} (b) by examining row-buffer conflict overhead across different row precharge times (tRP). 
While a smaller tRP reduces the overhead slightly, it 
does not
%is not sufficient to 
fully eliminate the problem. Our mapping strategy consistently reduces row-buffer conflict overhead, with increasing benefit as the timing overhead per conflict worsens.
}

\looseness=-1
To reduce conflict, we first try an interleaved strategy that maps consecutively allocated tensors to disjoint banks.
Assuming consecutive allocation dictates access concurrency, this approach heuristically places potentially concurrent tensors in different banks, reducing same-bank concurrent accesses and thus row-buffer conflicts (\mbox{\S\ref{sec:challenges}}).
%The approach assumes allocation consecutiveness dictates access concurrency to heuristically place tensors that may be accessed concurrently to different banks, reducing the concurrent accesses of tensors from the same bank and thus row buffer conflicts (\mbox{\S\ref{sec:challenges}}).
To balance per-bank capacity, 
% each tensor's bank count is determined by its size. 
the number of banks allocated for each tensor is determined based on the tensor size.
%Still, the result is insufficient. %not fully satisfying.
However, this is insufficient.
First, consecutively allocated tensors that are not concurrently accessed cause false positives:
if the actively accessed tensor is in only a subset of banks, TSVs that only serve other banks remain unused. %are unused during the time.
Second, 
concurrently accessed but non-consecutively allocated tensors cause false negatives, incurring
%when concurrent accessed tensors are not consecutively allocated, false negatives occur, causing 
nontrivial row-buffer conflict overhead when TSV bus sharing is limited (i.e., when available bandwidth $\geq$ %no less than 
12TB/s in \mbox{\Cref{fig:dram_bw}} (a)).

% Message/topic: Program-aware tensor2bank mapping reduces row change overhead by avoiding bank contention, enabling performance to scale with increased DRAM bandwidth.
{To maximize performance improvement with increased DRAM bandwidth, we design a software-aware tensor-to-bank placement strategy that detects concurrent accesses from the execution graph.
For each singular or fused operator, all its tensors are concurrent.
For two consecutive operators, the outputs of the preceding operator and inputs of the next are concurrent.
This detection eliminates the inaccuracies of heuristic approaches, minimizing the conflict overhead and fully utilizing all TSVs.}
% \hl{The accurate prediction minimizes the previous false positives and negatives, which minimizes conflict overhead while best utilizing all TSVs.}
% Thus, cores will not simultaneously access multiple tensors from the same bank, reducing the bank conflicts.
% \hl{Change this to decenter TSIM / make it more about IRL} To enforce this new placement strategy, the compiler should specify which operator will access a tensor via the \texttt{op\_id} parameter (i.e., operator index) when calling \mvdata{src\_tensor,dest\_tensor,op\_id}, so that tensors associated with the same or consecutive operator indices are placed into different banks.
In \Cref{fig:dram_bw} (a), this software-aware strategy achieves low row conflict overhead ($\le$14.8\% of total decode time) even at ultra-high DRAM bandwidth. 
% \hl{maybe rephrase to help?}
%allowing the overall performance to scale smoothly.

% which enables continued scaling to 16 TB/s and beyond, demonstrating effective mitigation of the row-change overhead.
% The compiler can leverage \pname{}'s \mvdata{} API to explicitly place tensor pieces that are expected to be accessed concurrently in different banks when possible (see \S\ref{sec:design:compiler-optimization}).
% \Cref{fig:dram_bw} shows that the program-aware placement can achieve low row-change overhead (up to \hl{X}\% of total time) even when the DRAM bandwidth is ultra high, allowing the overall performance to scale smoothly.

\begin{figure} 
% \vspace{-1ex}
    \centering
       \includegraphics[width=.9\linewidth]{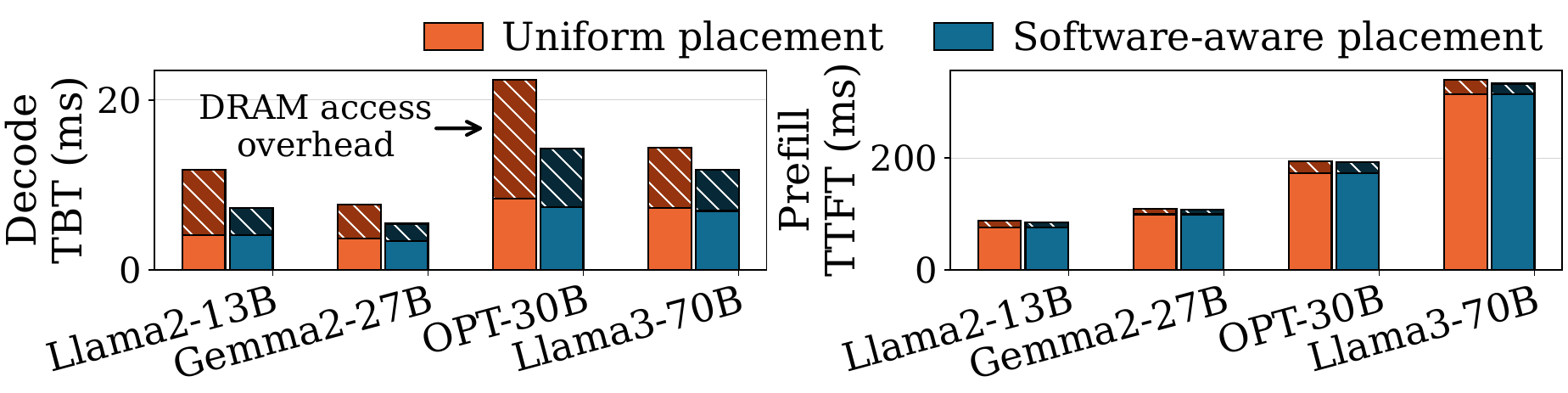}
\vspace{-2ex}
  \caption{LLM serving latencies when using different tensor-to-bank placement policies. The dark upper parts of columns represent DRAM access overhead.}
  \vspace{-1ex}
  \label{fig:sw_dram}
\end{figure}

% Message: 
While the new placement strategy scales LLM decode performance, it brings limited performance benefit to prefill, as shown in \Cref{fig:sw_dram} (a).
This is because prefill is compute-bound, so most
DRAM accesses can be hidden behind computation.
% We show in \Cref{fig:eval_all_decode} (a) that under this placement strategy, additional DRAM bandwidth greatly improves decode performance.
Similarly, while increasing total DRAM bandwidth consistently 
benefits LLM decoding performance 
%brings significant performance benefits to LLM decoding 
in \Cref{fig:eval_all_combined} (b), prefill remains insensitive. % to increased DRAM bandwidth.

\observation{C1}{
% OLD
% Increasing the DRAM bandwidth of a 3D AI chip makes it more difficult for DRAM banks to hide each other's row change overheads.
% This exposed overhead worsens as bandwidth further increases. 
% In our experiments, the row-change overhead accounts for up to \hl{X}\% of total execution time, which negates almost all of the potential performance benefits brought by the additional bandwidth.
%
% Increasing the DRAM bandwidth of a 3D AI chip makes it more difficult for DRAM banks to hide each other's row change overheads.
% It becomes more difficult to fully utilize \hl{achieve?} the DRAM bandwidth of a 3D AI chip as its peak bandwidth is increased. 
% leading to significant DRAM bus underutilization at high DRAM bandwidths.
As we scale DRAM bandwidth on 3D AI chips, bandwidth underutilization is a new bottleneck. It will be more difficult to minimize row-buffer conflicts and saturate the available bandwidth. We show that row-buffer conflicts can account for up to 43.35\% of the total execution time, which negates the potential performance benefit of increasing DRAM bandwidth. 
%As the DRAM bandwidth of a 3D AI chip increases, it becomes more difficult for DRAM banks to hide each other's bank conflict overheads.  
%In our experiments, this overhead accounts for up to \hl{X}\% of total execution time, 
%which negates the potential performance benefits brought by the increased DRAM bandwidth.
% This prevents high DRAM bandwidth from delivering any improvement in the overall performance.
% This exposed overhead worsens as bandwidth further increases. 
% Increasing the DRAM bandwidth on a 3D AI chip by adding more TSV links will prevent the DRAM row-change overhead from being hidden by link sharing.
% This exposed overhead leads to severe bandwidth underutilization, which worsens as the peak bandwidth further increases. 
% This exposed overhead decreases bandwidth utilization when more bandwidth is provisioned.
%which prevents additional DRAM bandwidth from improving performance past $\approx$10TB/s.
}

\observation{C2}{
% OLD
% Tensor-to-bank placement plays a critical role in reducing the row-change overhead.
% We recommend using our program-aware placement strategy, which prevents cores from concurrently accessing different tensors from one DRAM bank.
% This reduces the row-change overhead by up to \hl{XX}\% compared to the uniform placement strategy.
% A program-aware tensor-to-bank placement can reduce the row-change overhead by up to \hl{XX}\% compared to a uniform placement strategy, since it prevents cores from concurrently accessing different tensors in the same DRAM bank.
Tensor-to-bank placement plays a critical role in reducing row conflicts. We show that a software-aware placement strategy can reduce row conflict overhead by an average of 80.68\% compared to a uniform placement strategy, allowing memory-bound workloads to keep scaling performance with increased DRAM bandwidth.
%
% In our experiment, a program-aware placement strategy can reduce the row-change overhead by up to \hl{XX}\% compared to a uniform placement strategy, as it prevents cores from concurrently accessing different tensors from the same DRAM bank.
% % This allows the overall performance of memory-bound workloads to scale well with increased DRAM bandwidth.
% Once row-change overhead is alleviated, the performance of memory-bound workloads scales well as DRAM bandwidth increases.
%
% plays a critical role in reducing the row-change overhead.
% We recommend using our program-aware placement strategy, which revents cores from concurrently accessing different tensors from one DRAM bank.
% This reduces the row-change overhead by up to \hl{XX}\% compared to the uniform placement strategy.
}

\subsection{Scale FLOPS with Sustained DRAM Bandwidth Utilization}
\label{sec:explore:flops}

\begin{figure}[t]
    \centering
    % \vspace{-.5ex}
    \includegraphics[width=0.99\linewidth]{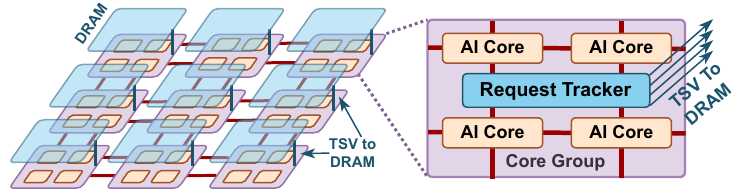}
    \vspace{-3.9ex}
    \caption{Synchronizing DRAM accesses with core groups.}
    % \hl{map them to 3D AI chip.}} q
    \label{fig:core-group}
    \vspace{-1ex}
\end{figure}
 % \hl{TODO replace, make inputs and outputs of request tracker clear}

\begin{figure}[t]
    \centering
    % \vspace{-.2ex}
    \includegraphics[width=.99\linewidth]{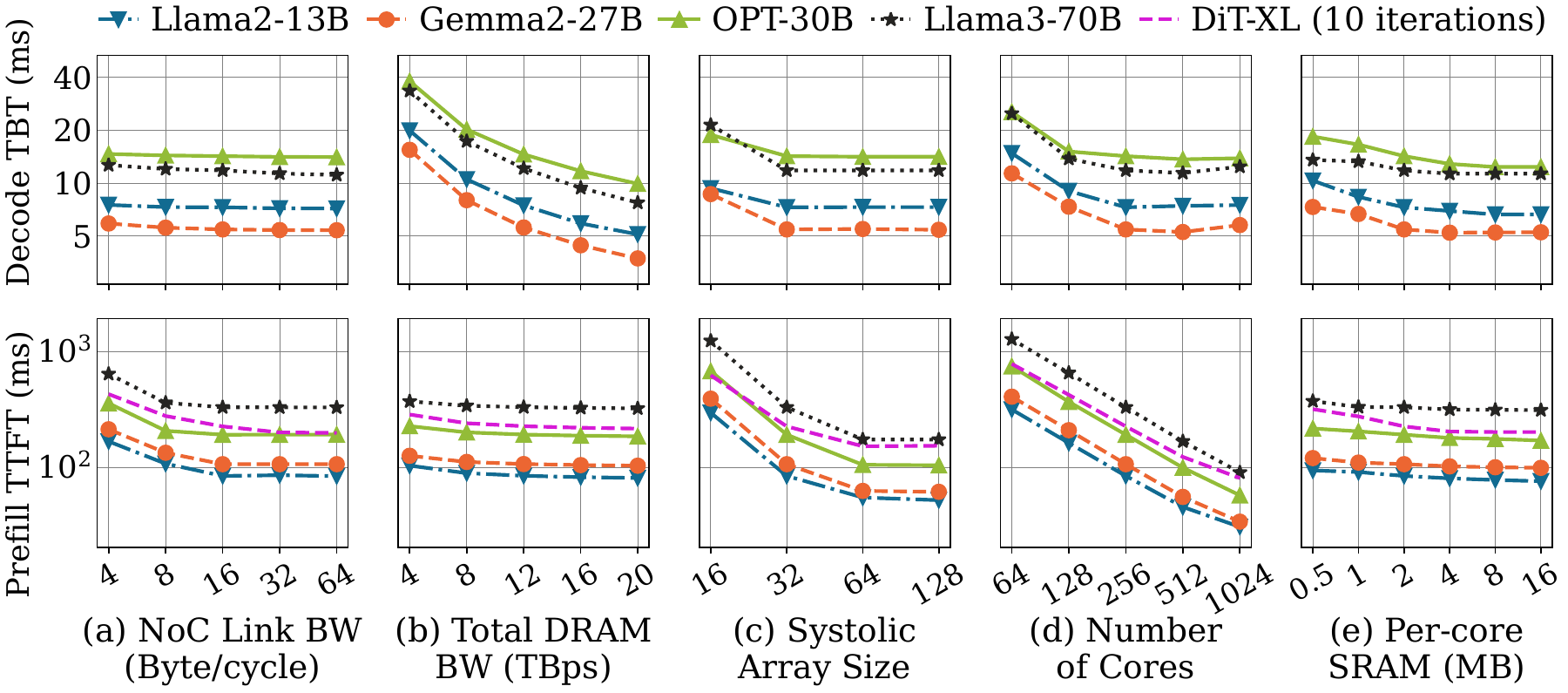}
    \vspace{-3.9ex}
    \caption{LLM decode/prefill time with different hardware.}
    \label{fig:eval_all_combined}
    \vspace{-1ex}
\end{figure}

% \begin{figure}[t]
%     \centering
%     \includegraphics[width=\linewidth]{figures/eval_lines_vit.pdf}
%     \vspace{-5ex}
%     \caption{DiT-XL inference latency with different hardware. \hl{change to 3*2}}
%     \label{fig:eval_vit}
%     \vspace{-2ex}
% \end{figure}

% Message: compute scaling important,  
Improving computation throughput is crucial to exploit the high memory bandwidth of 3D AI chips and speed-up compute-bound workloads.
However, simply increasing systolic array size or core counts may worsen DRAM bandwidth utilization, leading to worse-than-expected performance gains.
% However, simply enlarging each core's systolic array or increasing the number of cores may lead to significant underutilization of compute or DRAM bandwidth, preventing the additional resources from yielding performance improvement.
% However, simply scaling the size of each core's systolic array or increasing the number of cores on a chip leads to significant compute or DRAM bandwidth underutilization, preventing the scaling efforts from yielding commensurate improvements in end-to-end performance.
% However, naively scaling the size or quantity of cores leads to significant compute or DRAM bandwidth underutilization, preventing the scaling efforts from yielding commensurate improvements in end-to-end performance.
% We start with a detailed examination of the issues facing compute scaling on 3D AI chips. \hl{Update figure refs for this section once ready.}
% As such, architects must take measures to mitigate underutilization issues to enable 3D AI chips to scale (\hl{check}).
% We explain the issues in \hl{cite location} and present a solution in \hl{cite}.

\begin{figure}[t]
    \centering
    \vspace{-.5ex}
    \includegraphics[width=0.99\linewidth]{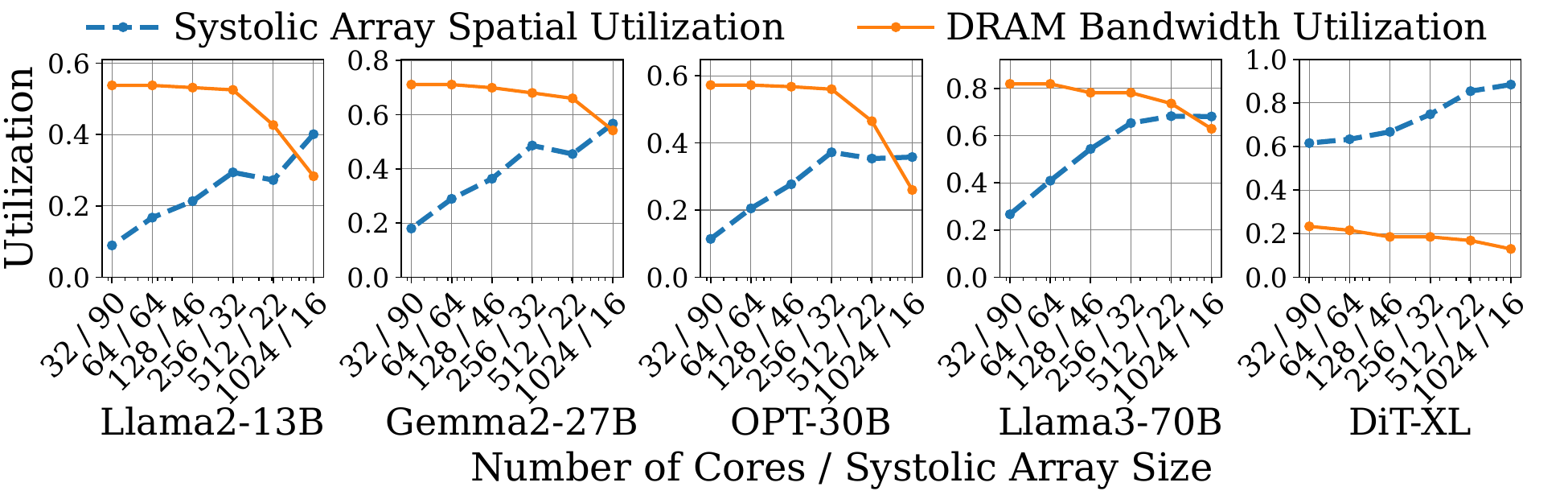}
    \vspace{-3.9ex}
    \caption{Spatial utilization of {SA}s and DRAM bandwidth utilization of LLM decode and DiT-XL serving, as we vary core count and {SA} size without core grouping.
    Each chip variant has the same total FLOPS and SRAM capacity.}
    % \hl{same total FLOPS} and .}
    \label{fig:eval_x_util}
    \vspace{-1ex}
\end{figure}

\begin{figure}[t]
    \centering
    \includegraphics[width=0.99\linewidth]{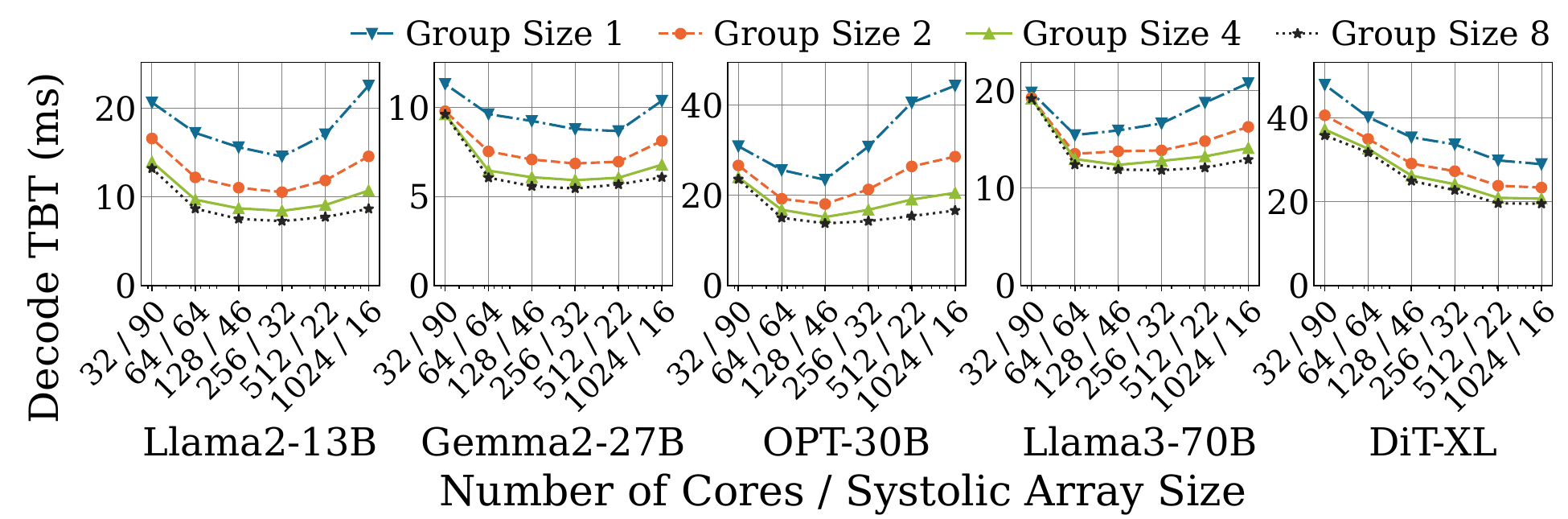}
    \vspace{-3.9ex}
    \caption{LLM decode and DiT-XL serving latency with various core group sizes, core counts, and {SA} sizes. Each variant has the same total SRAM capacity for fair comparison.}
    \label{fig:eval_smile}
    \vspace{-1ex}
\end{figure}

% Message: Large SA provide higher peak flops, but padding causes spatial underutilization; FLOPS are wasted.
\noindent\textbf{Large systolic arrays lead to low FLOPS utilization.}
One way to scale the peak FLOPS of a 3D AI chip is to enlarge the systolic arrays (SA) on each core.
% However, as shown in \Cref{fig:eval_all_combined} (c), increasing SA size yields diminishing returns beyond the size 32$\times$32 for both prefill and decode, since the additional FLOPS are wasted due to significant tensor padding.
% Operators mapped to SAs require padding when tile dimensions do not align with the SA size.
% This padding overhead increases with SA size, since the probability and potential magnitude of dimension mismatch will grow \cite{tpu}.
% When padding occurs, some processing elements on the SA waste their FLOPS on padded data, resulting in \textit{spatial underutilization}.
However, as shown in \Cref{fig:eval_all_combined} (c), increasing SA size beyond 32$\times$32 yields diminishing returns for both prefill and decode.
This is due to the growing misalignment between the operator tile and SA sizes. An operator must be padded to align with the SA shape before it can be executed by the SA.
This padding causes some processing elements in the SA to perform useless computations, which we refer to as
% waste their FLOPS on useless computations. 
% This padding causes some processing elements in the SA to waste their FLOPS on useless computations. 
% operators mapped to SAs require padding when their tile dimensions do not align with the SA dimensions.
% when operator tile dimensions do not align with the SA dimensions, the tile will be padded to the SA size, and as such, some processing elements will waste their FLOPS on non-useful computations.
\textit{spatial underutilization}.
% When padding occurs, some processing elements on the SA waste their FLOPS on padded data, resulting in \textit{spatial underutilization}.
The severity of this underutilization grows with SA size \cite{tpu}.
% The probability and magnitude of this underutilization will grow with SA size \cite{tpu}.
% This padding overhead increases with SA size, .
% In \Cref{fig:eval_x_util}, the dashed line shows the inverse relationship between SA size and spatial utilization during LLM decoding.
In \Cref{fig:eval_x_util}, the dashed line shows the increase in spatial utilization during LLM decoding as we decrease the SA size.
% The severity of this underutilization grows with SA size \cite{tpu}, as shown by the dashed line in \Cref{fig:eval_x_util}
% shows the inverse relationship between SA size and spatial utilization during LLM decoding.

\noindent\textbf{More cores lead to low DRAM bandwidth utilization.}
A 3D AI chip may also add more cores to increase its peak FLOPS.
% INGI change
However, increasing core count may reduce DRAM bandwidth utilization (\Cref{fig:eval_x_util}), degrading performance.
% However, as shown by the solid line in \Cref{fig:eval_x_util}, simply increasing core count may lower the DRAM bandwidth utilization, degrading the performance. 
% With more cores, operators are partitioned into a larger number of smaller tiles.
% , which harms DRAM performance.
\hlA{This is because increasing the number of cores decreases the ratio of banks to cores, which means more cores contend to access shared tensors from any given DRAM bank.}
% This will cause severe row buffer thrashing when more cores access a shared tensor at different pace (see \mbox{\S\ref{sec:challenges}}).} 
%the likelihood increases that some of these accesses become stragglers due to the non-uniform communication cost on a 3D AI chip.
\hlD{
The resulting row buffer conflicts (see \mbox{\S\ref{sec:challenges}}) cannot be fully resolved by request coalescing in the memory controller. 
% DRAM bandwidth utilization may still be poor.
Specifically, NoC contention can amplify the effects of non-uniform memory accesses, causing some cores to lag behind their peers.
Once execution progress across cores grows uneven, the memory controller can no longer reliably coalesce requests: a slow core's requests to a bank may not yet be in the queue when a fast core accesses the same bank. 
Thus, a separate structure is required to synchronize the execution and memory access progress of neighboring cores.
}

% \hlD{As discussed in \mbox{\S\ref{sec:challenges}}, this will lead to interleaved row accesses due to non-uniform NoC distance and exposure to NoC contention, increasing row conflicts.}
% Thus, the DRAM bandwidth utilization will drop, harming the performance of memory-bound workloads.

% % Second, with more tiles, the size of each tile becomes smaller, which causes finer-grained accesses to these smaller tensor pieces.
% Fine-grained accesses experience lower DRAM performance than long sequential ones, lowering the DRAM utilization.
% This underutilization harms the performance of memory-bound workloads such as LLM decode.

\noindent\textbf{Synchronize DRAM accesses with core groups.}
% \hl{need to explain why it doesn't scale infinitely -- talk about tradeoff of limited benefit with extra routing overhead etc.}
% \hl{NOTES:}
% Relaxed consistency, as in the NVIDIA CUDA model -- probably not necessary
% coalesce buffer is small, write back cache
% 
%
%Since we cannot achieve the additional peak FLOPS provided by scaling SA size, 
% When scaling the FLOPS of a 3D AI chip by adding more cores, we should address the increased row-buffer conflicts.
% -- but doing so increases DRAM bank conflict overhead. % To address this...
% -- but doing so requires addressing the increased DRAM bank conflicts.
% We address tis by forming \textit{core groups}: small groups of spatially proximate cores (e.g., 2-8) between which we coordinating their DRAM accesses.
To address these added row buffer conflicts when scaling cores, 
we group nearby cores into \textit{core groups}, as shown in \Cref{fig:core-group}. 
% Within each group, we use a hardware \textit{request tracker} to synchronize DRAM requests from different cores, which reduces the frequency of interleaved row accesses to alleviate row buffer thrashing.
Within a group, a hardware \textit{request tracker} synchronizes DRAM requests of different cores, reducing interleaved row accesses.

% \hl{GEMINI look at HERE!}
Our proposed request tracker synchronizes cores in a group by selectively stalling DRAM requests, since DRAM requests from cores whose execution progress is ahead of their peers may cause a DRAM bank to change its row early.
% , before other cores finish accessing the current row.
% \hl{This thrashes the row-buffer, which heavily penalizes slower cores that access the evicted row later, causing them to fall further behind.}
{When slower cores access the evicted row later, they experience higher memory access latencies due to row-buffer thrashing.
This causes them to fall further behind, resulting in longer end-to-end execution time.}
% Then the global execution time severely degrades as it is bounded by the slowest core.
% \hl{In fact, once a core begins to run}
To avoid this, the tracker identifies cores that are too far ahead of their peers and stalls their requests until other cores catch up.
% \hl{This is always beneficial to overall performance, since the net-effect is the acceleration of straggler cores at the cost of only small performance degradation for faster cores. Without such a mechanism, the stragglers would fall farther behind, degrading end-to-end performance.
% Such stall never harms the overall performance -- since global execution time is bounded by the slowest core, it is always worthy to accelerate slower cores by slightly stalling faster cores.}
% \hl{This mechanism consistently improves end-to-end performance
% Because global execution time is bounded by the slowest core, stalling faster cores incurs no overall penalty. 
% by accelerating straggler cores at the cost of only minor degradation to faster ones.
% This prevents stragglers from falling further behind, minimizing end-to-end execution latency.}
% This mechanism is always beneficial to overall performance; it accelerates straggler cores at the cost of only minor degradation to faster ones. Without such aid, the stragglers would become victims of the DRAM row thrashing and fall farther behind.
% In contrast, running row-buffer thrashing in the shared DRAM bank would incur severe latency penalties for both the run-ahead core and its lagging peers.
Since most ML compilers \cite{roller,tvm,xla,t10} evenly partition an operator across cores, each core executes identical tiles with identical memory access patterns.
{For DRAM banks that hold tensor edges, we zero-pad their tiles to match the shape in other banks.}
% ThuCores are considered ahead whenever they have made more DRAM requests than their peers, as we assume that cores in the same group have the same memory access pattern.
% To identify a run-ahead core, we exploit the identical memory access pattern between cores that execute tiles of the same shape.
% To identify run-ahead cores, we exploit the identical memory access pattern between cores; cores that have made more DRAM requests than their peers are considered run-ahead.
% This is because most SOTA compilers \cite{roller,tvm,xla,t10} evenly partition operators across cores, so all cores execute identical tiles with identical memory access and computation behaviors.
% so they have identical per-core tile shapes which result in identical memory access patterns.
% perators are homogeneously partitioned across cores, 
% OLD
% Unlike general workloads, where identifying these run-ahead cores would require complex tracking, AI workloads compiled by SOTA compilers \cite{roller,tvm,xla,t10} share a useful property: operators are homogeneously partitioned across cores, so they have identical per-core tile shapes which result in identical memory access patterns.
% We leverage a key property of ML compilers to simplify the identification of run-ahead requests: most SOTA ML compilers will always homogeneously partition an operator across cores \cite{roller,tvm,xla,t10}.
Thus, a core's execution progress compared to its peers is directly measurable based on the number of DRAM requests each core has sent to the tracker. The tracker only needs to enforce that no core's $(i+1)^{th}$ access is dispatched to DRAM before the $i^{th}$ requests of all cores in the group have been dispatched.
% their access.
% , which can be implemented with a negligible logic footprint.
% \hl{Since a tracker only has one 16-bit counter per core and a comparator tree of size $O(log^\text{group\_size})$, it has a negligible logic footprint.}
% \hl{Since a tracker only has one 16-bit counter per core and a comparator tree of size $O\left(\log G \right)$, with $G$ being the group size, it has a negligible logic footprint.}
% \hl{For a group size of $G$, the tracker only requires a single 1-bit counter per core, an $G$-input \texttt{AND} gate, thus it has a negligible logic footprint.}
%\hl{For a group size of $G$, the tracker state and control logic only requires around $50\times G$ transistors.}
% only requires a single 1-bit counter per core, an $G$-input \texttt{AND} gate, thus it has a negligible logic footprint.}
Combined with the tensor-to-bank placement strategy discussed in \S\ref{sec:explore:dram-bw}, it can avoid most unnecessary row conflicts.

\Cref{fig:eval_smile} shows how core groups benefit decode latency for different core counts and SA sizes. 
% Without core grouping (i.e., group size 1), designs at both ends of the core size spectrum yield poor decode performance due to significant FLOPS or DRAM underutilization.
Without core grouping, designs with large SAs or high core counts yield poor decode performance, due to significant FLOPS or DRAM underutilization (\Cref{fig:eval_x_util}).
With core groups, the performance of all core configurations improves, with more improvement for higher core counts, as core groups enable more efficient DRAM accesses.
In \Cref{fig:eval_smile}, for a 1024-core chip, a group size of 8 cores is up to 57\% faster than group size 1 for LLM decode. 
% As core groups enable more efficient DRAM accesses, 
% the performances of all core configurations improve, with more significant improvements for higher core counts. 
As we further increase group size (more than 8 cores), we observe negligible additional benefit. % in our experiment.
% \hl{TODO explain why we don't scale past 8}
% are often unnecessary since the synchronized accesses of 8 cores are already long enough to utilize the DRAM bandwidth well.
Using a proper core group size (i.e., 8), we can effectively scale the performance of a 3D AI chip with more cores, as shown in  \Cref{fig:eval_all_combined} (d).

\observation{D1}{
% , due to the increased risk of cores interleaving their accesses to different rows in a DRAM bank.
% Simply increasing the number of cores on a 3D AI chip will cause DRAM bandwidth underutilization.
% by increasing the DRAM row-change overhead. 
Simply increasing the number of cores on a 3D AI chip will cause more frequent row-buffer conflicts, leading to DRAM bandwidth underutilization.
%Compared to a chip with fewer cores, 
This may even cause decreased end-to-end performance. 
% Therefore, increasing the number of cores on 3D AI chip may lead to performance degradation.
% This may result in decreased end-to-end performance compared to a chip with fewer cores.
%
% At high core counts (i.e., $>$128), the overall performance of a 3D AI chip may degrade due to 
%
% Naively scaling the number of cores on a 3D AI chip may harm the overall performance of a 3D AI chip.
% The lack of coordination between DRAM accesses of different cores induces row change overhead, leading to DRAM bandwidth underutilization.
}
% To efficiently scale the compute capability of 3D AI chips by increasing core counts, we must mitigate the interference between multiple cores' DRAM accesses. 

\observation{D2}{
As we scale the core count, we can organize nearby cores in \textit{core groups} to reduce row-buffer conflicts. We can use a simple hardware logic (request tracker in Figure~\ref{fig:core-group}) to synchronize memory accesses from cores in the same group. With groups of 8 cores, we improve the end-to-end performance by up to 58\% (42\% on average) for a 3D AI chip with 1,024 cores. 
%To maintain high DRAM bandwidth utilization when scaling the core count, we add a small amount of on chip logic to synchronize repeated DRAM row accesses within small groups of cores, which we refer to as \textit{core groups}.} 
% divide cores into groups, and coalesce the DRAM accesses of cores in each group into larger contiguous accesses.
%In our experiments, grouping sets of 8 cores can improve the end-to-end performance by up to \hl{XX}\% for a 3D AI chip with 1024 cores.
% This allows core count to scale without inducing DRAM underutilization.
}
% make it clear what coallescing is...

% \observation{D3}{
% Based on our experiment, we recommend including 8 cores in each group.
% It can unlock up to \hl{X}\% end-to-end performance improvement for a chip with a high core count of 1024, while incurring a low hardware overhead of $\sim$2\%.
% Larger group sizes will allow limited additional benefits, since the coalesced accesses of 8 cores are already long enough to well utilize the DRAM bandwidth.
% }

% with small switches allows them to coalesce their memory accesses. Larger core groups make larger contiguous DRAM accesses, improving DRAM efficiency through greater spatial locality.
% These groupings allow us to scale core counts without inducing DRAM underutilization, leading to higher performance.
% }

\begin{figure}[t]
    \centering
    \includegraphics[width=.99\linewidth]{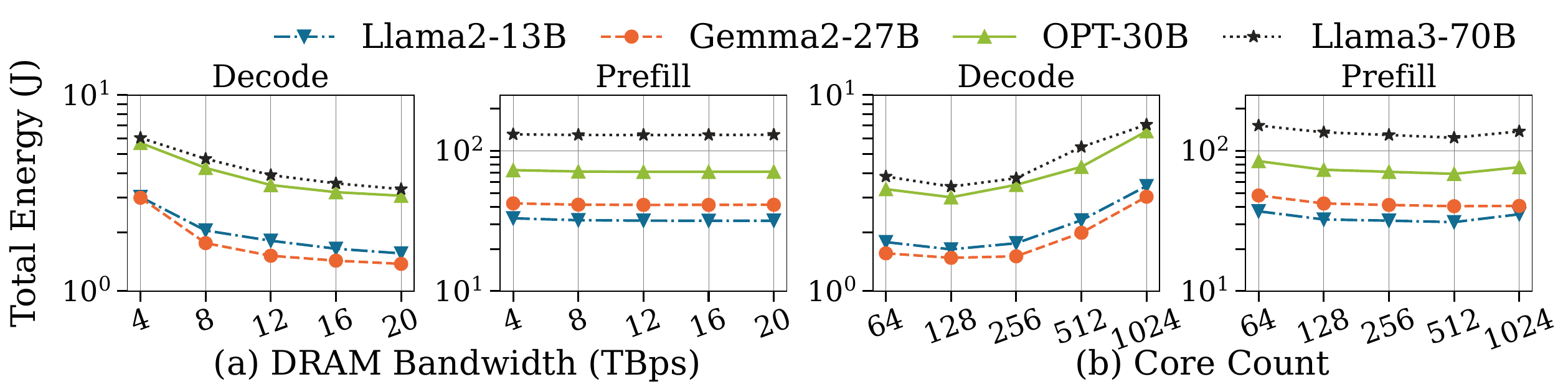}
    \vspace{-3.9ex}
    \caption{LLM decode (single iteration) and prefill energy consumptions for different hardware configurations.}
    % \hl{Emphasize decode is 1 iteration (for all figs)}} 
    \label{fig:dram-energy}
    \vspace{-1ex}
\end{figure}

\subsection{Scaling SRAM Capacity}
\label{sec:explore:sram}

In \Cref{fig:eval_all_combined} (e) we examine the impact of per-core SRAM size. % but for different reasons.
% However, their benefits come from different reasons.
Larger SRAM sizes can benefit LLM decoding;
%by improving the DRAM bandwidth utilization.
during execution, a 3D AI chip can use SRAM to store prefetched data from DRAM.
% Then, while the core is executing an operator 
Thus, larger SRAM space can improve the DRAM bandwidth utilization by enabling more prefetch opportunities.
Once the SRAM is sufficient ({8 MB} per core in our experiments) to fully utilize the available DRAM bandwidth, further scaling has no benefit.
% Then, while the core is executing an operator, a larger SRAM space can keep the DRAM utilized by prefetching data and keep the DRAM utilized for longer, until the execution is bound by memory again \cite{elk}.
% This improves the DRAM bandwidth utilization, which is critical for the decode performance.
% However, scaling SRAM will provide no benefit after exceeding 4 MB per core, since the DRAM is already busy all the time. 

Larger SRAM can benefit prefill performance by allowing larger tiles, which can better utilize the FLOPS on SAs.
However, since prefill is already compute-bound with high FLOPS utilization on SAs even when SRAM is small (67\% on average with 0.5MB SRAM), the improvement is limited.
% In \Cref{fig:eval_all_combined} (e), by increasing the per-core SRAM size from \hl{0.5MB to 16MB}, we only improve prefill performance by 29\% on average.
In \Cref{fig:eval_all_combined} (e), we show that doubling the per-core SRAM size improves prefill latency by 3.5\% on average.

\observation{E1}{
% Large SRAM improves decode performance by allowing high DRAM bandwidth utilization, and it improves prefill performance by allowing larger tiles. 
For memory-bound workloads, larger SRAM size can benefit the overall performance by improving DRAM bandwidth utilization, as it enables a larger window of data to be prefetched from DRAM.
% allows more data to be prefetched from DRAM each time. 
This trend stops when the prefetch window is large enough (8MB per core in our experiments) to saturate DRAM bandwidth.
% For prefill, the increased capacity enables larger tile sizes, reducing the required communication volume between tiles.
}

\observation{E2}{
% Large SRAM improves decode performance by allowing high DRAM bandwidth utilization, and it improves prefill performance by allowing larger tiles.
% Increasing SRAM capacity can improve both prefill and decode performance. 
% For decode, the increased capacity enables higher DRAM bandwidth utilization by allowing more data to be prefetched each time.
For compute-bound workloads, larger SRAM size brings limited benefits to the overall performance.
Scaling the per-core SRAM size by 32$\times$ delivers only 16.3\% performance improvement for LLM prefill.
This is because compute-bound workloads have already  utilized the FLOPS even when SRAM is small, leaving limited room for improvement.
% Although larger SRAM can improve the FLOPS utilization on SAs by enabling larger tile sizes,
% the effect is limited because the FLOPS utilization is already high for compute-bound workloads.
% Thus, architects may compensate for low NoC performance with more SRAM or vice versa.
}

\subsection{Energy Consideration of Scaling 3D AI Chip}
\label{sec:explore:energy}

\begin{figure}[t]
    \centering
    \includegraphics[width=\linewidth]{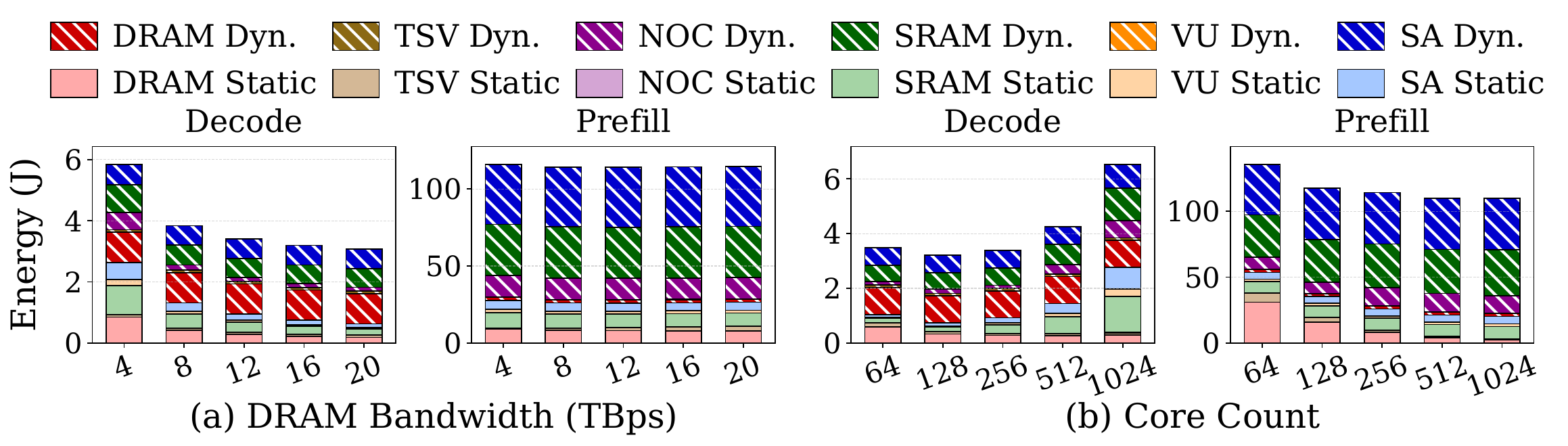}
    \vspace{-3.9ex}
    \caption{Component energy breakdown of different chip configurations for Llama3-70B decode iteration and prefill.}
    \label{fig:energy-breakdown}
    % \vspace{-1ex}
\end{figure}

% Different hardware resources also impact the overall energy consumption of the chip differently.
% Different on chip resources contribute differently to energy consumption, and have different scaling implications.
To understand the energy efficiency of 3D AI chips, we study the energy consumption as DRAM bandwidth and number of cores are varied.
% from the default configuration.
% In \Cref{fig:dram-energy}, we
% Understanding how different resources impact a chip's overall energy consumption can help guide architects when determining resource scaling priorities.
% \Cref{fig:dram-energy} (a) demonstrates that increasing the DRAM bandwidth improves overall energy efficiency.
% As shown in \Cref{fig:dram-energy} (a), \hl{increasing DRAM bandwidth can save energy for LLM decode by alleviating the memory bottleneck}: higher bandwidth decreases execution latency, thereby reducing the duration over which static power is dissipated.
%
% \noindent
% \textbf{Increasing DRAM bandwidth.}
%
Increasing DRAM bandwidth reduces the energy consumed by decode, but has no benefit on prefill (\Cref{fig:dram-energy}(a)).
% \Cref{fig:dram-energy}(a), shows that higher DRAM bandwidth improves decode's energy efficiency, but brings no benefits to prefill.
The energy consumption breakdown (\Cref{fig:energy-breakdown}(a))
shows that the added TSV links incur minimal static power overhead, and the overall dynamic energy consumption remains similar with increased DRAM bandwidth, due to the similar volume of data accessed.
% We break down the energy consumption of Llama3-70B in \Cref{fig:energy-breakdown} (a), and find that
% as we scale the DRAM bandwidth by adding TSV links, the additional TSVs incur little energy overhead, and the dynamic energy consumption does not change much with increased DRAM bandwidth, since the data access volume is similar.
%Also, since improved DRAM bandwidth will not change the total computation or data access volume of decode or prefill, we observe that their dynamic energy does not change much with bandwidth.
% Also, since improved DRAM bandwidth will not change the total computation and data access volume that needs to be performed during decode or prefill, we observe that their dynamic energy does not change much with bandwidth.
Decode's reduced energy consumption is due to the reduced static power consumption, explained by the significant decrease in latency additional bandwidth enables (see \Cref{fig:eval_all_combined}(b)).
Higher DRAM bandwidth does not reduce prefill latency, and does not reduce static power consumption, leaving energy unaffected.
% % Thus, the energy benefit experienced by decoding is largely attributed to its reduced static energy consumption.
% For LLM decoding, since scaling DRAM bandwidth greatly reduces latency (see \Cref{fig:eval_all_combined} (b)), 
% %without significantly increasing static power dissipation, 
% the static energy consumption is reduced. 
%which benefits the total energy reduction.   
%by a decode iteration, saving total energy.
% For decode, since scaling DRAM bandwidth greatly reduces latency (see \S\ref{sec:explore:dram-bw}), the total static energy dissipated during decode significantly reduces, saving the total energy.
% For prefill, however, higher bandwidth cannot significantly reduce its latency, so it experiences little benefit in static energy.
% Since dynamic energy also does not change much with increased bandwidth, the total energy of prefill experiences minimal change as DRAM bandwidth increases.
% For prefill, however, higher bandwidth cannot reduce its latency, so it experiences little benefit in static energy.
% Since dynamic energy also does not change much with bandwidth, the total energy of prefill experiences little variance. 

% In \Cref{fig:dram-energy}(b), we observe that scaling the core count brings limited energy efficiency benefit to prefill.
% When increasing core count, the energy consumption during prefill remains similar, while the energy consumed by decode worsens with large core counts (see \Cref{fig:dram-energy}).
More cores bring limited energy benefit for prefill, and do not benefit decode. (see \Cref{fig:dram-energy}).
% \Cref{fig:dram-energy}(b) shows that scaling core count hardly changes energy consumption during prefill, and that excessive core counts worsen (increases) energy consumption during decode.
% Also, an excessive core count may even degrade the energy efficiency for both prefill and decode.
Increasing core count consistently decreases the latency of prefill, 
but the decode latency improvement plateaus at a moderate core count (\Cref{fig:eval_all_combined}(d)).
As a result, the increased power consumption of the additional cores does not improve latency enough to consistently lower energy consumption past a threshold core counts (\Cref{fig:energy-breakdown}(b)).
%increasing the core count can greatly reduce the  static energy consumption of DRAM by reducing the execution time (see \Cref{fig:eval_all_combined} (d)). %during which static power is dissipated.
%When the core count is small ($\leq$256) 
%%and the execution latency is long, 
%the DRAM static energy contributes a substantial portion to the total energy, so it brings notable energy efficiency benefits for both prefill and decode.
% However, as core count grows further, these potential benefits diminish.
%
% \hl{Also, these benefits will be reversed by the energy overhead of the additional AI cores.}
% However, as we further increase the number of cores, the benefit will be offset by the energy overhead of additional cores.
%%%%%%%%%%%%%%%%%%%%%%%%%%%%% However, further increases to core count do not benefit energy overhead.
% First, the additional cores increase the static energy of the core components (SA, VU, and SRAM).
%For example, as the core count increases from 512 to 1024, their static power doubles.
For example, as core count doubles from 512 to 1024, the static power draw also doubles.
Decode latency is unchanged with these additional resources (\Cref{fig:eval_all_combined}), and prefill latency only decreases by 32\%, which is insufficient to offset the extra static power draw.

\observation{F1}{
Scaling DRAM bandwidth can improve energy efficiency for memory-bound workloads like LLM decoding, since it greatly reduces the execution time without increasing the static power of the chip. For compute-bound workloads like LLM prefill, scaling DRAM bandwidth brings little benefit.
% When the execution is bound by memory, scaling DRAM bandwidth improves energy efficiency by reducing the execution time without increasing static power consumption.
% Excessively scaling the core count, however, worsens energy efficiency, as it increases the chip power without decreasing the execution time.
}

% \observation{F2}{
% For compute-bound workloads, scaling core count brings little improvement in energy efficiency, since the reduced latency cannot offset the increased power.
% }

\observation{F2}{
Scaling the number of cores on a 3D AI chip may initially bring limited benefit to energy efficiency. However, as the core count further scales, the overall energy efficiency degrades, especially for memory-bound workloads, as its performance benefit cannot offset the increased power consumption. 
% For memory-bound workloads, increasing core count may significantly degrade energy efficiency. 
% the  of compute-bound workloads
% Memory-bound workloads may experience 
% Memory-bound workloads do not derive significant energy efficiency improvements 
}
%\hl{you should break this insights input multiple ones. This is actually an interesting insight, but do need profiling results to support it. }

% \observation{F2}{
% Increasing the core count when the workload is not compute bound ($<$256 cores in our experiment) will worsen energy efficiency, as it increases the chip power consumption without decreasing the execution time.
% Further scaling core count after the bottleneck has shifted from compute to memory ($>$256 cores in our experiment) will worsen energy efficiency, as it increases the chip power without decreasing the execution time.}
% Further increasing core count ($>$256 cores in our experiment) will worsen energy efficiency, as it increases the chip power without decreasing the execution time.
% Thus, to maximize the energy efficiency of 3D AI chips, architects can scale DRAM bandwidth and select the minimum core count that prevents a compute bottleneck.

% \observation{F2}{
% , as more cores increase both static and dynamic energy.

% The different hardware parameters have varied

% Continuing to scale core count after compute throughput matches the DRAM bandwidth will worsen energy efficiency...

\subsection{Operator-Level Breakdown}
\label{sec:explore:operator-breakdown}

\begin{figure}[t]
    \centering
    % \vspace{-.5ex}
\includegraphics[width=.99\linewidth]{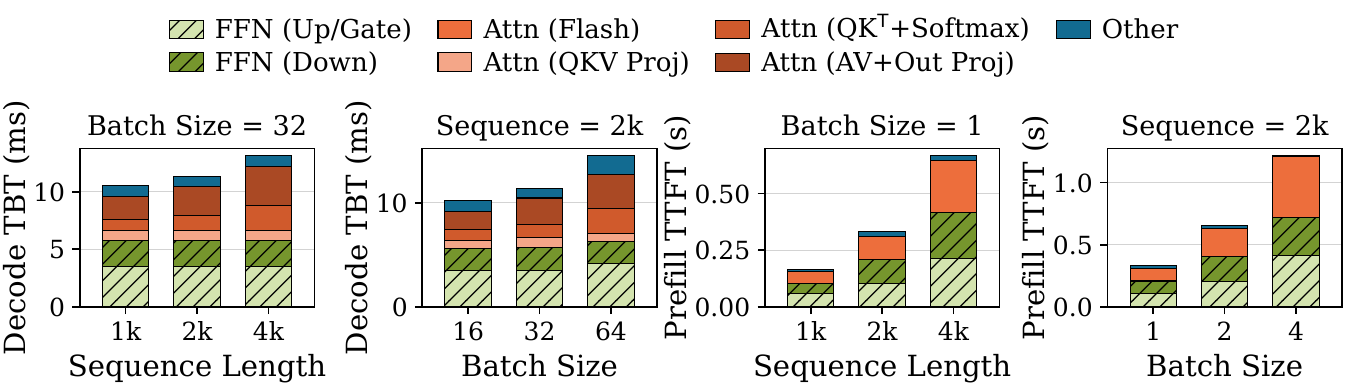}
    \vspace{-3.9ex}
    \caption{\hlA{Llama3-70B prefill/decode time breakdown by operator types, with various sequence lengths and batchsizes. The prefill enables FlashAttention\mbox{\cite{flashatt}} to save DRAM traffic.
    % , but it does not apply to decode due to existence of KV cache.
    }}
    \label{fig:op_break}
    % \vspace{-1ex}
\end{figure}

\hlA{
We show the operator level breakdown of LLM decode and prefill of Llama3-70B in \mbox{\Cref{fig:op_break}}.
For decode, the FFN latency is mostly constant as sequence length and batch size are varied. This is because decode is memory bandwidth bound, due to the weight loading. Therefore, adding extra computation %(i.e., increasing the number of tokens processed)
does not affect the overall execution time.
However, the attention operator overhead grows with both increased sequence length and batch size. 
This is because increasing either of these parameters will increase the number of KV cache entries, thus increasing memory bandwidth pressure for an already-memory-bound workload.
For prefill, both the FFN and attention computation time increase with sequence length and batch size, since prefill is already compute bound. 
%so increasing the total FLOP count by increasing the number of tokens processed will extend the time of all operators.
% We also present a sensitivity study over batch size and sequence length in  \mbox{\Cref{fig:op_break}}.
% TTFT increases approximately linearly with batch size across all models, as larger prompts and batches proportionally increase the amount of prefill computation. The impact is particularly pronounced for larger models such as Llama3-70B, whose higher parameter count amplifies the computational cost of processing long prompts and large batches.
% %
% In contrast, TBT is substantially less sensitive to sequence length. While longer contexts modestly increase TBT due to larger KV-cache accesses, the effect is significantly smaller than for TTFT. The exception is OPT-30B due to its high relative time spent on attention.
% %
% At small batch sizes, we also see that TBT increases only modestly with increased batch size. However, as batch size continues to grow, the KV cache loading latency becomes significant enough to offset the benefit of the batched FFN, and TBT begins to increase. more significantly. This effect is especially pronounced due to the relatively small size of the evaluated models and the un-optimized KV cache.
% Summarize all the hardware component contributions -- aggregate in S4.
% Summarize all sensitivity study results in S4 
}
% \vspace{-1ex}

%% file: related.tex
\begin{figure}[t]
    \centering
    \includegraphics[width=.99\linewidth]{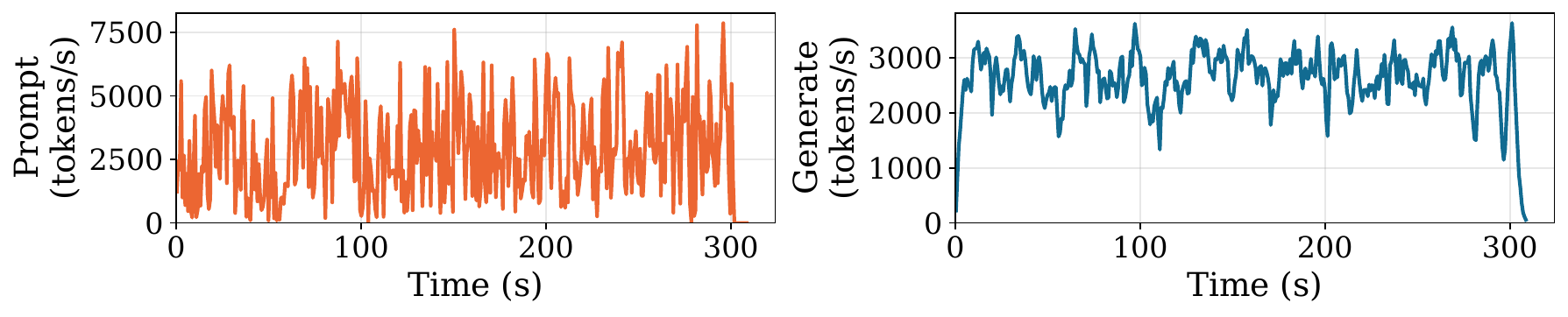}
    \vspace{-3.9ex}
    \caption{\hlA{Token throughput for Llama3-70B on a serving system with one prefill chip and one decode chip.}}
    \label{fig:real-workload-serving}
    \vspace{-1ex}
\end{figure}

\section{Discussion on \pname{} Extensibility}
\label{sec:discuss}

\noindent
\hlC{\textbf{Compiler support for 3D AI chips.}
As discussed in \mbox{\S\ref{sec:explore}}, compilers play a critical role in
unleashing the efficiency of 3D AI chips. Without a compiler capable of
producing optimized mappings, the takeaways in our design exploration may
change, though \pname{} could still be used to study 3D AI chips under that setting.}

% \hlC{We nonetheless expect compilers  for 3D AI chips to be realistic,
% since modern ML compilers already produce optimized, hardware-aware mappings for
% today's accelerators, and extending them is largely a matter of reusing existing
% mechanisms.} 
\hlC{However, as modern ML compilers already produce optimized, hardware-aware mappings for
accelerators, we expect comparable compiler support for 3D AI chips is highly feasible.}
% First, 
% Second, 
% as extending is largely a matter of reusing existing mechanisms.} 
\hlD{Moreover, some existing optimizations such as fusion \mbox{\cite{xla}} are hardware-agnostic and can be directly applied to 3D AI chips.
Others need only minor modifications: SPMD frameworks \mbox{\cite{jax2018github,alpa}} can partition computation across cores by treating each
core as a distributed node, and compilers for SRAM-based AI chips
\mbox{\cite{t10,waferllm}} can handle inter-core data mapping for a 3D AI chip, just as they map data to per-core SRAM in wafer-scale chips.}
\hlE{A few optimizations are truly unique to 3D AI chips, most notably the software-aware tensor-to-bank mapping we propose in \mbox{\S\ref{sec:explore:dram-bw}}, which compilers can adopt when mapping tensors on a 3D AI chip, to best utilize its high DRAM bandwidth.}

\noindent
\hlC{\textbf{Generality of our study.}
3D-stacked AI chips will likely continue to evolve in response to changing workloads.
% as the community better understands the requirements of target workloads. 
To account for this uncertainty, we evaluate a wide range of architecture parameters and study how design choices affect system behavior in \mbox{\S\ref{sec:explore}}. 
The sensitivity studies show that \mbox{\pname{}} is not tied to a single fixed architecture, but it can explore a broader design space. 
Moreover, regardless of the specific form future 3D AI chips take, compiler/hardware co-design will remain central to extracting performance from tightly coupled compute, communication, and stacked-memory resources. 
% We therefore expect \mbox{\pname{}} to remain relevant across a broad class of future 3D-stacked AI chip designs.
Thus as the architecture and workloads evolve, developers can utilize \mbox{\pname{}} to derive new relevant insights.
}

%\noindent
%\hlJ{I have no idea why you introduced a new simulator in your evaluation?! Don't do that, please, this will kill your paper.}
%\hlJ{We should not put these two paragraphs here...If you want to show an end-to-end performance, it should be an independent subsection.}

\noindent
\textbf{\hlA{Using \pname{} in end-to-end serving system.}}
\hlA{\pname{} can work as a chip simulation backend in serving system simulators to evaluate the end-to-end serving performance of 3D AI chips. \mbox{\Cref{fig:real-workload-serving}} shows \pname{} being used with LLMServingSim \mbox{\cite{cho2026llmservingsim}}, a cluster-level simulator to model an end-to-end PD-disaggregated environment running 
a trace generated from the ShareGPT dataset  \mbox{\cite{huggingfaceShibing624sharegpt_gpt4Datasets}}. 
It shows the performance behavior under realistic datacenter workloads.
The serving system employs continuous batching and prefix-caching to maximize throughput. We show the execution throughput
%execution across a wide range of batch sizes and sequence lengths 
as requests arrive and complete over time. 
}

\noindent
\hlE{\textbf{MoE support in \pname{}.}
The \pname{} interface is flexible enough to express MoE models with token routing via \mbox{\mvdata{}} and per-expert computing via \mbox{\comp{}}.
Users can first obtain the routing results (i.e., selected expert IDs) via a random function or by running the model.
Based on the expert tensor ID and the tensor placement, \pname{} can induce required \mbox{\mvdata{}} events. After that, \pname{} can be configured to enforce that these \mbox{\mvdata{}} events will not start until the routing operator finishes.}
% \vspace{-1ex}
% Notably, \pname{} may experience extended simulation time on MoE because dynamic routing reduces DRAM-access-pattern repetition, reducing the similar-event coalescing hit rate (\mbox{\S\ref{sec:design:e2e_simulation}}). 
% Additionally, MoE models are typically served using inter-chip expert parallelism to increase FFN batch size and decrease the model footprint. Without these optimizations, it is difficult to fully understand their performance.
% Thus, we leave the evaluation of MoE models to future evaluation.} 

\section{Related Work}
\label{sec:related}

\noindent\textbf{3D-stacked computing architectures.}
3D-stacked packaging is a mature technology that enables dies to be stacked vertically via TSVs \cite{xie2022stacking}.
It has been used to vertically stack compute dies, IO dies, or the on-chip cache on a chip~\cite{wuu20223d, amd-cdna4-whiteppaper, intel-clearwater-forest, han20233d, tain2025j3daitinydnnbasededge}. 
However, few studies have thoroughly investigated the 3D AI chip architecture that stacks DRAM banks on top of AI cores via numerous TSVs pins. 
%It has proven to be a promising approach to scaling the AI chip architecture. 
% Commercially, this technology has been used to increase on-chip cache capacity in CPU designs \hl{cite AMD}.
% traditional 3D stacking - should we also introduce term "die stacking"
%Existing 3D AI chip architectures use 3D integration to scale the number of CPU cores or on-chip cache~\cite{han20233d, tain2025j3daitinydnnbasededge}.
% \hl{Add NMP designs -- compare by saying we instead stack DRAM on the main accelerator to form a distributed memory space}
In this paper, we focus on such a 3D-stacked architecture with the goal of exposing and addressing its efficiency challenges. 
% We conduct a thorough study to explore the possible design choices to maximize the end-to-end efficiency of 3D AI chips.

\noindent\textbf{Techniques to break memory bandwidth wall.}
% \hl{group with memory wall - Memory technologies..} % HBM, 3D stacked memory 
% GPUs use HBM to get high memory bandwidth due to demands of AI workloads
% still facing a bottleneck, researchs explore solutions, one such solution is PIM
%While there are not yet any commercially available AI chips using 3D stacked DRAM, 
\looseness=-1
To overcome the memory bandwidth wall, PIM and near-memory-processing (NMP) architectures were proposed \cite{wang2023nicepim, kwon2022system, kim2021aquabolt, kim2016neurocube, h2llm, 3d-moe}. \hlE{They offload memory-intensive operations to the DRAM banks via the SIMT programming model \mbox{\cite{xie2023mpu, xie2022mpusim}}, allowing the high DRAM internal bandwidth to be exploited.}
% \hlcommon{A representative line of near-bank PIM pushes lightweight compute into the DRAM dies to harvest bank-internal bandwidth under a SIMT programming model \mbox{\cite{xie2023mpu, xie2022mpusim}}; in contrast, the 3D AI chip places full AI cores on a separate die and stacks DRAM banks over them, so its first-order challenges are non-uniform access to distributed banks over dedicated TSV buses and NoC contention rather than in-DRAM compute.} 
However, due to hardware resource constraints on memory devices, they cannot easily scale the compute capacity~\cite{PIM-CCA,kim2016neurocube}. \hlE{The 3D AI chip directly exposes more DRAM internal bandwidth to the compute dies in the form of high-density TSVs connected to distributed stacked DRAM banks on top of the compute dies.
This also allows the exposed bandwidth to scale linearly with the compute die area.}
However, it has unique efficiency challenges ($\S$\ref{sec:background}), which our study targets via fast, accurate hardware-software co-exploration. 
%and shares our findings throughout the paper. 

%often require explicit workload distribution between computation dies and PIM-enabled DRAMs, preventing existing LLM serving frameworks designed for GPUs/TPUs \cite{megatron, vllm, zheng2023efficiently, rajbhandari2020zero} from supporting PIM-enabled hardware.
%Existing works also explore NMP designs \cite{h2llm,3d-moe}.
%Stratum \cite{3d-moe} proposes a novel prototype design that stacks compute with Mono3D DRAM, but does not allow users to keep improving the design with software-hardware space exploration.
% Also, 
%H2-LLM \cite{h2llm} comprehensively explores how to efficiently map an LLM decoding dataflow to a 3D NMP chip.
%\pname{}, however, focuses on identifying and exploring a wider range of software/hardware factors that are crucial for the performance and efficiency of 3D AI chips. 
% due to these issues (describe a bit above) ... 3d stack/
% Some prior works have begun to examine 3D memory as a solution to the memory bandwith wayy \cite{3d-moe}, but these... \hl{todo}

% \noindent\textbf{Performance modeling for design space exploration.}
\noindent\textbf{Facilitating design space exploration.}
Design space exploration helps architects navigate large configuration spaces and understand complex performance/energy behaviors. % with different tradeoffs.
% Design space exploration is widely studied, and is commonly applied to problems where complex behavior makes selecting from a large number of choices difficult.
Typically, design  space exploration uses simulators and performance/energy models~\cite{parashar2019timeloop, li2019chip, wu2019accelergy, binkert2011gem5, accelsim, h2llm} to provide feedback on design choices.
In this paper, we develop \pname{}, a compiler-aware end-to-end simulation framework for 3D AI chips. Different from existing simulation tools (see Table~\ref{tab:compare_sim}), \pname{} can accurately reflect the unique features of the 3D-stacked AI chip architecture ($\S$\ref{sec:background:3d-arch}) and enable the exploration of both software and hardware optimizations.

%% file: conclusion.tex
\section{Conclusion}
\label{sec:conclusion}
%In this paper, we first identify unique performance challenges of 3D AI chips with stacked DRAM banks.
% To investigate the challenges, we developed \pname{}, a new simulator for fast yet accurate simulation of 3D AI chips.
%To navigate the challenges, 
In this paper, we develop \pname{}, a fast and  compiler-aware simulation framework that enables the exploration of 3D AI chips for LLM inference workloads. Our study reveals multiple insights that could shed light on the architecture design and implementation of new and emerging 3D AI chips. We will open source \pname{}. 
%Using \pname{}, we explore the design space of 3D AI chips, uncovering new software and hardware optimization opportunities for unlocking its potential in overcoming the memory bandwidth wall.
%We uncover new software and hardware optimization opportunities, unlocking their potential to overcome the memory bandwidth wall.

\section*{Acknowledgments}
We thank the anonymous reviewers at MICRO'26 for their insightful feedback. 
We thank Ingi Helgason, Jefferson Zhang, Saipranav Venkatakrishnan, and Yuqi Xue from the Systems Platform Research Group (Illinois PlatformX) at UIUC for proofreading our paper. We thank CloudLab~\cite{cloudlab} for providing compute resources that powered a substantial part of our exploration. This work was partially supported by the Hybrid Cloud and AI program at the IBM-Illinois Discovery Accelerator Institute (IIDAI), and NSF under the grants CAREER CNS-2144796 and CCF-2107470. 

%% file: appendix.tex
% Artifact Appendix for VoxelSim, following the cTuning AE template V20240722
% (https://github.com/ctuning/artifact-evaluation/blob/master/docs/template/ae.tex).
% Include this file in your paper (e.g. \input{appendix} after \appendix),
% or drop the body between the markers into the camera-ready source.

% Let long file names / paths typeset with \textstt break at _ - . / ?
% (prevents overfull lines and overlapping text in two-column layout).
\def\UrlBreaks{\do\/\do-\do_\do.\do=\do?\do\&}

\newpage
% IEEEtran gobbles the title of \section in the appendix; pass the title
% via \appendix[...] instead. acmart keeps the plain \appendix + \section.
\makeatletter
\@ifclassloaded{IEEEtran}{\appendix[Artifact Appendix]}{\appendix}
\makeatother

% Two-column appendix: allow looser line breaking so long \textstt names
% (scripts, flags, file names) do not overflow the column.
\sloppy
% \emergencystretch=1.5em

\makeatletter
\@ifclassloaded{IEEEtran}{}{\section{Artifact Appendix}}
\makeatother

%%%%%%%%%%%%%%%%%%%%%%%%%%%%%%%%%%%%%%%%%%%%%%%%%%%%%%%%%%%%%%%%%%%%%
\subsection{Abstract}

The artifact is \pname{}, a cycle-level simulator for LLM inference on
a 3D-stacked AI accelerator, together with the experiment drivers and plotting
scripts that reproduce the paper's key results (\Cref{fig:pareto,fig:sw_compiler,fig:eval_noc_topo,fig:dram_bw,fig:sw_dram,fig:eval_all_combined,fig:eval_smile,fig:dram-energy,fig:energy-breakdown,fig:op_break}).  \pname{} simulates systolic-array/vector compute, SRAM scratchpad,
3D-stacked DRAM, and on-chip network.
It reports latency, energy, power,
utilization, and area.
A one-command workflow
(\textstt{master\_runner.sh}) regenerates all simulation data, draws every
figure, and runs the Pareto-frontier design-space exploration on a commodity
Linux server; completed work can be cached, so interrupted runs resume
incrementally.

\subsection{Artifact check-list (meta-information)}

{\small
\begin{itemize}[leftmargin=*]
  \item {\bf Algorithm: } cycle-level simulation; coordinate-descent DSE
  \item {\bf Program: } Python~3 simulator and plotting scripts, Bash drivers
  \item {\bf Model: } LLM inference (llama2-13, llama3-70, opt-30, gemma2) and a vision model (dit-xl), decode and prefill
  \item {\bf Data set: } model specifications included in \textstt{models/}
  \item {\bf Run-time environment: } Linux (Ubuntu 20.04+), Python~3.10+
  \item {\bf Hardware: } x86-64 server, 32+ CPU cores (64 recommended), 128~GB RAM (256~GB recommended)
  \item {\bf Metrics: } latency (cycles/ms), energy (mJ), power (W), area (mm$^2$)
  \item {\bf Output: } 9 figure PDFs + 1 Pareto PNG in \textstt{figures/}
  \item {\bf Disk space required: } 30~GB (50~GB recommended)
  \item {\bf Time required to prepare workflow: } 10--15 minutes
  \item {\bf Time required to complete experiments: } 6--20 hours total (3--6\,h simulation data, $<$1\,min figures, 10--15\,h Pareto DSE)
  \item {\bf Publicly availability: } \url{https://github.com/yiqiliu2/voxelsim}
  \item {\bf Code licenses: } Apache-2.0
  \item {\bf DOI: } \url{https://doi.org/10.5281/zenodo.21433909}
\end{itemize}
}

%%%%%%%%%%%%%%%%%%%%%%%%%%%%%%%%%%%%%%%%%%%%%%%%%%%%%%%%%%%%%%%%%%%%%
\subsection{Description}

\subsubsection{How to access}

While the artifact can be downloaded from Zenodo at
\url{https://doi.org/10.5281/zenodo.21433909}, we recommend accessing the latest version from our GitHub:

\begin{Verbatim}[frame=single,fontsize=\footnotesize]
git clone https://github.com/yiqiliu2/voxelsim
cd voxelsim
\end{Verbatim}

The repository includes all necessary scripts and files. A full step-by-step guide is in the
repository \textstt{README.md}.

% \subsubsection{Hardware dependencies}

% A commodity x86-64 Linux server: 64+ CPU cores (128 recommended) and 128~GB
% RAM (256~GB recommended).  Prefill (full-sequence) simulations are the
% memory bottleneck; the pipeline auto-limits concurrency, and on machines with
% $\le$96~GB RAM the guide explains how to lower it
% (\textstt{--decode-parallel-limit}~2).

% \subsubsection{Software dependencies}

% \begin{itemize}
%   \item Linux (Ubuntu 20.04+), Bash
%   \item Python 3.10+ with \textstt{venv}
%   \item See \textstt{requirements.txt}
%   \item Optional: HotSpot, 3D-ICE, DSENT, ORION, and other existing simulators
%         (\textstt{bash} \textstt{scripts/setup_external_backends.sh}) for
%         thermal/NoC-power analysis.  \emph{Not required for AE} --- the
%         simulator falls back to built-in models and data.
% \end{itemize}

% \subsubsection{Data sets}

% No external data sets.  Model specifications are JSON files shipped in
% \textstt{models/}; pre-computed DRAM timing traces are shipped in
% \textstt{tsim_components/dram_cache/}.

% \subsubsection{Models}

% Four LLMs (llama2-13, llama3-70, opt-30, gemma2) in decode and prefill modes,
% plus one vision model (dit-xl) used by the Figures~15 and~17 rows, as specified
% in \textstt{models/}.

%%%%%%%%%%%%%%%%%%%%%%%%%%%%%%%%%%%%%%%%%%%%%%%%%%%%%%%%%%%%%%%%%%%%%
\subsection{Installation (5 minutes)}

\begin{Verbatim}[frame=single,fontsize=\footnotesize]
sudo apt-get update
sudo apt-get install -y python3.10-venv python3-pip
python3 -m venv venv
source venv/bin/activate
pip install -r requirements.txt
\end{Verbatim}

% This takes roughly 10--15 minutes.

%%%%%%%%%%%%%%%%%%%%%%%%%%%%%%%%%%%%%%%%%%%%%%%%%%%%%%%%%%%%%%%%%%%%%
\subsection{Experiment workflow}

One command runs everything end-to-end (sequential, resumable; completed simulations are cached and skipped):

\begin{Verbatim}[frame=single,fontsize=\footnotesize]
bash master_runner.sh
\end{Verbatim}

% \noindent
% It performs three stages:

\begin{enumerate}[leftmargin=*]
  \item \textbf{Simulation stage} (3--6\,h): run
        \textstt{run\_all\_modes.py} (parameter sweeps; compiler, dataflow,
        DRAM-mapping and row-conflict studies) plus the operator-breakdown
        sweep (\textstt{run\_seq\_batch\_sweep.py} and \Cref{fig:op_break}).    
        
        \newpage\vspace*{18pt}
        
  \item \textbf{Figure stage} ($<$1\,min): nine \textstt{draw\_*.py} scripts write
        the figure PDFs into \textstt{figures/}.
  \item \textbf{Pareto DSE stage} (10--15\,h): \textstt{dse\_pareto.py} runs
        area-constrained coordinate descent for decode/prefill, merging
        them into \textstt{pareto\_front.png} (\Cref{fig:pareto}).
\end{enumerate}

\noindent\textbf{Useful partial runs:} \textstt{master\_runner.sh} \textstt{--figures} (draw only),
\textstt{--dse} (DSE only), or \textstt{--dry-run} (data-completeness check).
A faster parallel alternative (\textstt{run\_figure\_data.sh}, 7 groups) and
per-mode/per-figure commands are documented in \S3 of \textstt{README.md}.

%%%%%%%%%%%%%%%%%%%%%%%%%%%%%%%%%%%%%%%%%%%%%%%%%%%%%%%%%%%%%%%%%%%%%
\subsection{Evaluation and expected results}

To verify for completion or check for progress, use:
\begin{Verbatim}[frame=single,fontsize=\footnotesize]
bash run_figure_data.sh --dry-run
\end{Verbatim}
After \textstt{master\_runner.sh} finishes, simulation logs will be under
\textstt{results/logs/} and figures under \textstt{figures/}:

\begin{itemize}[leftmargin=*]
  \item \Cref{fig:pareto}: \textstt{pareto\_front.png}
  \item \Cref{fig:sw_compiler}: \textstt{eval\_sw\_compiler.pdf}
  \item \Cref{fig:eval_noc_topo}: \textstt{eval\_noc\_topo\_combined.pdf}
  \item \Cref{fig:dram_bw}: \textstt{eval\_merged\_rowchange\_overhead\_*.pdf}
  \item \Cref{fig:sw_dram}: \textstt{eval\_sw\_dram\_map.pdf}
  \item \Cref{fig:eval_all_combined}: \textstt{eval\_lines\_all\_combined.pdf}
  \item \Cref{fig:eval_smile}: \textstt{eval\_core\_group\_decode\_smile\_curve.pdf}
  \item \Cref{fig:dram-energy}: \textstt{eval\_total\_energy\_mj\_total\_*.pdf}
  \item \Cref{fig:energy-breakdown}: \textstt{llama3-70\_energy\_breakdown\_absolute.pdf}
  \item \Cref{fig:op_break}: \textstt{eval\_op\_breakdown\_sweep.pdf}
\end{itemize}

%%%%%%%%%%%%%%%%%%%%%%%%%%%%%%%%%%%%%%%%%%%%%%%%%%%%%%%%%%%%%%%%%%%%%
\subsection{Experiment customization}

All drivers accept narrowing flags documented in \textstt{README.md}, such as \textstt{--modes} or \textstt{--prefill-modes} (mode subset),
\textstt{--model} or \textstt{--exclude-models} (model subset),
\textstt{--sweep-params} (parameter subset), and
\textstt{--decode-parallel-limit} (concurrency vs.\ host RAM usage).
The DSE accepts \textstt{--num-sweeps} and \textstt{--max-cycles} to trade
coverage for exploration time.
An experimental web UI (\textstt{web\_profiler/}) can browse
results and launch runs, but is optional and not part of this evaluation.

%%%%%%%%%%%%%%%%%%%%%%%%%%%%%%%%%%%%%%%%%%%%%%%%%%%%%%%%%%%%%%%%%%%%%
\subsection{Notes}

\begin{itemize}[leftmargin=*]
  \item Re-running any command resumes from cached outputs
        (existing logs skipped; DSE keeps a
        disk-backed cache).
  \item Run commands sequentially; do not launch the simulation groups as
        parallel background jobs.
  \item On RAM-constrained machines, follow the OOM recipe in \S4 of \textstt{README.md} (e.g., lower
        \textstt{--decode-parallel-limit}).
\end{itemize}

%%%%%%%%%%%%%%%%%%%%%%%%%%%%%%%%%%%%%%%%%%%%%%%%%%%%%%%%%%%%%%%%%%%%%
\subsection{Methodology}

Submission, reviewing, and badging methodology:

\begin{itemize}[leftmargin=*]
  \item \url{https://www.acm.org/publications/policies/artifact-review-and-badging-current}
  \item \url{https://cTuning.org/ae}
\end{itemize}

\newpage

% Existing open-source simulators for thermal, LLM service, DRAM, etc. Mandatory ones are included. To install optional ones, please see README.md.